\documentclass[11pt]{aastex}
\usepackage{emulateapj5,apjfonts}
\usepackage{onecolfloat}
\usepackage{epsf}

\shortauthors{Bell}
\shorttitle{Radio and Infrared Star Formation Rates}

\newcommand{\nd}{\nodata}
\newcommand{\ha}{{\rm H$\alpha$ }}

\newcommand{\hi}{{\rm H{\sc i} }}
\newcommand{\molhy}{{\rm H$_2$ }}
\newcommand{\hans}{{\rm H$\alpha$}}
\newcommand{\hii}{{\rm H\,{\sc ii} }}

\newcommand{\rf}{{radio--IR }}

\slugcomment{{\sc The Astrophysical Journal: } 
	accepted on December 5th, 2002 }

\begin{document}


\def\head{

\title{Estimating Star Formation Rates from Infrared
and Radio Luminosities: The Origin of the Radio--Infrared Correlation } 

\author{Eric F.\ Bell\altaffilmark{1}}
\affil{Steward Observatory, University of Arizona, 933 North Cherry Avenue, 
   Tucson, AZ 85721, USA}

\begin{abstract}
I have assembled a diverse sample of galaxies from the literature
with far-ultraviolet (FUV), optical, infrared (IR) and 
radio luminosities to explore
the calibration of radio-derived and IR-derived star formation (SF)
rates, and the origin of the radio-IR correlation.  
By comparing the 8--1000{\micron} IR, which samples dust-reprocessed starlight,
with direct stellar FUV emission, I show that the IR traces
most of the SF in luminous $\sim L_*$ galaxies but 
traces only a small fraction of the SF in 
faint $\sim 0.01 L_*$ galaxies.  If radio emission were a perfect
SF rate indicator, this effect would cause easily detectable 
curvature in the radio-IR correlation.  
Yet, the radio-IR correlation is nearly linear.  This implies that 
the radio flux from low-luminosity galaxies is substantially
suppressed, compared to brighter galaxies.  This is 
naturally interpreted in terms of a decreasing efficiency 
of non-thermal radio emission in faint galaxies.  
Thus, the linearity of the \rf correlation is a conspiracy:
both indicators underestimate the SF rate at low luminosities.  
SF rate calibrations which take into account this effect are
presented, along with estimates of the random and systematic
error associated with their use.  
\end{abstract}

\keywords{radio continuum: galaxies --- infrared: galaxies 
--- ultraviolet: galaxies --- dust, extinction --- 
galaxies: general --- cosmic rays }
}

\twocolumn[\head]
\altaffiltext{1}{Present address: Max Planck Institut f\"ur Astronomie,
K\"onigstuhl 17, D-69117 Heidelberg, Germany; \texttt{bell@mpia.de}}

\section{Introduction} \label{sec:intro}

Because ultraviolet (UV) and optical star formation (SF) rate indicators are 
so sensitive to dust 
\citep[see, e.g.,][]{k98,adel00,uit,gold02,bell02}, 
there has been 
much recent interest in using infrared (IR) and radio
luminosities in their stead 
\citep[see, e.g.,][]{blain99,flores99,haarsma00,hopkins01,mann02}.  
While IR emission
is straightforward to understand in the optically-thick
case for an intensely star-forming galaxy \citep{k98}, radio
emission is a highly indirect indicator of SF rate, relying
largely on the complex and poorly-understood physics of cosmic
ray generation and confinement \citep[see the excellent review by][]{condon92}.
Indeed, the strongest argument for radio luminosity as a SF rate
indicator has come from the astonishingly tight (a factor of two over 
5 orders of magnitude in luminosity) and arguably linear radio-IR correlation
\citep[e.g.,][]{dejong85,condon91,yun01}.  This close link between
the radio and IR luminosities of galaxies, even when 
normalized by galaxy mass \citep[e.g.,][]{fitt88,price92},
has often been used as a supporting argument for the 
efficacy and robustness of radio- and IR-derived SF rates.
In this paper, I compare UV, \hans, IR and
radio luminosities for a diverse sample of galaxies to 
demonstrate that neither the IR nor radio emissions
linearly track SF rate.  I argue that the tight, nearly linear
radio-IR correlation is a conspiracy: both the IR and radio luminosities
of dwarf galaxies significantly underestimate the SF rate.  Finally,
new SF rate calibrations which take into account this effect are
presented.

\subsection{The origin of IR and radio emission}

The primary prerequisite for an effective SF rate indicator is that it
reflects the mass of young stars in some well-defined way.
However, in practice, no SF rate indicator directly reflects the mass
of young stars.  It is useful at this stage to develop an intuition 
for the physical origin, strengths and limitations of IR and 
radio emissions as SF rate indicators.  For more in-depth discussion 
of these SF rate indicators, see \citet{k98} and \citet{condon92}.

\subsubsection{IR emission}

In systems with ongoing SF, the light from both newly-formed
and older stars can be absorbed by dust and reprocessed into the IR.  
There are thus two questions that should be addressed.
{\it i)} What are the relative contributions of old and young stars
to the IR luminosity?  {\it ii)} How much light is reprocessed
into the IR?  Put differently, what is the optical depth
of galaxies?  Because of my focus on the young stellar populations,
I will tend to focus on the optical depth of galaxies to light
from young stars.

The relative balance of dust heating by young and old stars 
in star-forming galaxies is a matter of some debate.  One observational
indicator of this balance is the temperature of the dust. 
Young stars in \hii regions heat up dust to relatively 
high temperatures (with a low 100{\micron} to 60{\micron} ratio of $\sim 1$).
Older stars in the field, and far-ultraviolet (FUV) light from field OB
associations (which have dispersed their natal clouds and so are 
relatively unattenuated in the FUV), heat the dust to much lower temperatures
\citep[100/60 $ \ga 5$; see, e.g.,][]{lonsdale87,buat96,walterbos96}. 
This difference between \hii region and diffuse dust temperatures
leads to a wide range in 100{\micron}/60{\micron} on galaxy-wide scales,
from $\sim 10$ for quiescent early-type spiral galaxies through 
to $\la 1$ for the most intensely star-forming galaxies.  This
suggests that earlier types are influenced more by old stellar
populations than later types; this is also supported by an analysis
of far-IR (FIR) and \ha data by \citet{sauvage92}.
For a `median' spiral
galaxy, the `cold' dust IR luminosity fraction is between 50\% and 70\%
\citep{lonsdale87,bothun89}. 
Despite this domination by cooler dust, 
more recently it has been argued
that the young, FUV-bright stars
provide the dominant contribution to the IR flux 
\citep[$\sim$70\%; see, e.g.,][]{buat96,popescu00,misiriotis01}.
This is because FUV light is absorbed much more efficiently than 
optical light per unit dust mass.  Thus, for a `median' spiral
galaxy, the IR luminosity comes from three components in 
roughly equal amounts: $\sim 1/3$ of the IR is from warm dust 
heated by FUV light from intense SF in \hii regions, another $\sim 1/3$ is 
cold dust heated by optical photons from the old and young stellar
populations, and the last $\sim 1/3$ of the IR is from cold
dust heated by FUV light from OB associations in the field 
\citep{buat96}.  
This interesting issue is discussed further in \S \ref{sec:optical}.

Given the apparent dominance of young stars in determining
the IR flux, it is appropriate to address the opacity of
dust to light from young stars.  Observationally,
there is a strong but scattered correlation between 
galaxy luminosity ($\sim$ mass) and dust opacity to 
UV or \ha light \citep{wang96,adel00,uit,hopkins01,sullivan01,buat02}.
Low-luminosity galaxies ($L/L_* \sim 1/100$) 
tend to have substantially less dust 
absorption and reddening than high-luminosity galaxies ($\sim L_*$).
Furthermore, these papers demonstrate that
low-luminosity galaxies have IR/FUV$\la 1$, meaning
that the IR emission of low-luminosity galaxies misses most of the SF.  
In contrast, many high-luminosity galaxies have IR/FUV$\gg 1$, implying
that the IR may be a relatively good SF rate indicator in this case
\citep[\S \ref{subsec:firfuv}]{wang96,buat02}.
This will have clear implications
for IR-derived SF rates, and correlations involving IR luminosities,
such as the \rf correlation.  This paper explores these implications in
detail.

\vspace{0.2cm}
\subsubsection{Radio emission}
\vspace{0.2cm}

Radio continuum emission from star-forming galaxies
has two components: thermal bremsstrahlung from ionized Hydrogen 
in \hii regions \citep[see, e.g.,][]{caplan86}, and 
non-thermal synchrotron emission from cosmic ray electrons
spiraling in the magnetic field of the galaxy
\citep[see, e.g.,][{ }for an excellent review]{condon92}.
Thermal radio emission has a spectrum $\propto \nu^{-0.1}$, whereas
non-thermal emission has a much steeper radio spectrum $\propto \nu^{\alpha}$,
where $\alpha \sim -0.8$ \citep[however, note that $\alpha$ can vary, and
even can vary with frequency;][]{condon92}.  
Because of this difference in spectral shape,
the relative contributions of the two emissions vary with 
frequency.  At lower frequencies $\la 5$\,GHz non-thermal 
radiation tends to dominate \citep[at 1.4\,GHz, the `standard model' of star-forming
galaxies attributes typically 90\% of the 
radio continuum flux of luminous spiral galaxies
to non-thermal emission;][]{condon92}.
Based on the standard model, thermal emission may dominate at frequencies 
$\ga 10$\,GHz \citep[see also ][]{price92}.
In addition, the relative fractions of thermal and non-thermal emission 
may depend on galaxy mass. Dwarf galaxies seem to have a lower
non-thermal to thermal emission ratio than normal spiral galaxies
\citep{klein84,klein91,klein91b,price92}, although estimating 
the balance of thermal and non-thermal radio emission is 
painfully difficult, and can be uncertain for even well-studied galaxies
at a factor of five level \citep{condon92}.  
This difference between dwarf and larger galaxies is often interpreted as a higher 
efficiency of cosmic ray confinement in physically larger (or more
massive) galaxies \citep[e.g.,][]{klein84,chi90,price92}.
For interesting discussions about the relative balance of 
non-thermal and thermal emission see \citet{condon92} and \citet{niklas97}.

\subsection{The Radio--IR correlation}

Given the complexity of the emission mechanisms of radio continuum
and IR light, it seems to be a miracle that the two fluxes are 
tightly correlated, with a scatter of only a factor of two.
Yet, when examined closely, the \rf correlation 
betrays the richness of the astrophysics which determine galaxies'
radio and IR luminosities.

The slope of the \rf correlation seems to depend on
galaxy luminosity.  Samples which are richer in relatively
faint galaxies ($L_{\rm IR} \la 10^{10} L_{\sun}$) 
tend to have steep \rf correlations in the sense that
$L_{\rm radio} \propto L_{\rm IR}^{\gamma}$ and $\gamma > 1$ 
\citep[e.g.,][]{cox88,price92,xu94}, whereas samples with a better 
representation of highly luminous galaxies 
($10^{10} L_{\sun} \la L_{\rm IR} \la 10^{12.5} L_{\sun}$)
tend to have slopes close to unity \citep[e.g.,][]{condon91,yun01}.
The differing behavior of galaxies as a function of luminosity
is beautifully illustrated in Fig.\ 5 of \citet{yun01} and Figs.\
1 and 2 of \citet{condon91}.
In addition, the slope depends on the radio frequency.  At low
radio frequencies $\la 5$\,GHz the slope tends to be steeper
than unity, whereas for higher frequencies the slope approaches
unity \citep[wonderfully illustrated in Fig.\ 2 of][]{price92}.

Workers in this field have typically sought to explain
the luminosity-dependent slope in terms of heating of dust by
older stellar populations, or non-thermal/thermal radio effects.
\citet{fitt88} and \citet{devereux89}
both subtracted off plausible contributions from old stellar populations
(using either FIR color $\theta$ or total IR luminosity as 
the constraint), which they found `linearized' the \rf correlation.
\citet{condon91} compared IR/radio with optical $B$/radio, finding
that IR-overluminous galaxies were overluminous in optical 
$B$-band ($\sim 4400${\AA}), which was interpreted as indicating
contributions from old stellar populations.  \citet{xu94} presented
a model which described the non-unity slope and some of the scatter 
of the \rf correlation in terms of the contributions of old stellar 
populations.  Similarly, a number of studies have investigated the 
r\^ole of non-thermal/thermal emission on the \rf correlation.
\citet{price92} and \citet{nikradfir} find that thermal radio
continuum (which directly reflects the SF rate) correlated linearly
with the IR luminosity; however, non-thermal emission had a steeper
correlation with IR luminosity with $\gamma \sim 1.3$.  
Taken together, the steepening of the \rf correlation 
at low IR luminosities, and with decreasing radio frequency, have been
interpreted as reflecting increasingly large contributions from 
old stellar population heating of the IR towards low IR luminosities, 
and non-thermal radio emission
which is non-linearly related to the SF rate.

\subsection{The goal of this paper}

In contrast with the commonly accepted picture, 
I argue that these interpretations of the \rf correlation 
are incomplete because they neglect the effect of dust opacity
on the IR emission of star-forming galaxies 
\citep[note that][briefly discussed the role of 
dust opacity, but not in a luminosity-dependent sense]{lisenfeld96}.  
The argument can be (but has not been, as yet) pieced together 
from results in the literature.
Empirically, high-luminosity galaxies are optically thick 
to FUV light, and so their IR emission reflects the SF rate
reasonably well.  In contrast, low-luminosity galaxies 
have low IR/FUV; therefore, their IR emission underestimates
the SF rate substantially \citep{wang96}.  Yet the \rf correlation 
is more or less linear \citep[e.g.,][]{yun01}.  Therefore, the radio emission
must be suppressed for low-luminosity galaxies.  This offers 
independent support to the argument that low-luminosity
galaxies tend to have substantially suppressed non-thermal radio
emission \citep[e.g.,][]{klein84,klein91,price92}.  Thus,
the \rf correlation is linear not because both 
emissions reflect SF rate perfectly, but because both
radio and IR emissions underestimate the SF rates of low-luminosity
galaxies in coincidentally quite similar ways.

In this paper, I assemble a sample of star-forming 
galaxies with FUV, IR and radio data to quantitatively
explore this basic argument for the first time.  
The result that low-luminosity galaxies have 
IR and radio emissions that underestimate their SF rates
is not new \citep[e.g.,][]{wang96,klein84,dale01}.  However, the 
assembly of an extensive star-forming galaxy sample with FUV, IR and radio 
data, the quantitative exploration of the consequences of this result on
the \rf correlation, and the presentation of SF rate calibrations
which take into account this effect, are new.

The plan of this paper is as follows.
I first investigate, in detail, dust opacity 
indicators, and trends in dust opacity with galaxy luminosity, 
in \S \ref{sec:sample}.  The galaxy sample is also introduced there.
In \S \ref{sec:fuv}, the \rf correlation is constructed, and 
the effect of dust opacity on the \rf correlation is estimated.
In \S \ref{sec:optical}, the 
effect of optical light from old stellar populations is
discussed.  In \S \ref{sec:radio}, deviations from the 
expected trends in the radio--IR correlation are used
to investigate the relationship between radio emission and 
SF rate.  In \S \ref{sec:disc}, new IR and radio SF rate calibrations
are presented and discussed.  In \S \ref{sec:conc}, 
I summarize the conclusions of this study.  In Appendix \ref{sec:app},
the FUV, optical, IR and radio data are discussed in more detail, and 
I present a table of galaxy photometry.  In Appendix \ref{app:model},
I present and discuss in detail a model for a luminosity-dependent
FUV optical depth.  
Sections 2.1--2.3, \S 4, and the appendices are less central to my discussion
of IR/radio SF rates, and may be skipped by casual readers.
A distance scale 
compatible with $H_0 = 75$\,km\,s$^{-1}$\,Mpc$^{-1}$ is assumed, and unless 
stated otherwise, I correct FUV and optical data for galactic foreground
extinction using \citet{sfd}.

\section{Understanding Dust Opacity in the Local Universe} \label{sec:sample}

In order to understand the implications of a correlation between
dust opacity and luminosity, it is important to understand both the 
overall amount of opacity and the increase in the amount of opacity
with luminosity in the wavelength regions that contribute
the most to the heating of dust.  Radiative transfer modeling 
coupled with observations \citep[e.g.,][]{buat96,fluxrat} 
suggest that the bulk of the energy that goes into heating 
the dust comes from non-ionizing FUV light, 
between 1216{\AA} and $\sim$3000{\AA}.  Clearly, then, the 
vital question that must be addressed is that of 
the optical depth of dust to FUV light in a wide range of
galactic environments.  

\subsection{Estimating the FUV optical depth of galaxies} \label{subsec:indic}

There are three established methods for estimating the 
FUV attenuation\footnote{Attenuation differs from extinction in 
that attenuation describes the amount of light lost because of dust at a given
wavelength in systems with complex star/dust geometries
where many classic methods for determining extinction, such as
color excesses, may not apply.} in star-forming galaxies.
\begin{itemize}
\item UV spectral slopes were found to correlate strongly with 
	optical and FUV extinction, as estimated using a variety of
	observational techniques, for {\it starburst galaxies} 
	\citep{calzetti94,calzetti95,meurer99}.  Because of its
	observational efficiency, this method has been extensively
	utilized at high redshift 
	\citep[see, e.g.,][{ }and references therein]{adel00}. However,
	it has recently been shown that UV spectral slopes are 
	poor attenuation indicators for other types of galaxy
	\citep{bell02,gold02}.  Thus, 
	I will not use this indicator in this paper.
\item Total \hi and/or  \molhy column density has been used
	to estimate dust content, and therefore FUV extinction 
	\citep[e.g.,][]{buat89}.  However, a number of factors, 
	such as metallicity (through the dust-to-gas ratio), 
	dust/star geometry, or extinction curve will introduce
	considerable scatter into any correlation between gas density
	and extinction.  This was confirmed by \citet{buat92} and
	\citet{xu97}.  Thus, I will 
	not use gas density-derived extinctions in this paper.
\item The TIR/FUV ratio, where TIR is the total 8--1000{\micron} luminosity
	and FUV$ = \lambda F_{\lambda} = \nu F_{\nu}$ at 
	$\sim 1550${\AA} (in this particular case) 
	is, in principle, an excellent indicator
	of the amount of FUV extinction.  
	This indicator of the direct vs.\ obscured light from young
	stars is a robust estimator of the FUV attenuation 
	$A_{\rm FUV}$, and is relatively unaffected by changes in 
	dust extinction curve, star/dust geometry and SF history
	\citep{fluxrat}.  
	The main limitations of this method are
	{\it i)} that the r\^ole of older stellar populations in 
	heating the dust is neglected \citep[although it can be accounted
	for by using a more realistic method to estimate $A_{\rm FUV}$,
	such as the flux-ratio method;][]{fluxrat}, and {\it ii)} that  
	some asymmetric star/dust
	geometries affect TIR/FUV (e.g., for a system with a dust
	torus, TIR/FUV would overestimate the FUV extinction and total 
	SF rate if viewed pole-on, and would 
	underestimate the total SF rate if viewed edge-on).  
	Despite its limitations,
	I will use this method in this paper, not least because
	a greater understanding of the IR emission is one of the central
	goals of this work.  This attenuation indicator has been used
	extensively before by, e.g., \citet{buat92}, \citet{adel00}, 
	\citet{buat02}, and 
	\citet{bell02}, and is directly related to the IR excess of
	\citet{meurer99}.
\end{itemize}

\subsection{The Sample} \label{subsec:sample}

\begin{figure*}[tb]
\epsfxsize=\linewidth
\epsfbox{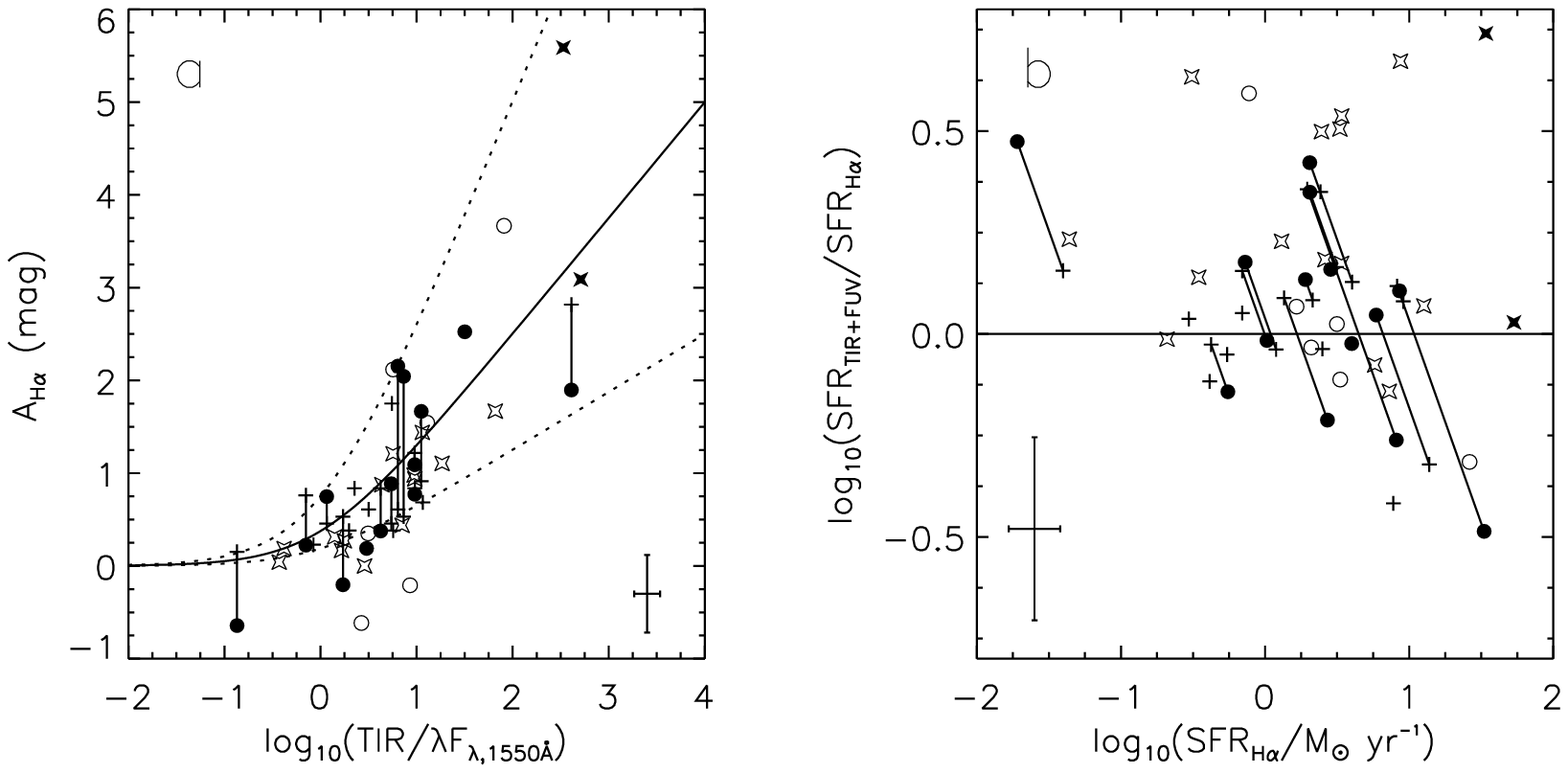}
\caption{\label{fig:hauv} Intercomparison of attenuation in the 
FUV and at \hans.  Panel {\it a)} shows the \ha attenuation
$A_{\rm H\alpha}$ against TIR/FUV.  Circles denote normal galaxies
with (uncertain) thermal radio-derived \ha attenuations \citep[filled circles are
UIT and FAUST galaxies, open circles are from][]{rifatto3}.
Crosses denote normal galaxies with Balmer-derived \ha attenuations.
Open stars denote starbursting galaxies and filled stars denote
ULIRGs (both have Balmer-derived \ha attenuations).  The
solid line is the relationship between the two that would be 
expected if TIR/FUV were a perfect indicator of FUV attenuation, and
assuming that the \ha attenuation is 1/2 of the FUV attenuation.  The 
dotted lines show the relationship if the \ha were 1/4 of, or the same as,
the FUV attenuation.  Panel {\it b)} shows the ratio of the TIR$+$FUV
SF rate and the \hans-derived SF rate, against the \hans-derived
SF rate.  The symbols are the same as in panel {\it a)}.  The solid
line denotes equality.  In both panels the typical errors are 
shown, and thermal/Balmer measurements for the same galaxy are 
connected. }
\end{figure*}

Because of my focus on exploring the r\^ole of dust opacity
and its effects on the \rf correlation, I have selected a sample of 249
galaxies that for the most part have {\it both} FUV and IR luminosities in 
the literature.  The sample properties, a more in-depth discussion 
of the systematic and random errors, and a table of the relevant
data is presented in Appendix \ref{sec:app}.  Here, I briefly
discuss only the most important points.

Normal, star-forming spiral and irregular galaxies were taken
from \citet[{ }100 galaxies]{rifatto3}, the {\it Far Ultraviolet Space Telescope} 
\citep[FAUST;][{ }75 galaxies]{deharveng94} and the 
{\it Ultraviolet Imaging Telescope} \citep[UIT;][{ }37 galaxies]{uit}.
FUV flux uncertainties from \citet{rifatto3} may be as large 
as 0.19 dex (larger galaxies, with substantially larger 
uncertainties, were removed from this sample).  Flux
uncertainties for galaxies from \citet{deharveng94} and \citet{uit}
are lower, $\sim 0.08$ dex.  Intensely star-forming galaxies
have also been added to the sample.
Starbursting galaxies \citep[{ }22 galaxies]{calzetti94,calzetti95} have
$10\arcsec \times 20\arcsec$ FUV fluxes from the 
{\it International Ultraviolet Explorer} (IUE).  To limit
the effects of aperture bias, I use only the FUV data for the 14 starburst
galaxies with optical diameters $\le 1.5\arcmin$.
Eight larger starbursts are included in this study, but are assumed to 
have no FUV data (i.e.\ only the optical, IR and radio data are used).
The typical measurement accuracy of the FUV fluxes is $\la$ 0.08 dex; clearly,
the systematic aperture bias is more of a concern.
Seven ultra-luminous infrared galaxies (ULIRGs) \citep{gold02} were
added to the sample, with a typical FUV accuracy of 
$\la$ 0.12 dex.  
Eight Blue Compact Dwarves (BCDs) have been added to the sample
also to check for consistency with other galaxy types 
\citep{hopkins02}.
FUV fluxes are quoted at wavelengths within 100{\AA} of 1550{\AA}:
the error introduced by assuming that they are all at 1550{\AA} is $\sim 6$\%.
In the remainder of this paper, these $\sim$1550{\AA} fluxes are
denoted as `FUV' fluxes or `1550{\AA}' fluxes.  
Note also that galaxies classified as Seyferts in 
the NASA/IPAC Extragalactic Database have been removed from 
the sample.

IR data at 12--100{\micron} was taken from the
{\it Infrared Astronomical Satellite} (IRAS) for 245 galaxies, and is 
accurate to better than 20\% in both a random and systematic 
sense \citep{rice88,soifer89,moshir90,tuffs02}.  Total
IR 8--1000{\micron} (TIR) and 42.5--122.5{\micron} (FIR) fluxes 
were derived from the IRAS data, and are accurate to 
$\sim$30\% (see the discussion in Appendix \ref{sec:app}).  
In this paper, I adopt the TIR 8--1000{\micron} fluxes, in
order to more accurately probe the true relationship between
the amount of light reprocessed by dust into the IR with the radio
emission \citep[e.g.,][find a `normal' FIR-to-radio ratio
for the starbursting SBS\,0335$-$052 but a large TIR-to-radio
ratio because of a large population of hot dust]{dale01}.
The 42.5--122.5{\micron} FIR fluxes are only used as 
a consistency check; all the results in this paper apply to both
TIR and FIR fluxes, taking into account that FIR$\sim 0.5$\,TIR
(see Appendix \ref{sec:app} for more details).

Optical data were carefully taken from the literature, using
the NASA/IPAC Extragalactic Database and NASA's Astrophysics Data System.
Optical data for 247 galaxies was taken from a variety of sources
and is accurate to $\la 0.2$\,mag in most cases, and to 
$\la 0.5$\,mag in all cases.  Radio data for 166 galaxies at 1.4\,GHz were, for the most part, 
taken from the NRAO VLA Sky Survey \citep[NVSS;][]{c98}.
NVSS data were taken for 159 galaxies from \citet{condon02}, 
\citet{hopkins02}, and \citet{g99i,g99ii} in that order of preference.  
Additional data at frequencies between 1.4 and 1.5\,GHz (translated
to 1.4\,GHz assuming a $\nu^{-0.8}$ non-thermal spectrum) were 
taken from other sources for seven galaxies which were not in 
the above catalogs, but were important to have in the sample because
of their properties (ULIRGs or interacting pairs), or because
they had measured thermal radio fractions.  The radio data were extensively
and exhaustively cross-checked with many other radio catalogs, and 
were found to agree to within 20\% in most cases.  Galaxies with 
highly contentious radio fluxes (by more than a factor of three) 
were removed from the sample. 

How does sample selection affect my results?  Clearly, the sample
is selected very inhomogeneously to have FUV, IR and (as much as possible)
radio data.  This makes the effects of sample selection difficult
to assess.  I would argue that the effects of sample selection 
are minimal in this paper, partially because of the inhomogeneously-selected
sample.  In particular, care was taken to include both normal 
and starbursting galaxies across a wide range in luminosities, limiting
that particular source of bias.  Furthermore, the trends (or lack thereof)
explored in this paper are established over 4--5 orders of magnitude 
in galaxy luminosity, and are impressively {\it quantitatively} consistent
with other datasets which were selected in totally independent
ways \citep[see, e.g.,][]{wang96,yun01,price92}.  Taken together, 
this argues for a minimal r\^ole for selection effects in driving 
the results of this paper, although further work with independently-selected
samples in the future (for example, from the {\it Galaxy Evolution Explorer}
or {\it Space Infrared Telescope Facility}) will prove to be 
the ultimate test of selection effects and systematic error in 
this and other investigations of the \rf correlation.

\subsection{Comparing TIR/FUV with \hans-derived extinctions} 
	\label{subsec:hafuv}

\begin{figure*}[tb]
\epsfxsize=\linewidth
\epsfbox{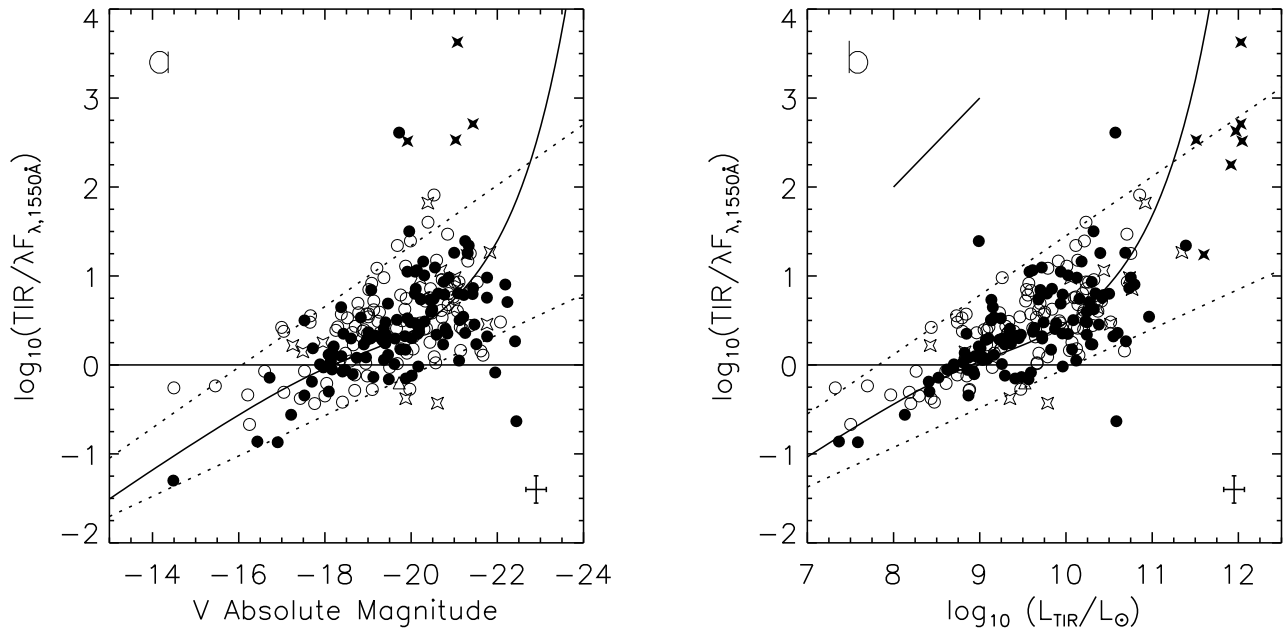}
\caption{\label{fig:firfuv} Trends in TIR/FUV with galaxy luminosity.  
Panel {\it a)} shows TIR/FUV as a function of optical $V$ band 
luminosity.  Panel {\it b)} shows TIR/FUV as a function of TIR
luminosity.  Symbols are as in Fig.\ \ref{fig:hauv}.  A single BCD
is also plotted \citep[with FUV data from][]{rifatto3} as an open triangle.
Dotted lines delineate the locus of the data points in panel
{\it a)}, and are translated into panel {\it b)} using the 
$V$ band--TIR correlation discussed in the text.  The solid
line is the model FUV opacity--luminosity correlation (driven 
by the gas density--luminosity and metallicity--luminosity
correlations).  The short solid line in panel {\it b)} is 
the effect of an order of magnitude increase in TIR luminosity
at fixed FUV luminosity.  Non-detections are not plotted
here; see, e.g., Fig.\ 1 of \protect\citet{wang96}
for a version of panel {\it b)} which includes TIR non-detections
(showing that galaxies have TIR/FUV $\sim 0.1$ down to 
$L_{\rm TIR} \sim 10^6\,L_{\sun}$). }
\end{figure*}

The crucial parameter of interest at this point is the opacity
of dust to the FUV light of a given galaxy.  Therefore, the vital
question that must be addressed is to what level can IR/FUV be said
to represent the true FUV opacity?  This question is difficult
to address directly; however, a number of papers have examined
IR/FUV indirectly, in some detail.
\citet{meurer99} show that it correlates well
with a number of other extinction indicators for starburst galaxies.
\citet{fluxrat} use radiative transfer models to show IR/FUV's robustness
at a theoretical level.
\citet{lmc} shows that it correlates well with other extinction 
indicators for Large Magallanic Cloud \hii regions.  Finally, 
\citet{buat02} show correlations between IR/FUV and Balmer-derived
\ha attenuation, and show a relatively good correspondence between 
IR$+$FUV SF rates and attenuation-corrected \hans-derived SF rates.

A galaxy's true SF rate is 
impossible to measure without using detailed, complete stellar color-magnitude
diagrams.  Thus, I must approach this question using an indirect 
two-pronged approach \citep[similar to that of][]{buat02}, where I 
inter-compare SF rate indicators.  First, I compare \ha attenuation
with TIR/FUV for galaxies with estimates of \ha attenuation 
as a sanity check.  Then, I compare SF rates derived using 
extinction-corrected \ha against TIR$+$FUV (essentially extinction-corrected
FUV) to assess how well the SF rates match.

I derive \ha attenuations 
in two ways.  {\it i)} The ratio of thermal radio to \ha light is a known 
constant, to first order, therefore deviations in that ratio give a 
robust constraint on the \ha attenuation.  The thermal radio fraction 
is estimated by fitting the radio spectral energy distribution with 
contributions from thermal and non-thermal emission.  However, the 
non-thermal emission dominates at most radio frequencies, making
a robust and reliable determination of thermal radio flux 
highly challenging at this time \citep{condon92}.  
{\it ii)} The Balmer decrement (\hans/H$\beta$) is again constant to first
order in the absence of dust, and is easier to measure, 
but suffers from optical depth effects
\citep[see, e.g.,][{ }for fuller 
discussions of these issues]{caplan86,uit,lmc}.
Despite the substantial limitations of both techniques, I choose to 
compare the TIR and FUV with attenuations derived using both approaches
because the goal is to assess the efficacy of 
TIR/FUV: corrupted \ha attenuation
estimates will only make the TIR/FUV look worse.

I take Balmer decrements for 
14 starburst galaxies with diameters $< 1.5{\arcmin}$ 
from \citet{calzetti94}, and supplemented these with
Balmer decrement measurements for two of Goldader et al.'s ULIRGs
\citep{wu98}.  For normal galaxies, I use 
thermal radio-derived \ha extinctions, and some Balmer 
decrements which have been averaged over a number of \hii regions
in each galaxy \citep[taken from][]{uit}.
Thermal radio fluxes were taken from \citet{niklas97} for 6
Rifatto et al. galaxies, and from \citet{uit} for 12 UIT and two FAUST
galaxies \citep[most of their thermal radio fractions, were, in turn,
from][]{niklas97}.  

The results are shown in Fig.\ \ref{fig:hauv}.  Panel {\it a)} of 
Fig.\ \ref{fig:hauv} shows the comparison of \ha attenuation and
TIR/FUV (this is similar to panel {\it b)} of Bell \& Kennicutt's
(2001) Fig.\ 4, and Buat et al.'s (2002) Fig.\ 2).  In common
with those studies
and \citet{calzetti94}, I find that \ha attenuation 
and TIR/FUV are correlated with scatter, and that the \ha attenuation 
is $\sim 1/2$ of the FUV attenuation (though with much scatter).  
Note that the expectation from a dust foreground 
screen model is that \ha attenuation
would be $\sim 1/4$ of the FUV attenuation. 
This discrepancy of a factor
of two from the screen model is consistent with the interpretation
of \citet{calzetti94}, who postulate that nebular line emission is 
attenuated by roughly twice as much dust as the stellar continuum
\citep[see also][]{charlot00}.

Of course, it is not clear, {\it a priori}, what a correlation between
\ha and FUV attenuations really tells us.  One can easily imagine
pathological dust geometries which will essentially decouple
\ha and FUV attenuation.  A complementary, and perhaps more stringent, 
test is to compare attenuation-cor\-rec\-ted \hans-derived SF rates
with SF rates determined from the combined TIR$+$FUV emission (essentially
the same as extinction correcting FUV with TIR/FUV).  Statistically, 
these SF rates should be equal, even though the timescales of 
\ha and FUV emission differ by nearly an order of magnitude (5 Myr vs.\
50 Myr).  This comparison is conservative: although I know
that the \ha extinction corrections are deficient
in both random and systematic ways, I nevertheless attribute
any mismatches to TIR$+$FUV in an effort to constrain 
the accuracy of the TIR$+$FUV methodology.

A comparison of TIR$+$FUV SF rates and attenuation-cor\-rec\-ted
\hans-derived SF rates is shown in panel {\it b)} of 
Fig.\ \ref{fig:hauv}.  SF rates are estimated using the 
SF rate conversion factors given by \citet{k98}.
Normal galaxies with thermal radio-derived
\ha attenuations (circles) have statistically equal SF rates derived
from the TIR$+$FUV and \hans, with less than a factor of two scatter.
Normal galaxies with Balmer-derived \ha attenuations (crosses) have 
statistically equal SF rates (TIR$+$FUV vs.\ \hans) also, with 
a factor of 1.5 scatter.
Starburst galaxies and ULIRGs (stars) have SF rates which are 
a factor of two higher in TIR$+$FUV than in the Balmer attenuation-corrected
\hans, with again less than a factor of two scatter.  It is 
unclear, at this stage, why starbursting galaxies appear to 
have lower Balmer-corrected \hans-derived SF rates (compared to 
the TIR$+$FUV case) than normal galaxies.  This offset was 
also observed by \citet{buat02}.  This may be an aperture
effect (FUV and extinction-corrected \ha are in the {\it IUE} 
aperture, whereas the TIR is total), although there is no trend
in TIR$+$FUV/\ha with galaxy size.  Alternatively, 
it is possible that differences
in star/dust geometry could cause an effect of this
type \citep{buat02}, as there are strong suspicions that
the star/dust geometries of the two galaxy types are different
\citep{bell02}.  It is also possible that integrated 
galaxy spectra (as used for starbursts and ULIRGs) systematically
underestimate the true \ha attenuation because of radiative
transfer effects and/or contamination from diffuse ionized gas.
Without more Balmer decrement and thermal radio data
for a reasonably-sized sample of starburst and normal galaxies
it is impossible to unambiguously track down the origin of 
this factor-of-two discrepancy.

Either way, this comparison is extremely encouraging: assuming very
conservatively that {\it all} of the scatter in TIR$+$FUV vs. 
\ha SF rates is from the TIR and FUV (and not from the \ha extinction
correction, intrinsic differences in FUV/\ha because of bursts of 
SF, mismatches in the FUV, IR and \ha SF rate calibrations, etc.), 
I find that TIR/FUV reflects
the attenuation in the FUV to better than a factor of two in both 
a random and a systematic sense, and is perhaps much more accurate.
\footnote{Later on I examine the r\^{o}le of old stellar populations
in heating the dust, correcting the TIR for a contribution from the 
optical $V$ band light from a galaxy.  Including this effect
in this analysis does not significantly affect the conclusions; TIR/FUV
is still found to be a good attenuation indicator to much better
than a factor of two in a systematic and random sense.}

\subsection{Trends in TIR/FUV} \label{subsec:firfuv}

I have argued that TIR/FUV is the FUV attenuation indicator of
choice on both modeling and observational bases.
Now, following, e.g., \citet{wang96}, \citet{buat99}, and
\citet{adel00} I proceed to explore TIR/FUV for my diverse sample of galaxies.
I show the correlation between TIR/FUV and optical luminosity, and TIR/FUV
and TIR luminosity, in panels {\it a)} and {\it b)} of 
Fig.\ \ref{fig:firfuv}.  There is a scattered but strong correlation between
the ratio of total TIR 8--1000{\micron} to FUV (defined as 
$\lambda F_{\lambda}$ at $\sim 1550${\AA}) and luminosity
in either the optical $V$ band or in the IR.  

The dotted lines 
encompass the majority of the points in panel {\it a)}, 
and translate into panel {\it b)} using the 
least-squares regression of TIR on $V$-band absolute magnitude:
$\log_{10} ({\rm TIR}/L_{\sun}) = 9.83 - 0.511 M_V$.
The solid line shows a highly simplistic model which 
links $V$-band luminosity and the optical depth of 
dust in the FUV.  The main assumptions are that:
{\it i)} the dust-to-gas ratio is proportional only
to the metallicity, and {\it ii)} the dust optical depth
is proportional to the dust per unit area, which 
therefore is proportional to the gas surface density.
The dust optical depth increases with 
galaxy luminosity because of the typically higher gas densities
and metallicities of more luminous galaxies.  
Curvature in the model behavior primarily comes from 
my somewhat crude derivation of the gas density--luminosity
correlation (which is bootstrapped from the gas fraction--luminosity
and stellar surface density--luminosity correlations).  The model is 
discussed in more detail in Appendix \ref{app:model}.

It is clear that TIR/FUV increases, on average, by over 1.5 orders 
of magnitude between low-luminosity galaxies
at $V \sim -16$ ($L \sim 1/100 L_*$) and high-luminosity 
galaxies at $V \sim -22$ ($L \sim 3 L_*$).  
These data are quantitatively consistent with (largely) independently-selected
samples of galaxies \citep[e.g.,][]{wang96,buat99,adel00}.
The main advantage of this sample is its size: it is slightly larger
than the local samples of 
\citet{wang96}, \citet{buat99}, and \citet{adel00} combined.
It is interesting
to note that this increase in dust opacity is reasonably well-tracked, 
in the mean, by the simple model which was presented above.  Furthermore,
while TIR/FUV$ \gg 1$ for most high-luminosity galaxies,
for lower-luminosity galaxies TIR/FUV$ \la 1$, meaning
that many low-luminosity galaxies are optically thin
in the FUV.  Thus, the IR
luminosity in low-luminosity galaxies will underestimate the
SF rate substantially (remember that TIR$+$FUV is not a bad proxy 
for SF rate; \S \ref{subsec:hafuv}).  

\section{The Effects of Dust Opacity on the Radio--IR Correlation} 
\label{sec:fuv}

Assuming that radio is a `perfect'
SF rate indicator (i.e.\ radio $\propto$ SF rate), the 
systematic depression of IR emission in low-luminosity galaxies
should be easily visible in the 
\rf correlation, because of its tightness.  In this section, 
I assess the effects of the optically-thin low-luminosity
galaxies on the \rf correlation.

\subsection{The Radio--IR correlation} \label{sec:rf}

\begin{figure}[tb]
\vspace{-0.5cm}
\hspace{-0.5cm}
\epsfbox{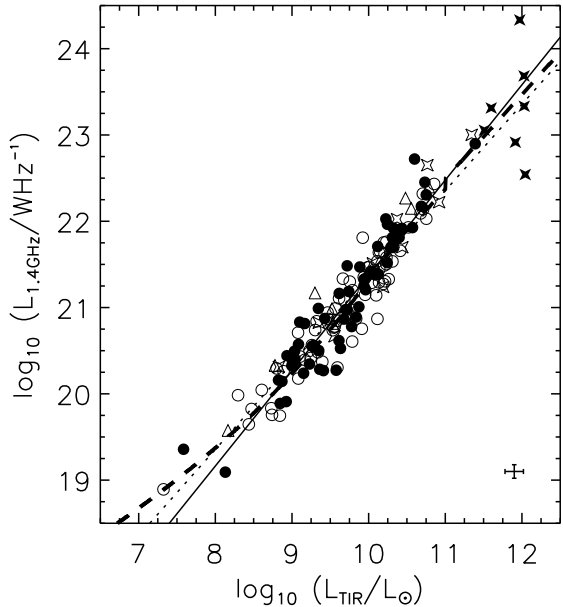}
\vspace{-0.2cm}
\caption{\label{fig:radfir}
The \rf correlation for a total of 162 galaxies.  
Normal, star-forming galaxies are
plotted using circles \citep[filled circles are from UIT and FAUST, and 
open circles are from][]{rifatto3}.
Intensely star forming galaxies are denoted by stars
(filled stars are ULIRGs and open stars are starbursts).
A comparison sample of BCDs
are shown as open triangles.  A representative
error bar is shown in the bottom right-hand corner.  
Forward and bisector fits to the data are shown
by dotted and solid lines respectively.  The thick dashed line
shows the trend predicted by the final 
SF rate calibrations (see \S \ref{sfr}).
}
\end{figure}

\begin{figure*}[tb]
\epsfxsize=\linewidth
\epsfbox{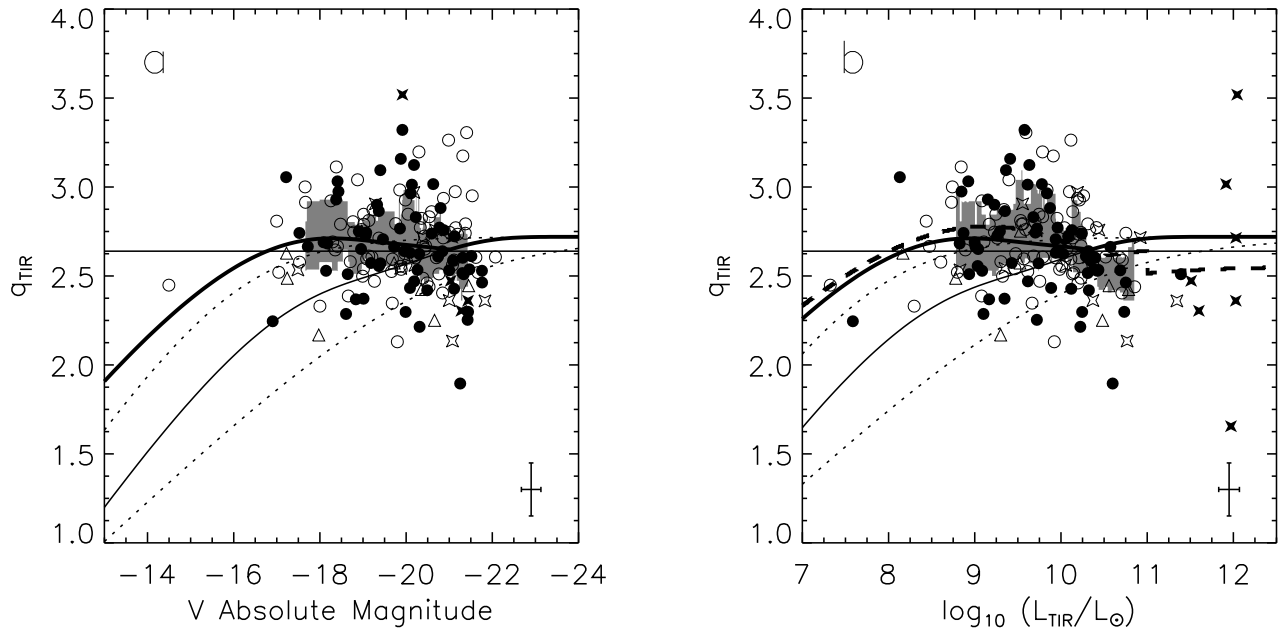}
\caption{\label{fig:qtrend} Trends in $q_{\rm TIR}$ with galaxy luminosity.  
Panel {\it a)} shows $q_{\rm TIR}$ against $V$-band absolute magnitude, 
and panel {\it b)} shows $q_{\rm TIR}$ against TIR luminosity.  
Symbols are the same as in Fig.\ \ref{fig:firfuv}.  
The shaded area shows the upper and lower quartiles as a function 
of luminosity: this shows, in a less noisy fashion, any trends between
$q_{\rm TIR}$ and galaxy luminosity.
The effect of trends in TIR/FUV with luminosity are plotted
as thin dotted (the limits on TIR/FUV as a function of luminosity)
and thin solid (the model TIR/FUV with luminosity) lines.  If radio
SFR were a perfect SF rate indicator, there should be a trend in 
$q_{\rm TIR}$ which follows the general trend of the thin dotted and solid lines. 
The thick solid line shows the final model presented
in \S \ref{sec:radio}.  The thick dashed line in panel {\it b)}
shows the trend predicted by the final 
SF rate calibrations (see \S \ref{sfr}). }
\end{figure*}

The \rf correlation for all of the sample galaxies 
\citep[plus BCDs from][{ }shown as open triangles, to check for
consistency with other galaxy types]{hopkins02}
is shown in Fig. \ref{fig:radfir}, where the TIR
8--1000{\micron} extrapolated flux is being used.  
It is clear that the \rf correlation is both 
superbly tight (the scatter in the TIR/radio
ratio is 0.26 dex, or less than a factor of two) and nearly linear.
A forward fit, the ordinary least squares regression
of TIR on radio, yields a slope of $1.05 \pm 0.04$ (this
type of fit is suitable for, e.g., predicting the radio 
flux given the TIR flux).  A bisector fit, the average of 
the forwards and backwards fits, yields a slope of
$1.10 \pm 0.04$ \citep[this type of
fit is more suitable in cases where the intrinsic correlation
is being sought, and measurement errors are dominated by 
intrinsic scatter;][]{isobe90}.  These data are
consistent with the much larger sample of 1809 galaxies studied by
\citet{yun01}: they recover a forward fit slope of 
$0.99 \pm 0.01$ and a scatter of 0.26 dex for a comparison
of 60{\micron} and 1.4\,GHz fluxes.  Furthermore, they
find a tendency for low-luminosity galaxies to be somewhat underluminous
in the radio, which I also recover (this is the main effect
which drives the slightly steeper slope of the bisector fit).

A few points deserve mention at this stage.  Firstly, there is
a somewhat increased dispersion for very high luminosity galaxies.  This 
is consistent with a number of other studies
\citep[e.g.,][]{condon91,yun01,bressan02} and is discussed later
in \S \ref{sec:hilo}.  Secondly, I use
TIR 8--1000{\micron} extrapolated flux.  In this respect, I differ
from most other studies which plot either the 60{\micron} luminosity
\citep[e.g.,][]{yun01}, or the FIR 42.5--122.5{\micron} luminosity 
\citep[e.g.,][]{cox88,condon91,xu94}.  This difference in IR
luminosity estimation technique does not change the slope or scatter of the
\rf correlation significantly (the forward fit slope for the 
60{\micron} case is 1.01$\pm$0.04, 
42.5--122.5{\micron} case is 1.04$\pm$0.04, and the scatter is 0.25 dex
in all cases).

\subsection{Consequences of trends in IR/FUV with galaxy 
luminosity} \label{sec:conseq}

A complementary way of examining the \rf correlation is 
by constructing the TIR/radio ratio $q_{\rm TIR}$.  The quantity
$q_{\rm TIR}$ is defined as:
\begin{equation}
q_{\rm TIR} = \log_{10} \left( \frac{\rm TIR}{3.75 \times 10^{12} 
	{\rm W\,m^{-2}}} \right) - \log_{10} \left( 
	\frac{S_{\rm 1.4\,GHz}}{\rm W\,m^{-2}\,Hz^{-1}} \right) ,
\end{equation}
where $S_{\rm 1.4\,GHz}$ is the 1.4\,GHz radio flux
\citep[e.g.,][]{condon91}.  I define
$q_{\rm TIR}$ as the ratio of the {\it total} 8--1000{\micron} IR luminosity
to the radio power, as opposed to the 42.5--122.5{\micron} FIR
luminosity which is usually used in defining $q$.  
The median value of $q_{\rm TIR}$ is 2.64$\pm$0.02 for 162 galaxies with 
IR and radio data and no signs of AGN, and the 
scatter is 0.26 dex.  For reference, the median $q$ value defined using
the 42.5--122.5{\micron} flux is 2.36$\pm$0.02, with a scatter of 0.26 dex, 
in excellent agreement with the mean $q$ of 2.34$\pm$0.01 and scatter
of 0.26 dex of \citet{yun01}.  

I show the trends in $q_{\rm TIR}$ with galaxy luminosity in Fig.\
\ref{fig:qtrend}.  Panel {\it a)} shows $q_{\rm TIR}$ as a function 
of $V$-band absolute magnitude, and panel {\it b)} shows $q_{\rm TIR}$ against
TIR luminosity.  The shaded region shows the `running' upper
and lower quartiles of the
data\footnote{For a given galaxy's luminosity, the $q_{\rm TIR}$ values of the $\pm$10
galaxies in the luminosity ranked list were extracted.  The upper and
lower quartile were calculated, and plotted as the shaded region.}.
There are only gentle trends, if any, in $q_{\rm TIR}$ with galaxy luminosity, 
such that lower luminosity galaxies have somewhat higher values of 
$q_{\rm TIR}$ (this is particularly visible in the $q_{\rm TIR}$--TIR luminosity relation).  
This slight tendency for lower-luminosity galaxies to have somewhat
higher $q_{\rm TIR}$ is what drives the slight non-linearity in the 
bisector fit of the \rf correlation, and was also seen by 
\citet{yun01} in their sample of 1809 galaxies.  

This slight trend towards higher $q_{\rm TIR}$ at lower luminosity, 
or lack of trend, is in stark contrast to the trends in 
$q_{\rm TIR}$ which would be expected {\it if radio were a perfect
SF rate indicator}.  In Fig.\ \ref{fig:firfuv}, the dotted lines
outlined the locus of the majority of the data points, and the 
solid line described the overall trend in TIR/FUV with galaxy luminosity
reasonably well.  The thin solid and dotted lines in Fig.\ \ref{fig:qtrend}
are the mapping of the trend in TIR/FUV with luminosity 
onto $q_{\rm TIR}$, assuming only that the radio $\propto$ SF rate (the 
details of the translation are discussed in Appendix \ref{app:model}).
If radio $\propto$ SF rate, then $q_{\rm TIR}$ should decrease by 
at least 0.2 dex over the luminosity range over which there are 
decent statistics.  This decrease is not seen: in fact, a slight
increase in $q_{\rm TIR}$ with decreasing luminosity is observed.  
Given {\it i)} the robustness of TIR/FUV as
an attenuation indicator (\S \ref{subsec:hafuv}) and {\it ii)} the 
strength of the trend in TIR/FUV (\S \ref{subsec:firfuv}) and {\it iii)}
the fact that this trend has been observed by many other workers
\citep[e.g.,][]{wang96,buat99,adel00}, it is inescapable that
$q_{\rm TIR}$ must decrease with decreasing luminosity.  The fact that it
doesn't is a clear argument that {\it radio luminosity does
not directly reflect SF rate}.

\section{The Contribution from Optical Light from Old Stars} 
\label{sec:optical}

There is an important source of uncertainty which has been
neglected, however.  While it is argued that the bulk of the 
light which is reprocessed into the IR comes from the FUV 
\citep{buat96,misiriotis01}, there is nonetheless the potential for
a significant contribution from older stellar populations.

\subsection{A simple model for IR emission} \label{subsec:model}

To decompose the IR emission of the sample galaxies into contributions
from old and young stellar populations, I use a simple model
to interpret the FUV, $V$-band and IR data.
In essence, the energy in the FUV, $V$-band and IR is balanced 
(assuming a constant FUV to $V$ band dust opacity ratio) to
estimate the fraction of FUV and $V$ band light reprocessed into
the IR.  Thus, this approach is conceptually similar to (but more simple
than) the model explored by \citet{buat96}.

Simplistically, I assume that all of the light in the FUV
comes from the young stellar population, and that all the light in 
the optical $V$ band comes from the older stellar population.
I then link the optical depth in the $V$ band to the 
optical depth in the FUV; $\tau_V = c \tau_{\rm FUV}$, where
$c$ is a constant.   For 
Milky-Way type dust and the \citet{calzetti94} attenuation curve 
the ratio between $V$ band optical depth 
and $\sim$1550{\AA} optical depth is 0.4, and for Small Magellanic
Cloud Bar-type dust the ratio is closer to 1/4.  However, 
dust is preferentially clumped around younger stars 
\citep[e.g.,][]{calzetti94,zaritsky99,zaritsky02}, which 
would tend to decrease $c$.  Taken together, a value of 
$c \la 0.3$ is reasonable; I choose $c = 0.25$.
Note that adopting a higher value of $c = 0.4$ does
not significantly affect any of my conclusions (the average 
contribution from $V$-band light rises from 
31\% with a 16\% scatter to 44\% with a 18\% scatter).  

The observed luminosities are
$L_{\rm FUV,obs} = \lambda F_{\rm \lambda,FUV}$ and 
$L_{V,{\rm obs}} = \lambda F_{\lambda,V}$, where $F_{\lambda}$ is the 
observed monochromatic luminosity at a given wavelength.  Given 
the above assumption that $\tau_V = c \tau_{\rm FUV}$ and 
denoting the FUV optical depth as $\tau$ for brevity,
the intrinsic luminosities are related to the observed ones
by $L_{\rm FUV,obs} = e^{- \tau} L_{\rm FUV,intrinsic}$ and 
$L_{V,{\rm obs}} = e^{- c \tau} L_{V,{\rm intrinsic}}$.
Thus, the energy absorbed, and re-emitted into the IR is:
\begin{equation}
L_{\rm TIR} = (1-e^{- \tau}) L_{\rm FUV,obs} e^{\tau} +
	(1-e^{-c \tau}) L_{V,{\rm obs}} e^{c \tau}.
\end{equation}
This equation was then solved using an IDL implementation of 
Brent's method \citep[{ }p.\ 352]{press92} 
to find the root of the equation, given 
the observed $L_{\rm TIR}$, $L_{\rm FUV,obs}$ and $L_{V,{\rm obs}}$.

As examples, it is interesting to take S0--Sa galaxies and Scd--Sm galaxies
from \citet{popescu02}.  Their S0--Sa template has a FUV:$V$:TIR ratio
of 1:25:3 (roughly), corresponding to a fraction from $V$ band light of 86\%
(calculated by multiplying the $V$ luminosity by $c = 0.25$, and dividing by
the sum of itself and the FUV flux; in this case 
$0.25 \times 25 / [0.25 \times 25 + 1] \sim 0.86$).
In contrast, their Scd--Sm template has a FUV:$V$:TIR ratio of 
1:2:1, corresponding to a fraction from $V$ band light of 33\%.  
Notwithstanding the fact that not all $V$ band light is generated by
old stellar populations, this simple analysis fits in well
with what is known about the relative old stellar heating
fraction as a function of galaxy type \citep[e.g.,][]{sauvage92}.

\subsection{Comparison with dust temperatures} \label{subsec:comp}

\begin{figure}[tb]
\vspace{-0.5cm}
\hspace{-0.5cm}
\epsfbox{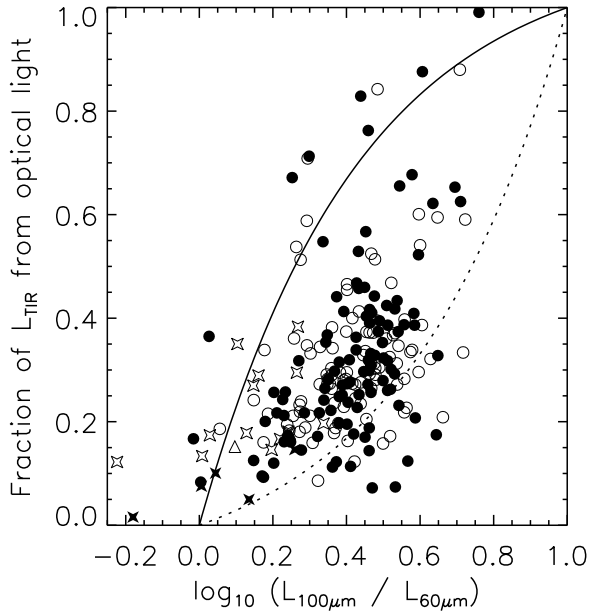}
\vspace{-0.2cm}
\caption{\label{fig:60100}
Comparison of FUV/optical/TIR-derived old stellar
population contribution to the TIR luminosity against
dust temperature as probed by the 100{\micron}-to-60{\micron}
ratio.  Symbols are as in Fig.\ \ref{fig:firfuv}. 
Overplotted are the expected relationship between the
fraction of luminosity from old stars at 60{\micron} against
dust temperature (dotted line) and the 
fraction of luminosity from old stars at 100{\micron} against
dust temperature (solid line).
}
\end{figure}

\begin{figure*}[tb]
\epsfxsize=\linewidth
\epsfbox{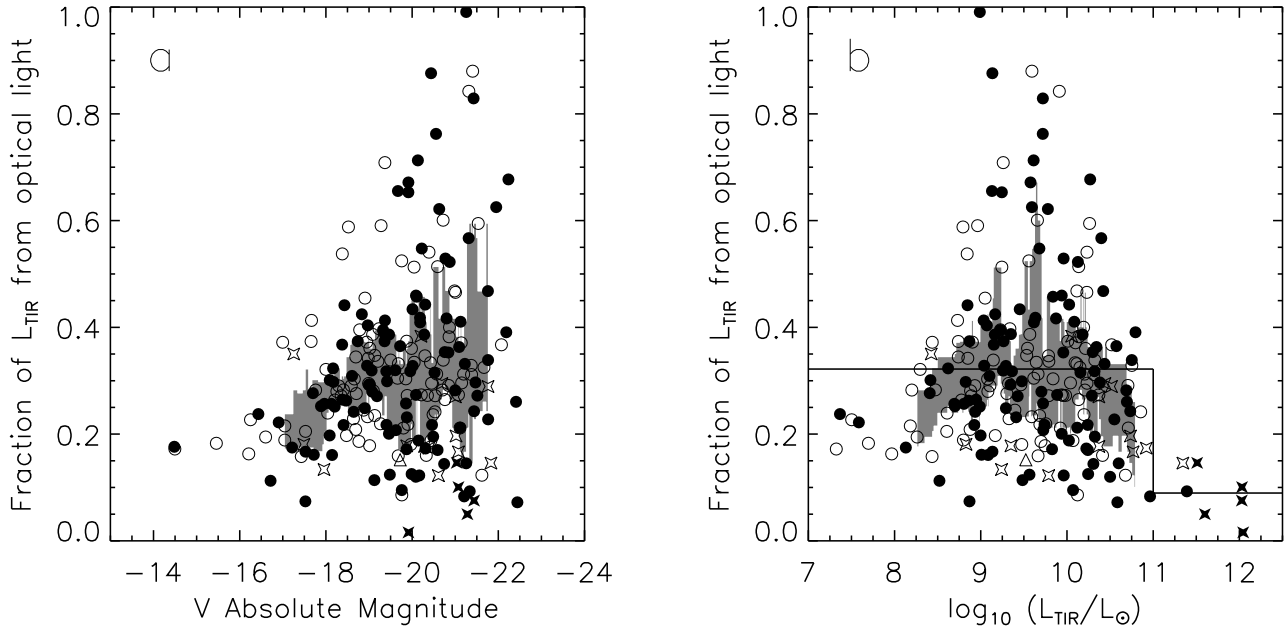}
\caption{\label{fig:oldtrend} Trends in contribution
to TIR luminosity from old stellar populations with galaxy luminosity
in the $V$-band (panel {\it a}) and in the TIR (panel {\it b}).
The symbols are as in Fig.\ \ref{fig:firfuv}.  The shaded area
shows the upper and lower quartile of the old stellar population 
contribution as a function of luminosity.
The solid line in panel {\it b)} is the average contribution 
from old stellar populations to $L_{\rm TIR}$ for galaxies
above and below $10^{11} L_{\sun}$.     }
\end{figure*}

\begin{figure*}[tb]
\epsfxsize=\linewidth
\epsfbox{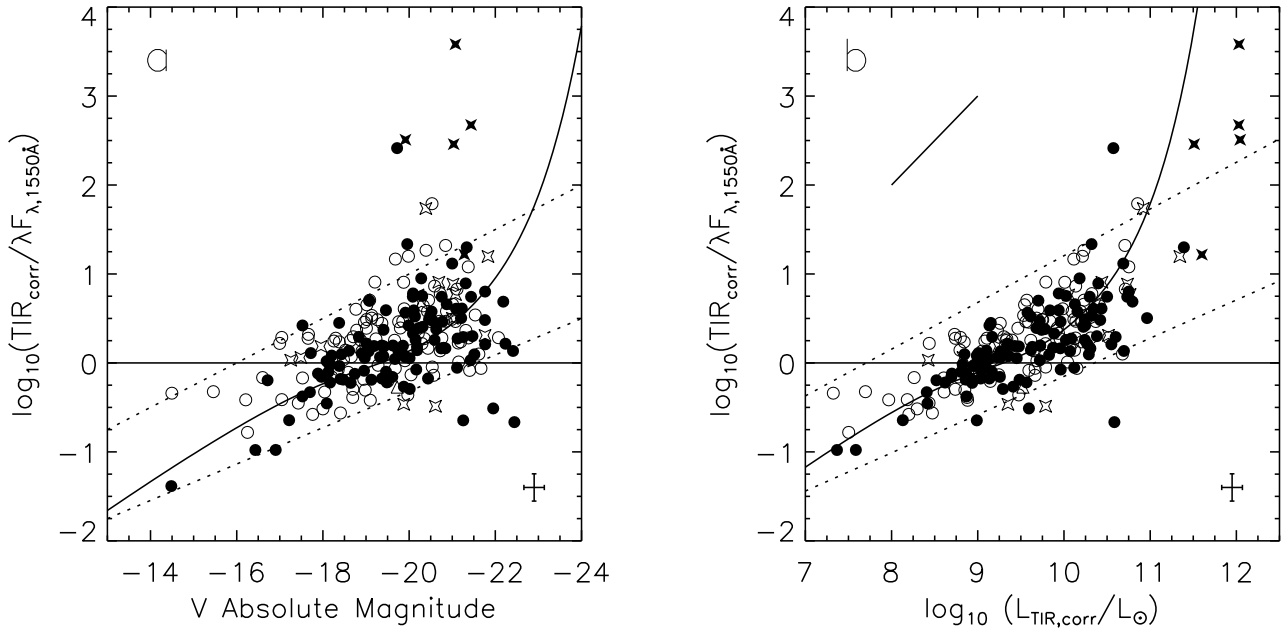}
\caption{\label{fig:firfuv2} Trends in TIR/FUV with galaxy luminosity,
corrected for the contribution from old stellar populations.
The panels and symbols are as in Fig.\ \ref{fig:firfuv}.   }
\end{figure*}

\begin{figure*}[tb]
\epsfxsize=\linewidth
\epsfbox{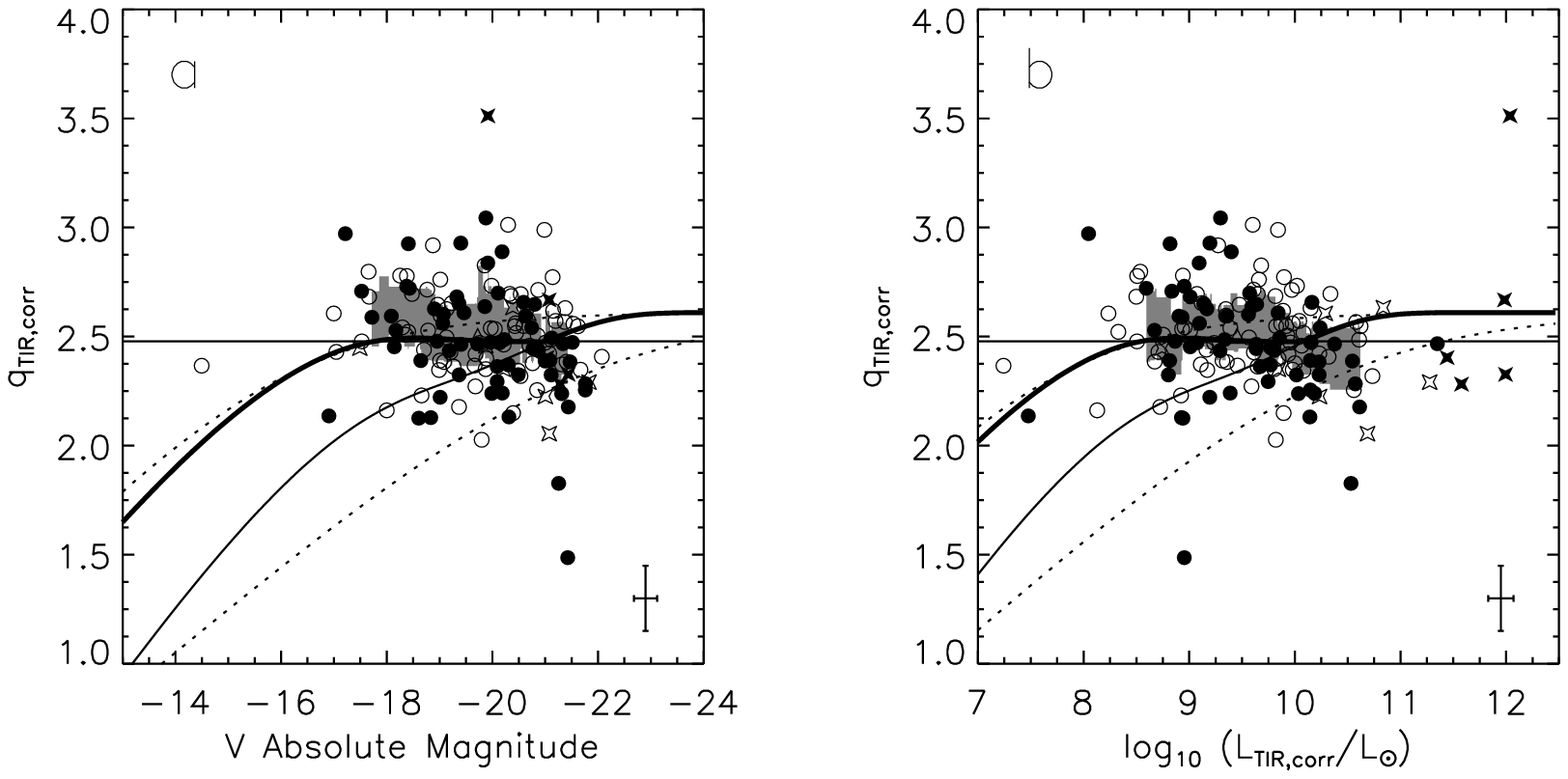}
\caption{\label{fig:qtrend2} Trends in $q_{\rm TIR}$ with galaxy luminosity,
where $q_{\rm TIR}$ has been corrected for the contribution from older
stellar populations.  The panels and symbols are the same as
in Fig.\ \ref{fig:qtrend}.   }
\end{figure*}

It is instructive to compare the fraction of $L_{\rm TIR}$ which
is plausibly associated with old stellar populations (by comparison
of the amount of optical, FUV and TIR light) with the dust 
temperatures (which are often used as indicators of
the contribution of older stellar populations to the 
IR luminosity).  In Fig.\ \ref{fig:60100}, 
it is clear that there is a scattered
correlation between the fraction of the $L_{\rm TIR}$ which is
from optical light (which I associate then with old stellar populations)
and the 100{\micron}-to-60{\micron} flux ratio
(where both are expressed as flux per unit frequency).  Galaxies with rather
larger contributions from $V$-band light tend to have somewhat
larger 100/60 than galaxies with a small contribution from $V$-band light.

It is interesting to estimate the fraction of IR light from 
old stellar populations which is implied by the 100/60 observations.
Estimation of a {\it total} fraction of the 8-1000{\micron} luminosity from
old stars using the FIR color-based technique is challenging, because
of the contributions from the mid-infrared and from wavelengths
longer than $\sim 120\micron$: thus I show the contributions 
at 60{\micron} (dotted line) and 100{\micron} (solid line) from the 
old population as a rough guide.  
The cold population was assumed to have 100/60 of 10, and 
the warm dust a 100/60 of 1, following \citet{fitt88}.  
Fig.\ \ref{fig:60100} shows that there is good overall 
agreement between the expectations of the 
FUV/optical/TIR energy balance estimate of the contribution of 
old stellar populations, and the FIR color.  Interestingly,
this method {\it independently} gives further credence to FIR color-based
analyses, at least at the factor-of-two level.

\vspace{0.2cm}
\subsection{Correcting for the contribution of old stars as a function
	of luminosity}
	\label{subsec:acctold}
\vspace{0.2cm}

I show trends in the contribution to $L_{\rm TIR}$ made by 
old stellar populations as a function of $V$-band and 
TIR luminosity in panels {\it a)} and {\it b)} of 
Fig.\ \ref{fig:oldtrend} respectively.  It is 
clear that the scatter in the contribution of old populations
is large at most galaxy luminosities.   However, in
panel {\it a)} of Fig.\ \ref{fig:oldtrend} there
is a clear general trend of increasing old stellar population 
contribution with increasing $V$-band luminosity (albeit with large scatter).
One could argue that this trend is a selection effect as galaxies
with larger $V$-band luminosity may have larger $V$-band/FUV luminosity,
and therefore have a larger old fraction.  However, the 
correlation between old fraction and dust temperature (Fig.\ \ref{fig:60100})
argues against this interpretation, as the trend in old fraction 
would persist even if 60/100 were shown against $V$-band luminosity 
(these are, of course, independent).  Thus, this reflects the
real and well-known observation that more optically-luminous
galaxies tend to have rather older stellar populations, in 
the mean \citep[e.g.,][]{peletier98,papii,boissier01,sd}.

Panel {\it b)} of Fig.\ \ref{fig:oldtrend} shows the fraction
of $L_{\rm TIR}$ from old stellar populations against TIR luminosity.
At very low luminosities, the old fraction increases with increasing
galaxy luminosity.  However, at $L_{\rm TIR} \sim 10^{10} L_{\sun}$ the
old fraction decreases with increasing $L_{\rm TIR}$.  This reflects
the increasing fraction of dusty, intensely star-forming galaxies 
towards the highest $L_{\rm TIR}$.  The scatter around this
general behavior is large, however.  

In Fig.\ \ref{fig:firfuv2} I show the effect that the old stellar
population correction has on trends in TIR/FUV with 
galaxy luminosity.  There are some modest changes: as expected,
the trend in TIR/FUV with $V$-band luminosity slightly flattens.  
However, on the whole, there is very little change in TIR/FUV with
luminosity.  This is, to a certain extent, for an obvious reason.  A
30\% change in TIR luminosity is not going to significantly affect
a trend which sees a factor of $\ga 30$ 
increase in TIR/FUV with a factor of 1000
luminosity increase.  In order to significantly affect this
trend, the old fraction would have to increase from essentially
0 to $\ga 95$\% over the luminosity range of interest, with relatively
little scatter.  This kind of behavior 
is clearly ruled out by the observations.
Thus, the conclusion that dust opacity
should leave an easily observable signature in the \rf correlation 
remains unchanged.  

I check this directly in Fig.\ \ref{fig:qtrend2}.  When corrected
for the contribution of older stellar populations, $q_{\rm TIR}$ decreases
by a median amount of 0.16 dex while the scatter decreases slightly to 0.25 dex.
The relative constancy of $q_{\rm TIR}$ with luminosity persists 
(compare Figs. \ref{fig:qtrend} and \ref{fig:qtrend2}).
Again, there is a slight hint of a slightly higher $q_{\rm TIR}$ for lower-luminosity
galaxies.  Also, the `bump' in $q_{\rm TIR}$ at $L_{\rm TIR} \sim 10^{10} L_{\sun}$
which was reasonably apparent in panel {\it b)} of Fig.\ \ref{fig:qtrend}
has been largely eliminated by the correction for the old
stellar population.  This `bump' was from a larger old
fraction in earlier-type galaxies
with reasonably high $V$-band luminosities but lower SF rates (see panel
{\it b)} of Fig.\ \ref{fig:oldtrend}).  At lower
luminosities, later-types dominate, and at higher IR luminosities,
ULIRGs and star-bursting galaxies tend to dominate.  Thus, the 
reasonably complete removal of this `bump' feature can be 
taken as independent evidence that the correction for the 
effects of older stellar populations is doing its job reasonably well.

It is clear then that {\it neither} the TIR/FUV vs.\ luminosity
correlation {\it nor} the $q_{\rm TIR}$ vs.\ luminosity correlation are
significantly affected by the contribution of old stellar populations.
Furthermore, this conclusion does not depend on the technique
used to estimate the contribution from the old stellar populations, as
demonstrated by the correlation between 100/60 and the fraction of
light from $V$-band light.

\section{Understanding the Radio Emission from Galaxies} 
\label{sec:radio}

In \S\S \ref{sec:fuv} and \ref{sec:optical}, I demonstrated
that, if radio emission $\propto$ SF rate, then the TIR-to-radio
ratio $q_{\rm TIR}$ should decrease by at least a factor of two when going
from $\sim L_*$ to $\sim 0.01 L_*$ galaxies, owing to the 
effects of dust optical depth.  Furthermore, accounting for 
old stellar populations does not affect this result.  
One is therefore left in the situation where a factor-of-two 
offset has to be there, but it isn't seen.  This implies
that the {\it radio emission of low-luminosity galaxies is 
suppressed, by at least a factor of two}.  

Furthermore, it is the non-thermal radio emission which
must be suppressed in low-luminosity galaxies.
The radio emission from normal (non-active) galaxies
comes from two sources.  Thermal radio emission from 
ionized hydrogen directly tracks the SF rate (because
the amount of ionized hydrogen reflects the ionizing
luminosity of the very young stellar populations which are 
rich in massive stars).
In contrast, it has been suggested for nearly 20 years that the 
non-thermal synchrotron emission of low-luminosity galaxies can be 
significantly suppressed \citep[{ }although the thermal contribution
can be very challenging to reliably estimate; Condon 
1992]{klein84,klein91,klein91b,price92}.  
This can be explained in a number of ways, as the 
physics which links the SF rate with non-thermal emission 
is complex, and involves the cosmic ray production rate, 
galaxy magnetic field strength, and galaxy size to name just a few
of the many variables \citep{chi90,helou93,lisenfeld96}.
For example, \citet{chi90} discuss a model
in which the non-thermal emission from low-luminosity galaxies 
is strongly suppressed, because most of the cosmic-ray electrons
escape from the galaxy due to their small sizes (although the 
size of the effect that they predict is a factor of 3--5 in excess
of the trend allowed by these observations).

Because of the complex and uncertain physics involved, I do not attempt
to construct detailed theoretical model for the non-thermal 
radio emission.  Rather, I use the data to guide me 
in constructing how non-thermal radio emission must track the 
SF rate \citep[cf.][]{price92}.  
I parameterize the total radio emission as
$R = (n + 0.1) \eta \psi$, where $R$ is the radio flux at 
1.4\,GHz, $\psi$ is the SF rate, $\eta$ is the constant
of proportionality linking the SF rate and radio flux for 
$\sim L_*$ galaxies, and $n$ is the relative amount of 
non-thermal emission. For a $\sim L_*$ galaxy, 90\% of the radio 
flux at 1.4\,GHz is non-thermal 
\citep{condon92}, and 10\% is thermal (which gives the 
value of 0.1).  I allow $n$ to decrease as a function
of galaxy luminosity: 
\begin{equation} \label{n}
n  = \left\{ \begin{array}{ll}
   0.9 & L > L_* \\
   0.9 ( L / L_* ) ^{0.3} & L \le L_*,
   \end{array} \right.
\end{equation}
where $L_*$ is taken to be $V = -21$.  The resulting 
relationship between $q_{\rm TIR}$ and luminosity is shown in 
Figs.\ \ref{fig:qtrend} and \ref{fig:qtrend2} as thick
solid lines: this variation in non-thermal radio emission
accounts reasonably well for the lack of a trend in 
$q_{\rm TIR}$ with luminosity.

Remarkably, given the uncertainties inherent to decomposing 
the contribution of thermal/non-thermal emission from radio 
spectra alone, the kind of suppression of non-thermal radio
emission which is required to produce a luminosity-independent $q_{\rm TIR}$ is
consistent with an independent analysis by \citet{price92}.
They used multi-frequency radio data to
construct the \rf correlation at a number of frequencies.  
They found that the \rf correlation at high frequency (where
thermal radio emission dominates) was nearly linear, and the
\rf correlation at lower frequencies (where non-thermal 
emission dominates) was steeper.  They suggested that 
thermal emission $\propto$ the SF rate $\psi$, but suggested that 
non-thermal emission $\propto \psi^{1.2}$.  Thus the 
non-thermal-to-SF rate ratio varied as $\psi^{0.2}$.  Thus, 
for a decrease in galaxy luminosity by a factor of 100, the 
non-thermal-to-SF rate ratio decreases by a factor of 2.5.
Equation \ref{n} predicts a factor of $\sim$3 decrease in the
non-thermal-to-SF rate ratio for the same luminosity range.  
Thus, the conclusion that low-luminosity galaxies have lower
non-thermal contributions has been established in two independent
ways: through multi-frequency radio observations 
\citep[e.g.,][{ }which, however, depend on uncertain radio spectral 
fitting]{klein91b,price92} and through a lack of curvature 
in the \rf correlation (this work, which does not depend on 
the fitting of radio spectra in any way).   

\begin{figure}[tb]
\vspace{-0.5cm}
\hspace{-0.5cm}
\epsfbox{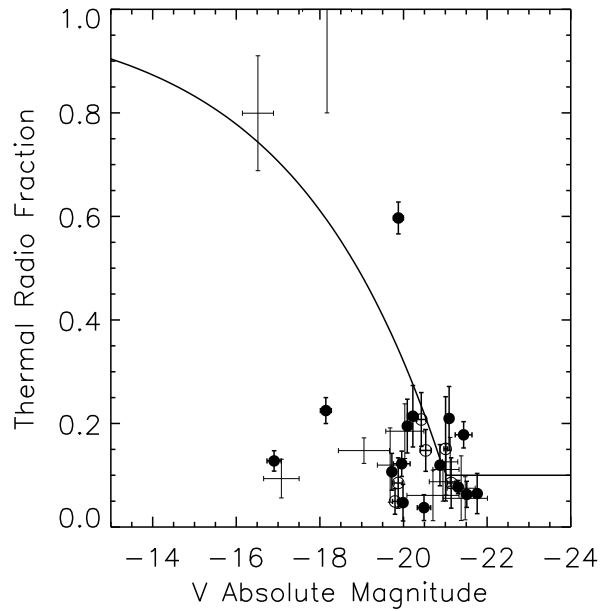}
\vspace{-0.2cm}
\caption{\label{fig:therm}
Comparison of thermal fraction at 1.4\,GHz with $V$-band 
absolute magnitude.  Symbols are as in Fig.\ \ref{fig:firfuv}.
An additional seven non-active galaxies from \citet{niklas97} and 
five galaxies from \citet{klein91b} are also included, plotted
as error bars.  Equation \ref{n} is overplotted as a solid line.
}
\end{figure}

A prediction of this model is an overall correlation between
galaxy luminosity and thermal radio fraction.  Testing this
prediction in detail is challenging because of observational
difficulties.  Typically, 
four or five radio fluxes at a range of frequencies are used
to derive a non-thermal slope \citep[which may, itself, vary with 
radio frequency;][]{condon92} and thermal fraction, as well as the overall
normalization \citep{condon92,niklas97}.  These difficulties lead to 
a large uncertainty in the thermal fraction, which for the well-documented
case of M82, leads to a factor of five discrepancy between 
different estimates of thermal fraction \citep{condon92}.
Notwithstanding these difficulties,
I plot my prediction for thermal radio fraction
from Equation \ref{n} against observed thermal radio fractions for 
my sample galaxies and a further dozen non-active galaxies 
from the \citet{klein91b} and \citet{niklas97} in Fig.\ \ref{fig:therm}.  
Clearly, most of the thermal fraction
determinations are for galaxies within 1 magnitude
of $L_*$, and are consistent with a value of 0.1 
\citep[as argued by][]{condon92}.  
There are only a few low-luminosity galaxies in this sample, 
and they have a large scatter between thermal fractions of 0.1 and 1.
This huge scatter should not be over-interpreted, as 
there are hints that it is largely intrinsic. For example,
\citet{yun01} show that the scatter
at the low-luminosity end of the \rf correlation is larger than for 
$\sim L_*$ galaxies, and it is not unreasonable to attribute some of that
scatter to the radio (and therefore non-thermal) 
flux.\footnote{I do not see this effect in 
Figs.\ \ref{fig:radfir}, \ref{fig:qtrend} or \ref{fig:qtrend2}, which
is likely due to small number statistics.}  Given the modest sample size, 
the observational difficulty, and possibly substantial intrinsic scatter it 
is fair to say that my non-thermal radio calibration 
is not inconsistent with the thermal radio fraction data, and 
is strongly supported by the linearity of the \rf correlation 
and the frequency dependence in the \rf correlation 
as reported by \citet{price92}.

\section{Discussion} 
\label{sec:disc}

\subsection{Implications for Radio- and IR-derived SF rates} 
\label{sfr}

In this paper, I have brought together a diverse and wide
range of literature data and interpretation into 
a coherent picture.  On one hand, I examine the luminosity
dependence in FUV  
\citep[e.g.,][]{wang96,buat99,adel00,buat02} and \ha attenuation
\citep[e.g.,][]{hopkins01,sullivan01}, showing that 
low luminosity galaxies with $L \sim L_*/100$ are optically-thin
in the FUV, whereas $\sim L_*$ galaxies are optically-thick in the FUV
(this result is robust to the inclusion of the
effects of old stellar populations).  On the other hand, I 
show that the \rf correlation is nearly linear \citep[e.g.,][]{yun01},
and shows no sign of the expected depression of the TIR luminosity
of the largely optically-thin dwarf galaxies.  This means that 
the radio is also suppressed in dwarf galaxies \citep[cf.][]{klein91}.
Since the thermal radio emission $\propto$ SF rate, the non-thermal
must depend non-linearly on SF rate: \citet{price92} find that
non-thermal radio emission $\propto \psi^{1.2}$, which is consistent
with my data.  Thus, the \rf correlation is linear not because
both radio and IR emissions track SF rate, but rather because both
radio and IR emissions fail to track SF rate in independent, but
coincidentally quite similar, ways.

In this section, I use this increased understanding of the 
\rf correlation to derive TIR and radio SF rate calibrations which
take into account the broad-brush suppression
of TIR and radio emission from low-luminosity galaxies.
Furthermore, because the scatter in TIR/FUV and $q_{\rm TIR}$ are
well-characterized, the scatter in these SF rate calibrations
will be quantified.
For the purposes of a simple-to-apply SF rate 
calibration, and given that the scatter in FUV opacity
and $q_{\rm TIR}$ at a given luminosity is considerable, it is sufficient
to calibrate for first-order variations in opacity and non-thermal
radio emissivity in a very simple fashion (not using, e.g., the 
model for dust opacity developed in Appendix \ref{app:model}).  
Furthermore, I will 
present calibrations of the SF rates simply in terms of 
the TIR or radio luminosity.  This allows 
workers to derive SF rates from one flux
alone while still being able to account for the reduced
efficiency of TIR and radio emission from low-luminosity galaxies.
If more luminosities are available
(e.g.\ TIR, radio, optical and FUV), then a fuller analysis of the
data would clearly prove superior to these simple-minded calibrations.

\subsubsection{Calibrating IR-derived SF rates}

In \S \ref{subsec:hafuv}, 
I showed that attenuation-corrected \hans-derived
SF rates and TIR$+$FUV SF rates were consistent to
at least a factor of two, when the SF rate
calibrations of \citet{k98} were adopted.  Thus, I
adopt the `starburst' calibration of IR luminosity
presented by \citet{k98} for luminous $L \ga L_*$ galaxies.
A reasonably acceptable fit to panel {\it b)} of 
Figs.\ \ref{fig:firfuv} and \ref{fig:firfuv2} is 
TIR/FUV$ \sim \sqrt{L_{\rm TIR} / 10^9 L_{\sun}}$.
Thus, adopting a Salpeter IMF from 0.1 to 100 $M_{\sun}$ following
\citet{k98}, the SF rate $\psi$ is:
\begin{equation}
\psi (M_{\sun} {\rm yr^{-1}}) = 1.72 \times 10^{-10} 
		L_{\rm TIR} (1 + \sqrt{10^9/L_{\rm TIR}}), 
\end{equation}
where $L_{\rm TIR}$ is in solar luminosities and 
is calculated between 8--1000{\micron} (where I have adopted
a solar luminosity of $3.9 \times 10^{26}$\,W).  Note however that
there is a $\sim 0.5$\,dex scatter about this correlation, which
will translate into $\pm 50$\% scatter in the SF rate calibration 
at $10^9 L_{\sun}$, and a $\pm 20$\% scatter at $10^{11} L_{\sun}$.

This calibration does not account for contributions from 
old stellar populations.  Fig.\ \ref{fig:oldtrend} demonstrates
that a correction for old stellar populations will be statistical at
best.  However, it is possible to correct for the mean
contribution from old stellar populations in a relatively robust manner.
At TIR luminosities 
$\le 10^{11} L_{\sun}$, the mean fraction and scatter are 32\% $\pm$ 16\%, 
and the mean fraction and scatter are 9\% $\pm$ 5\% 
at TIR luminosities  $> 10^{11} L_{\sun}$ (see the solid 
line in panel {\it b)} of Fig.\ \ref{fig:oldtrend}).  Thus, the final 
calibration, correcting for old stellar populations, is:
\begin{equation} \label{efir}
\psi (M_{\sun} {\rm yr^{-1}}) = 
\left\{ \begin{array}{ll}
	1.57 \times 10^{-10} L_{\rm TIR} 
	(1 + \sqrt{10^9/L_{\rm TIR}}) & L_{\rm TIR} > 10^{11} \\
	1.17 \times 10^{-10} L_{\rm TIR} 
	(1 + \sqrt{10^9/L_{\rm TIR}}) & L_{\rm TIR} \le 10^{11}. 
   \end{array} \right.
\end{equation}
Expected scatter around this correlation is at least 50\% at 
$10^9 L_{\sun}$, and 25\% at $10^{11} L_{\sun}$.  There are
data down to $L_{\rm TIR} \sim 10^7 L_{\sun}$; however, this calibration
should be applied with {\it extreme} caution at such low luminosities
because of the $\ga 10 \times$ extrapolation involved.  There
may be as much as a factor of two uncertainty globally, although
the overall calibration uncertainty is probably somewhat less
(see \S \ref{subsec:hafuv}).  Obviously, there will be uncertainties
because of IMF, etc.\ \citep[see][{ }for more discussion]{k98}.
It is important to realize that this calibration is certainly suspect on 
a galaxy-by-galaxy basis: indeed, Fig.\ \ref{fig:oldtrend} shows that there are
galaxies with anywhere between 1\% and 99\% of their IR emission 
from old stellar populations.

It is interesting to note in passing that the above calibration is 
within a factor of two of a constant TIR conversion 
factor of $1.72 \times 10^{-10}$ for galaxies with luminosities
in excess of $\ga 3 \times 10^8 L_{\sun}$.  
On one hand, the SF rate calibration 
is reduced by 10\%--30\% by the contribution of old stellar populations.
However, on the other hand, the reduction in dust opacity with decreasing
luminosity cancels out the effects of old stellar populations 
to first order until one reaches luminosities
$\la 10^9 L_{\sun}$, where the opacity is so low that the heating of dust
by old stars does not significantly help.  This argument was essentially
made by \citet{inoue02} from a more model-based standpoint.

\subsubsection{Calibrating Radio-derived SF rates}

In order to calibrate the radio flux in terms of SF rates, 
it is necessary to {\it i)} estimate the zero point of the 
SF rate scale, and {\it ii)} estimate the effect of increased 
suppression of non-thermal radio emission for low-luminosity galaxies,
making sure to re-cast the result in terms of radio flux.

Following the above, I make the assumption that $L \ga L_*$ galaxies
lose no cosmic rays, and have non-thermal radio emission 
that directly tracks the SF rate.  Thus, I choose to calibrate the 
radio SF rate to match the TIR SF rate for $L \ga L_*$ galaxies.
The geometric mean radio power per solar luminosity of TIR for 
$L_{\rm TIR} \ge 2 \times 10^{10} L_{\sun}$ 
galaxies is $3.12 \times 10^{11}$\,W\,Hz$^{-1}$\,$L_{\sun}^{-1}$,
corresponding to a $q_{\rm TIR}$ of 2.52.  
Thus, in the limit of high SF rate,  
a radio flux at 1.4\,GHz of 
$1.81 \times 10^{21}$\,W\,Hz$^{-1}$ is 
predicted per $1\,M_{\sun}\,{\rm yr^{-1}}$.  This is around 
a factor of two higher than the Milky Way-normalized 
radio SF rate calibration of \citet{condon92} adapted to
my adopted IMF \citep{haarsma00}, which is well within 
the factor-of-two uncertainties in the assumptions 
underpinning the two independent calibrations \citep[{ }also found this
offset between TIR-normalized and Milky Way-normalized radio
SF rate calibrations]{condon02}.

Adopting this zero point and the variation in non-thermal
radio emission from Equation \ref{n}:
\begin{equation} \label{erad}
\psi (M_{\sun} {\rm yr^{-1}}) = 
\left\{ \begin{array}{ll}
   5.52 \times 10^{-22} L_{\rm 1.4\,GHz} & L > L_c \\
   \frac{5.52 \times 10^{-22}}{0.1 + 0.9 (L/L_c)^{0.3}} L_{\rm 1.4\,GHz} &
	L \le L_c,
   \end{array} \right.
\end{equation}
where $L_c = 6.4 \times 10^{21}$\,W\,Hz$^{-1}$ is the radio luminosity 
at 1.4\,GHz of a $\sim L_*$ galaxy
($V = -21$, or $L_{\rm TIR} \sim 2 \times 10^{10} L_{\sun}$).
The scatter in $q_{\rm TIR}$ of 0.26 dex implies a factor-of-two uncertainty
in the application of this calibration on a galaxy-by-galaxy basis.
The increased scatter for both very high and very low luminosity galaxies 
\citep[Fig.\ \ref{fig:radfir} and][{} respectively]{yun01} implies somewhat 
larger uncertainties for very high and very luminosity galaxies, perhaps
as large as a factor of 5, to be conservative.  
Furthermore, the data run out at luminosities below
$3 \times 10^{19}$\,W\,Hz$^{-1}$; this calibration should be used
with {\it extreme} caution below this luminosity.  An idea of the 
systematic calibration uncertainty is given by the factor of two
offset between my and Condon's overall zero point.

The expected \rf correlation, and expected trends in $q_{\rm TIR}$
with IR luminosity as given by these TIR and radio calibrations
are shown as the thick dashed lines in Figs.\ \ref{fig:radfir} and
\ref{fig:qtrend}.  It can be seen that the above calibrations produce a 
nearly linear \rf correlation, while fully taking into account the 
non-linear effects of dust opacity, old stellar populations
and the non-linear dependence of non-thermal radio flux on SF rate, 
at least over the range over which I have data.

\subsubsection{A Pinch of Salt}

It is worth discussing briefly some of the limitations and caveats
of the above SF rate calibrations (Equations \ref{efir} and \ref{erad}).
\begin{itemize}
\item AGN were explicitly excluded from this sample.  Obviously,
	IR and radio luminosities
	will overestimate the SF rate 
	if {\it any} calibration is blindly applied to
	samples of galaxies which contain AGN.
\item The scatter in the above calibrations is at the factor-of-two
	level in terms of both systematic and random errors.  
	Furthermore, individual galaxies can deviate substantially from
	the mean behavior (e.g., galaxies with 99\% of their IR
	reprocessed from the optical, or low-luminosity galaxies with 
	thermal radio fractions which scatter considerably from the 
	expected trend).  Thus, these calibrations should not blindly
	be applied on an individual galaxy basis.  A comprehensive
	multi-wavelength analysis is required to significantly 
	constrain the SF rate of an individual galaxy.
\item This sample was selected extremely inhomogeneously.  Specially-selected
	samples \citep[e.g., UV-selected samples;][]{adel00,sullivan00}
	may be biased (for example towards low dust opacity) and 
	may present different behaviors from this diverse local sample.
\item Blind application of these calibrations as a function of lookback
	time may be inappropriate.  For example, it is uncertain how
	dust opacity, the contribution of old stellar populations to 
	dust heating, or cosmic ray retention depend on redshift.
	In this context, comprehensive multi-wavelength SF rates 
	from a variety of sources (such as the rest-frame FUV, 
	optical emission lines, IR and radio) may help to reduce
	the unavoidable systematic uncertainties that plague these
	kinds of analyses.
\end{itemize}

\subsection{Increased Scatter at Low and High IR Luminosities} 
\label{sec:hilo}

I found that there was an increased scatter in the \rf correlation 
at high IR luminosities \citep[Fig.\ \ref{fig:radfir}, and][]{yun01}.  
In addition, low IR luminosity galaxies tend to scatter more around
the \rf correlation \citep[e.g.,][]{condon91,yun01}, although
this dataset does not show this effect, perhaps because
of small number statistics.
This increase in the scatter
is intrinsic: the errors are $\ga 3 \times$ 
smaller than the scatter of the data.

\citet{bressan02} recently discussed the increase of scatter for 
intensely star-forming, high IR luminosity galaxies.  They presented
a comprehensive model which includes stellar population synthesis, 
dust radiative transfer, and a simplified model of radio emission 
from cosmic rays generated by supernovae.  They predict a strong
evolution in IR-to-radio ratio $q$ with time after an intense burst of 
star formation: essentially, the timescale for IR emission is shorter
than the timescale for radio emission, leading to variations with a
total range of $\Delta q \sim 1$ over $10^8$ yr timescales.  This 
scatter matches well the observed scatter in ULIRGs.  

An increased scatter at lower luminosities could partly be 
due to optical depth effects: low-luminosity
galaxies are largely transparent in the FUV, meaning that changes
in dust opacity translate directly into large changes in $q_{\rm TIR}$.
However, a number of recent studies 
\citep[e.g.,][]{dohm98,sullivan00,kauffmann02}
have suggested significant variations in SF rate over $\sim 10^8$ yr 
timescales for at least some lower-luminosity galaxies.  These variations 
would lead to scatter in $q_{\rm TIR}$, from mismatches between the 
IR and radio emission timescales \citep{bressan02}.  Interestingly, 
these variations in SF rate would lead to significant variation 
in the thermal radio fraction \citep[see Fig.\ 5 of][]{bressan02}, 
as the thermal radio emission tracks the SF rate over $\sim 5$ Myr timescales,
whereas the non-thermal emission arguably 
tracks the SF rate over $\sim 10^8$ yr
timescales.  This could well explain much of the scatter
seen by \citet{yun01} at low luminosities, and underlines
the need for thermal radio fractions for a large sample of 
low-luminosity star-forming galaxies.

\section{Conclusions} 
\label{sec:conc}

I have assembled a diverse sample of galaxies from the literature
with FUV, optical, IR and radio luminosities to explore the 
calibration of radio- and IR-derived SF rates, and the origin
of the \rf correlation.  My main conclusions are as follows.

In order to establish the efficacy of IR/FUV as an extinction 
indicator, I compare \ha and 8--1000{\micron} 
TIR/FUV properties of a subsample of my galaxies.
I find that \ha and FUV attenuations loosely correlate with each other, 
with the \ha attenuation being roughly half of the FUV attenuation.
A foreground screen model would predict an offset
of a factor of a quarter.  This lends support
to the claim of \citet{calzetti94} that the nebular extinction 
is roughly twice that of the stellar population of the galaxy.
Furthermore, when SF rates derived from TIR$+$FUV and attenuation-corrected
\ha are compared, I find that they 
agree to better than a factor of two (random and systematic).
This strongly argues that TIR/FUV will give FUV attenuation estimates which
are accurate to a factor of two, and probably much better.

Having established the efficacy of TIR/FUV as a FUV
attenuation indicator, I explored trends in TIR/FUV with 
galaxy luminosity.  This ratio increases on average by over a factor of 30 
between low-luminosity galaxies ($L \sim 1/100 L_*$) 
and high-luminosity galaxies ($L \sim 3 L_*$).  
Low-luminosity galaxies have TIR/FUV$ \la 1$, 
meaning that they are {\it optically thin} in the FUV.
Interestingly, the gross, overall trend in TIR/FUV is naturally
interpreted in terms of increasing gas surface density and 
galaxy metallicity with increasing galaxy mass.

Like \citet{yun01}, I find a nearly linear
\rf correlation, with perhaps a slight 
tendency for faint galaxies to have a somewhat higher TIR-to-radio
ratio $q_{\rm TIR}$ than brighter galaxies.  
However, the strong and robust increase in TIR/FUV with luminosity would, if 
radio were a perfect SF rate indicator, produce a clear and 
easily measurable decrease
in $q_{\rm TIR}$ for fainter galaxies.  The data show the opposite
(or no) trend, clearly demonstrating that {\it radio 
is not a perfect SF rate indicator}.
Accounting for the effects of older stellar populations
using a simple FUV/optical/IR energy balance model (which 
is consistent with FIR color-based methods) does not change
this key result.

In order to cancel out the trend in $q_{\rm TIR}$ from 
optical depth effects, the non-thermal emission 
must be suppressed by about a factor of 2--3
in $\sim L_*/100$ galaxies relative to $\sim L_*$ galaxies.
This result was also reached
independently, using a totally different dataset and method, 
by \citet{price92}.
Thus, the linearity of the \rf correlation is a conspiracy:
both radio and IR underestimate the SF rate for 
low-luminosity galaxies.  

I present SF rate calibrations which simultaneously
reproduce the linearity of the \rf correlation, and 
take account of the reduced non-thermal and IR emission 
in lower-luminosity galaxies.
However, there
is considerable scatter in the SF rate calibrations, which
can exceed a factor two at low galaxy luminosities.  This 
highlights the possible influence of selection effects in 
interpreting the IR or radio emission from distant galaxies.
Another challenge for those wishing to estimate the SF rates of
distant galaxies is the non-trivial physics that links
IR/radio and SF rate.  For example, the evolution of dust 
opacity, the importance of old stellar populations, and 
the evolution of the efficiency of cosmic ray confinement are 
all essentially unconstrained as a function of lookback time. 
This adds considerable systematic uncertainty to our 
understanding of galactic SF rates in the distant Universe.

\acknowledgements

I thank Betsy Barton Gillespie, Andrew Hopkins, 
Rob Kennicutt, Casey Papovich and J. D. Smith for useful discussions and
their comments on the manuscript.  The anonymous referees are
thanked for their feedback.
This work was supported by NASA grant NAG5-8426 and NSF grant AST-9900789.  
This work made use of NASA's Astrophysics Data System 
Bibliographic Services, and the NASA/IPAC Extragalactic Database 
(NED) which is operated by the Jet Propulsion Laboratory, 
California Institute of Technology, under contract with the 
National Aeronautics and Space Administration.

\appendix
\section{A. Multi-wavelength photometry} \label{sec:app}

My galaxy sample was primarily selected to have published FUV photometry at 
$\sim 1500${\AA}.  Here, I describe the sources of the FUV, optical, 
IR, radio, and other data, and the error estimates for each 
type of data.  I also present my galaxy luminosities, as an aid to 
workers in the field, in Table \ref{fluxes}.

\subsection{FUV Data}

\begin{figure}[tbh]
\vspace{-0.5cm}
\hspace{-0.5cm}
\epsfbox{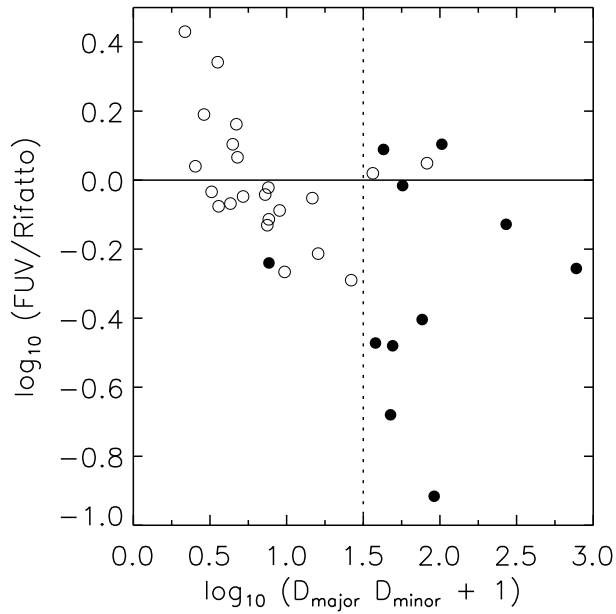}
\vspace{-0.2cm}
\caption{\label{fig:consistent} 
A comparison of FAUST (22 galaxies; open circles) and UIT 
(11 galaxies; filled circles)
FUV flux with those from \citet{rifatto3}.  The solid line
denotes no offset between fluxes, and the dotted line
represents the size cut applied to galaxies
from \citet{rifatto3}: galaxies with
$\log_{10} (D_{\rm major} D_{\rm minor} + 1) \ge 1.5$ are 
omitted from the sample.
}
\end{figure}

The FUV data at wavelengths $\sim 1550${\AA}
were taken from a variety of sources:
UIT fluxes at an average wavelength of 
1567{\AA} for normal spiral and dwarf galaxies from 
\citet{uit}, 1495{\AA} 
IUE fluxes for 
UV-bright starbursts from \citet{calzetti94,calzetti95},
1650{\AA} STIS data from the {\it Hubble Space Telescope} for 
a sample of ULIRGs from 
\citet{gold02}, 1650{\AA} fluxes from
normal galaxies from the FAUST experiment by
\citet{deharveng94}, and large-aperture 1650{\AA} UV
fluxes from a variety of UV experiments which were 
homogenized and extrapolated to total by 
\citet{rifatto2,rifatto3}. 

Error estimates for the different sources of data are as follows.
\begin{itemize}
\item UIT and FAUST were shown to be consistent to within 
20\% ($=$0.08 dex) by \citet{uit}.
\item IUE fluxes for this sample
are also accurate to $\sim 0.08$ dex \citep{kinney93}.
\item The STIS fluxes for the seven ULIRGs have a mean accuracy
	of $\sim 0.12$ dex \citep{gold02}.
\item Fluxes from \citet{rifatto3} were compiled from
	the literature, and so have variable quality.
In Fig.\ \ref{fig:consistent}, I show a comparison of 
FAUST and \citet{rifatto3} fluxes (open circles), and 
UIT and \citet{rifatto3} fluxes (filled circles) as a function 
of galaxy area.  
\citet{rifatto3} sometimes substantially
over-extrapolate the FUV fluxes of large galaxies: this conclusion 
was also reached by \cite{uit} when they compared UIT and \citet{rifatto3}
fluxes with OAO, SCAP and FOCA data (their Fig.\ 1).  
Accordingly, all galaxies from
\citet{rifatto3} with $\log_{10} (D_{\rm major} D_{\rm minor} + 1) \ge 1.5$
are removed from the sample (29 galaxies).  
The remaining 21 galaxies from \citet{rifatto3}
with UIT or FAUST data and 
$\log_{10} (D_{\rm major} D_{\rm minor} + 1) < 1.5$ are consistent with a
scatter of 0.19 dex.  While substantial, this scatter is consistent with 
the combined error estimates, and is substantially smaller than the 
intrinsic scatter in, e.g., the TIR/FUV correlation with total TIR
luminosity, or $V$ band absolute magnitude.  In Fig.\ \ref{fig:consistent},
there also appears to be a trend between the FAUST/UIT fluxes and 
\citet{rifatto3} fluxes for 
$\log_{10} (D_{\rm major} D_{\rm minor} + 1) < 1.5$.  A correction of 
either the \citet{rifatto3} data, or the FAUST/UIT data for this trend 
simply increases scatter in TIR/FUV at a given TIR or $V$-band luminosity,
and is therefore not applied.
\end{itemize}

The IUE spectroscopic aperture is $10\arcsec \times 20\arcsec$
in size.  Accordingly, to minimize aperture effects
I choose to use the FUV data for starburst galaxies with 
optical diameters $\le 1.5\arcmin$ only.  Tests have shown that this 
diameter cutoff excludes all of the galaxies with truly obvious
aperture effects (e.g., gross mismatches between FUV attenuations
measured by TIR/FUV, which are susceptible to aperture mismatch, and
UV-color based attenuations, which are impervious to aperture worries),
while keeping a reasonable sample size.  
Larger starbursts are included in this study, but are assumed to 
have no FUV data (i.e.\ only the optical, IR and radio data are used).

The data come from wavelengths between 1450{\AA} and 1650{\AA}. 
I make the simplifying assumption that the fluxes are all 
measured at 1550{\AA}.
A typical star-forming galactic spectrum has
$F_{\lambda} \propto \lambda^{-1}$ at these wavelengths \citep{bell02}.  Thus,
the error introduced by this assumption is $\sim 6$\%.

As is obvious from the above discussion, there is some overlap
between the different samples.  I prefer FAUST/UIT FUV photometry
above the \citet{rifatto3} photometry in all cases.  There were
two galaxies which overlapped between the \citet{rifatto3}
sample and the starburst sample.  While the \citet{rifatto3} 
total FUV fluxes would probably be preferred (because of aperture
effects in the IUE starburst data), I choose to use the 
IUE data for those two galaxies so that the \hans$+$Balmer decrement
data and FUV measurements have matched apertures.  The measurements
were consistent to within 0.3 dex at any rate, and adopting
\citet{rifatto3} fluxes for these two galaxies does not affect
the results.

\subsection{Optical Data}

Optical data are principally taken from the Third Reference Catalogue
of Bright Galaxies \citep[RC3;][]{rc3}.  Values of $V_T$, the total
$V$ band magnitude not corrected for extinction or inclination, were
taken, when available.  If necessary, $V$-band apparent magnitudes were
estimated from $B$ and $R$-band data from the ESO-LV Catalogue
\citep{esolv} assuming that $V$ is the average of $B$ and $R$ magnitudes
(assuming an error of 0.2 mag).
As a last resort, $V$ band fluxes were estimated from
RC3 $B_T$ values, assuming $B - V \sim 0.6$ for normal galaxies (assuming 
an error of 0.5 mag).  A more homogeneous and accurate optical magnitude
would be ideal; however, magnitudes of this accuracy will be adequate
for this study.

\subsection{IR Data}

The IR data were taken from (in order of preference) 
the catalog of IRAS observations of large
optical galaxies \citep{rice88}, the IRAS Bright Galaxy Sample
\citep{soifer89}, and the IRAS Faint Source Catalog
\citep{moshir90}.  Internal accuracy between the different
catalogs is typically better than 10\%, and the true uncertainties
(including zero point and calibration uncertainties) 
are $\sim 20$\% \citep[estimated by comparing 
ISO/IRAS cross-calibration; e.g.,][]{tuffs02}.

The 12{\micron}, 25{\micron}, 60{\micron} and 
100{\micron} fluxes are used to estimate total
IR fluxes in two ways.  The integrated 
42.5--122.5{\micron} emission is well-approximated
(to a few percent) by the FIR estimator of
\citet{helou88}:
\begin{equation}
{\rm FIR} = 1.26 \times 10^{-14}(2.58 S_{60 \mu {\rm m}} + S_{100 \mu {\rm m}})
\,{\rm W\,m^{-2}},
\end{equation}
where $S_{60 \mu {\rm m}}$ and $S_{100 \mu {\rm m}}$ are the 60{\micron} 
and 100{\micron} fluxes in Jy.  However, this definition of 
the FIR flux omits contributions shortwards of 42.5{\micron} and 
longwards of 122.5{\micron}: both spectral regions contribute
significantly to the total IR energy budget.  Following
\citet{fluxrat}, I estimate the total IR flux from 8{\micron} 
to 1000{\micron} by direct integration of the 12--100{\micron}
fluxes, and by extrapolating the flux longwards of 
100{\micron} using the 60{\micron} and 100{\micron} to define
the temperature of a modified blackbody curve with a 
$\lambda^{-1}$ emissivity.  If 12{\micron} or 25{\micron}
do not exist for a galaxy (89 and 87 galaxies out of 245, respectively), 
they are estimated using
$f_{12 \mu {\rm m}}({\rm Jy}) = 0.0326 f_{100 \mu {\rm m}}({\rm Jy})$ 
or $f_{25 \mu {\rm m}}({\rm Jy}) = 0.131 f_{60 \mu {\rm m}}({\rm Jy})$,
which were derived from the rest of the sample and are good to 
30\%.  The contribution to the total IR from the mid-IR component
is $\sim 20$\%, and so the total uncertainty introduced by this 
approximation is not large, $\la 10$\%.   Total IR (TIR) fluxes defined
in this way are typically a factor of two higher than 
the FIR estimator, with an obvious dependence on the 
60{\micron} to 100{\micron} ratio.  For reference, the TIR fluxes
are $\sim$30\% larger than the $F_{ir}$ 8--1000{\micron} 
estimator of \citet{sanders96}. 
I have used 150--205{\micron} {\it Infrared Space
Observatory} (ISO) measurements of 38 Virgo Cluster galaxies \citep{popescu02}
to check the extrapolation of the 12--100{\micron} 
fluxes using the $\lambda^{-1}$ emissivity.
I find that TIR estimates where the 12--170{\micron}
data are used are 10\% larger than TIR estimates 
where the 12--100{\micron} data are used, with a 10\% scatter.
Furthermore, this offset is 
{\it independent of dust temperature} between 
$\log_{10} (L_{\rm 100 \mu m} / L_{\rm 60 \mu m})$ values of 
0.1 to 0.8 (which covers most of my main sample galaxies).
Because the offset is modest and because the systematic 
errors in IR calibration are the same size or larger \citep{tuffs02},
I have chosen to leave my TIR values uncorrected for this offset.
Note that none of my conclusions are significantly affected by 
this 10\% offset: the TIR calibrations remain unchanged because
the calibration is, to first order, 
model-based and the radio calibration would 
increase by 10\% (in the sense that per unit radio flux, the SF
rate will be increased by 10\%) because the same radio flux
must reflect the 10\% increase in SF rate implied by the larger
TIR flux.   In this paper, I explore mostly the
ratio of TIR (8-1000{\micron}, extrapolated) fluxes
with radio; however the FIR to radio ratio is 
also briefly explored to check for consistency with the literature.
Errors in FIR and TIR are somewhat larger than the individual
flux errors, and are $\sim 30$\%.

\subsection{Radio Data}

Radio data for 166 galaxies at 1.4\,GHz were, for the most part, 
taken from the NRAO VLA Sky Survey \citep[NVSS;][]{c98}.
NVSS data were taken for 159 galaxies from \citet{condon02}, 
\citet{hopkins02}, and \citet{g99i,g99ii} in that order of preference.
Internally, the consistency between these studies of the NVSS was
$\la 10$\% for larger fluxes, and $\la 1$\,mJy for fainter sources.
Comparison of the NVSS fluxes with literature galaxy fluxes from
other, independent data was somewhat less clean: fluxes were 
repeatable to 20\% or so, degrading considerably for fainter
sources.  Noting the superior uniformity and resolution of 
the NVSS, I adopt error estimates of 10\% and 2\,mJy, 
to be added in quadrature, for these 159 galaxies.  I removed by 
hand a few galaxies which had highly discrepant 
(off by more than a factor of three) NVSS vs.\ literature
fluxes, where it was unclear which radio flux was more accurate, in the
interests of being as conservative as possible.
Additional data at frequencies between 1.4 and 1.5\,GHz (translated
to 1.4\,GHz assuming a $\nu^{-0.8}$ non-thermal spectrum) were 
taken from other sources for seven galaxies which were not in 
the above catalogs, but were important to have in the sample because
of their properties (ULIRGs or interacting pairs), or because
they had measured thermal radio fractions.  These radio data were extensively
cross-checked with many other radio catalogs, and 
were found to agree to within 20\% in most cases.  

\subsection{Other Data}

Exquisite distances are not central to the purpose of this paper,
as most of the diagnostics of the behavior of the \rf correlation 
are ratios of fluxes.  Nevertheless, in order to construct $L_{\rm TIR}$ and
$V$-band absolute magnitudes, distances are required to the sample
galaxies.  
Distances were taken from a variety of sources, and were scaled
roughly to reflect $H_0 = 75$\,km\,s$^{-1}$\,Mpc$^{-1}$ and a distance
to the Virgo and Ursa Major Clusters of 20\,Mpc 
\citep[e.g.,][]{shanks97,sakai00}.  Typical distance uncertainties of 
25\% will produce absolute magnitude errors of $\sim$0.5 mag, 
dwarfing the $V$-band apparent magnitude and TIR flux errors in most cases.

The galactic foreground extinction has been calculated using the 
models of \citet{sfd}.

\ha fluxes and attenuation estimates
were required in \S \ref{subsec:hafuv} in order
to establish the efficacy of TIR/FUV as a FUV attenuation 
indicator.  \ha fluxes for galaxies with \ha attenuation 
estimates in the literature were taken from a variety of sources.  
Inter-comparison of the fluxes indicates a $\sim$20\% uncertainty.
Thermal radio fractions
were taken from \citet{uit} for the UIT and FAUST galaxies
\citep[which in turn mostly come from][]{niklas97}, and from
\citet{niklas97} directly for the \citet{rifatto3} galaxies.
Balmer decrements were taken from \citet{calzetti94} and 
\citet{wu98} for starburst galaxies and ULIRGs respectively.
Average Balmer decrements for \hii regions in the UIT and FAUST 
galaxies are taken from \citet{uit}.
Thermal radio fractions and
Balmer decrements are difficult to do external comparisons on 
in detail, because
of their rarity in the literature, but there are substantial error
bars attached to each type of estimate because of 
the difficulty of disentangling the thermal and dominant non-thermal
contributions to the radio flux \citep[discussed extensively by 
e.g.,][]{condon92,niklas97} on one hand, and aperture mismatches and 
stellar absorption corrections, coupled with 
astrophysical uncertainties such as optical depth effects, on the other hand
\citep{caplan86}.

\section{B. Linking galaxy luminosity with optical depth} \label{app:model}

Observed attenuation--luminosity correlations 
\citep[e.g.,][]{wang96,buat99,hopkins01} are relatively
easy to understand, at least at a qualitative level, using
some simple arguments.  The basic argument
is that the metals-to-dust ratio is constant.  
Then the optical depth $\tau$ depends
only gas column density and gas metallicity $Z$.  
In turn, the gas column density varies as a function
of galaxy luminosity (tending to be rather higher for
more luminous galaxies), and the metallicity is higher
for more luminous galaxies.  Both effects tend to drive
a higher dust opacity in more luminous galaxies.   

The dust opacity $\tau$ at $\sim 1550${\AA} is 
given by $\tau = n_d C_{\rm ext}$
where $n_d$ is the number of dust grains along the line of sight
per unit cross section, and $C_{\rm ext}$ is the extinction
cross section \citep{whittet}.  The dust absorption 
cross section (which I use instead of the extinction cross section
because as much light will be scattered into the line
of sight as out of it) at 1550{\AA} is 
$\kappa_{\rm abs} = 4.2 \times 10^4$\,cm$^2$\,g$^{-1}$ \citep{li01}.
Using this, 
I obtain $\tau = 8.4 \Sigma_{\rm dust}$ where 
$\Sigma_{\rm dust}$ is in M$_{\sun}$\,pc$^{-2}$.  
Assuming that the dust-to-metals ratio 
is constant $\sim 0.2$ 
\citep[assuming solar metallicity and a gas/dust of 226;][]{sodroski97}, 
$\Sigma_{\rm dust} = 0.2 Z \Sigma_{\rm gas}$, where
$\Sigma_{\rm gas}$ is the gas density in M$_{\sun}$\,pc$^{-2}$.
However, trends in $\Sigma_{\rm gas}$ with luminosity are poorly
studied.  Therefore, I choose to approximate $\Sigma_{\rm gas}$ 
with the gas-to-stellar mass ratio multiplied by the stellar surface density
$\Sigma_*$.  Then, $\Sigma_{\rm dust} = 0.2 Z f_g / (1-f_g) \Sigma_*$, where 
$f_g$ is the gas fraction.  This crude estimation of the
gas densities from the gas fraction plus the stellar densities 
is clearly an over-simplification; however, it is clear 
that the general trend of increasing dust opacity with 
increasing luminosity is a robust one as {\it both}
the typically increasing metallicity and gas density
with galaxy luminosity (at least for star-forming galaxies)
will cause the dust opacity to 
increase with luminosity.
I now address each luminosity-dependent variable in turn.

The origin of the strong
metallicity--luminosity correlation 
\citep[e.g.,][]{skh89,ve92,zkh94,vanzee97,papii}
is not fully understood, but it is argued 
that both the greater degree of astration in more luminous
galaxies, and galaxy mass-dependent metal loss play a r\^ole
\citep[see, e.g.,][{ }for a thorough treatment of the issues]{pagel98}.
Following Fig.\ 16 of \citet{papii}, I adopt the following
metallicity--luminosity correlation:
$\log_{10} (Z/Z_{\sun}) = - 0.1875*V - 3.875$, where 
$Z_{\sun}$ = 0.02 is the solar metallicity, $V$ denotes
$V$-band absolute magnitude, and I assume $V - K \sim 3$.

There is a decreasing gas fraction
with increasing galaxy luminosity \citep[e.g.,][]{mcgaugh97,papii},
which is easily interpreted in terms of astration. More luminous 
galaxies tend to have formed stars with greater efficiency 
\citep[arguably because of their typically higher gas surface 
densities although other mechanisms 
are plausible; see, e.g.,][]{papii,ferreras01}
and have older stellar populations and lower gas fractions, albeit with 
a large scatter.  
I adopt a gas fraction--luminosity correlation:
$f_g = 0.5 + 1/\pi \arctan [0.5 (V + 19)]$, where $f_g$ is the gas fraction.
This is consistent with the correlation presented by \citet{papii}.

Lastly, the stellar
surface density $\Sigma_*$ is known to vary smooth\-ly with 
galaxy mass \citep{dejong00,sd} roughly as $\log_{10} \Sigma_* = 
2.34 - 0.213 (V+21)$, assuming that the stellar M/L in $V$-band
is roughly constant (which is wrong at only a factor of 3 level).  
Therefore, $\tau = 1.7 \eta Z f_g / (1-f_g) \Sigma_*$, where
$\eta$ is a constant of order unity that is tuned to fit the 
observed trend in TIR/FUV with luminosity: this constant 
allows me to fit out any of the crude modeling 
assumptions and accounts for the effects of star/dust
geometry on the opacity.  Note that star/dust geometry effects
will tend to drive $\eta$ below unity, as a star/dust mix attenuates
light less per unit mass than the screen model that I have assumed.
Given that the relationship between FUV attenuation
$\tau$ and TIR/FUV is
$\tau \sim 2.5 \log_{10} (1+{\rm TIR/FUV})$, 
it follows that $\log_{10} ({\rm TIR/FUV}) = \log_{10} (10^{0.4 \tau} - 1)$.
In order to match the data reasonably well (the solid curve in 
Fig.\ \ref{fig:firfuv}), I adopt $\eta = 0.7$.

In order to connect the TIR/radio ratio $q_{\rm TIR}$ with 
TIR/FUV, I assume that SF rate $\propto$ TIR$+$FUV, and
that radio is a perfect SF rate indicator.  Then, 
$q_{\rm TIR} = - \log_{10} \{1 + 1/(10^{0.4 a \tau} - 1) \} + q_0$, 
where $q_0$ is the intrinsic TIR/radio ratio which is
tuned to match the observed $q_{\rm TIR}$ at $\sim L_*$.

In \S \ref{sec:optical}, I correct the data for the effects of old
stellar populations.  In this case, I set $\eta = 0.5$, and
change the value of $q_0$ to match the observed
$q_{\rm corr}$ (the average $q_{\rm TIR}$, once the TIR has been corrected
for old stellar populations) at $\sim L_*$.

It is interesting to briefly note that the $\tau$ depends largely
on gas column density, modulated mostly by metallicity.  In the above
model, the gas column density was derived in a very statistical 
fashion by adopting a stellar surface density--luminosity correlation (which
has much intrinsic scatter), and a gas fraction--luminosity correlation
(which also has a lot of intrinsic scatter).  Thus, a 
very testable generic prediction of this type of model is 
that dust opacity should correlate well with a combination of 
gas column density and metallicity, with substantially less
scatter than the correlation between opacity and 
galaxy luminosity.  Testing this prediction in detail is far beyond 
the scope of this work, but it is interesting to note
that low surface brightness galaxies, with a very low gas column 
density, tend to be relatively dust-free \citep[e.g.,][]{matthews}, 
and ULIRGs, with very high
gas column densities, tend to be very dusty.  In addition,
the good correlation between FUV extinction as measured by 
TIR/FUV with total gas density \citep{buat92,xu97} is consistent
this this scheme.

\onecolumn
\clearpage

\begin{deluxetable}{lccccccccccccc}
\tablewidth{650pt}
\tablenum{A1}
\tabletypesize{\scriptsize}
\rotate
\tablecaption{Galaxy Luminosities {\label{fluxes}}} 
\tablehead{
\colhead{Galaxy} & \colhead{FUV} & \colhead{TIR} &
 \colhead{FIR$_{\rm Helou}$} & \colhead{100/60} & \colhead{1.4\,GHz} & 
\colhead{$f_{\rm thermal}$} & \colhead{$V$} & \colhead{$A_V$} & 
\colhead{$D$} & \colhead{\ha} & \colhead{$A_{\rm H\alpha,Balmer}$} &
\colhead{Type} & \colhead{References} \\
\colhead{Name} & \colhead{($\log_{10} [W/${\AA}$]^{-1}$)} & 
\colhead{($\log_{10} [W]^{-1}$)} & 
\colhead{($\log_{10} [W]^{-1}$)} & \colhead{} & 
\colhead{($\log_{10} [W/$Hz$]^{-1}$)} & \colhead{} & \colhead{(mag)} &
\colhead{(mag)} & \colhead{(Mpc)} & \colhead{($\log_{10} [W]^{-1}$)} & 
\colhead{(mag)} & \colhead{} }
\startdata
          NGC 598 & 32.35 $\pm$ 0.04 & 35.63 & 35.36 &  3.00 & 20.40 $\pm$ 0.04 &  \nd              & $-$18.95 $\pm$ 0.10 & 0.14 &   0.8 &  \nd  & \nd  & Scd & 1a,2a,4a \\
          NGC 628 & 33.00 $\pm$ 0.04 & 36.55 & 36.26 &  3.15 & 21.34 $\pm$ 0.04 &  \nd              & $-$20.84 $\pm$ 0.10 & 0.23 &  10.0 & 34.17 & 0.84 & Sc  & 1b\tablenotemark{a},2a \\
           DDO 81 & 32.09 $\pm$ 0.06 & 34.72 & 34.39 &  4.41 & 19.09 $\pm$ 0.08 &  \nd              & $-$17.22 $\pm$ 0.21 & 0.12 &   3.1 & 32.88 & \nd  & Sm  & 1b\tablenotemark{a},2a \\
           DDO 50 & 31.85 $\pm$ 0.02 & 34.18 & 33.91 &  2.28 & 19.36 $\pm$ 0.06 & 0.128 $\pm$ 0.020 & $-$16.90 $\pm$ 0.17 & 0.11 &   3.1 & 32.64 & 0.15 & Im  & 1b\tablenotemark{a},2a \\
         NGC 5457 & 33.45 $\pm$ 0.01 & 36.88 & 36.60 &  2.87 & 21.69 $\pm$ 0.04 & 0.063 $\pm$ 0.025 & $-$21.51 $\pm$ 0.10 & 0.03 &   7.4 & 34.49 & 0.53 & Scd  & 1b\tablenotemark{a},2a \\
         NGC 5055 & 32.54 $\pm$ 0.04 & 36.72 & 36.36 &  3.94 & 21.39 $\pm$ 0.04 & 0.119 $\pm$ 0.040 & $-$20.87 $\pm$ 0.10 & 0.06 &   7.6 & 33.95 & 1.22 & Sbc  & 1b\tablenotemark{a},2a \\
         NGC 3034 & 31.36 $\pm$ 0.15 & 37.16 & 36.88 &  1.06 & 21.93 $\pm$ 0.04 & 0.107 $\pm$ 0.036 & $-$19.72 $\pm$ 0.09 & 0.54 &   3.3 & 34.12 & 2.82 & I  & 1b\tablenotemark{a},2a \\
         NGC 3351 & 31.94 $\pm$ 0.06 & 36.20 & 35.98 &  1.99 & 20.62 $\pm$ 0.05 &  \nd              & $-$20.13 $\pm$ 0.10 & 0.09 &   9.0 & 33.67 & 0.69 & Sb & 1b\tablenotemark{a},2a \\
         NGC 2403 & 32.36 $\pm$ 0.03 & 35.85 & 35.58 &  2.88 & 20.55 $\pm$ 0.04 &  \nd              & $-$19.06 $\pm$ 0.08 & 0.13 &   3.0 & 33.69 & 0.38 & Scd & 1b\tablenotemark{a},2a \\
         NGC 5236 & 32.81 $\pm$ 0.02 & 36.75 & 36.44 &  2.40 &  \nd             &  \nd              & $-$20.52 $\pm$ 0.04 & 0.22 &   3.7 & 34.29 & 1.75 & Sc & 1b\tablenotemark{a} \\
           DDO 75 & 31.02 $\pm$ 0.04 & 32.91 & 32.63 &  2.63 &  \nd             &  \nd              & $-$14.48 $\pm$ 0.11 & 0.15 &   1.4 & 32.09 & \nd  & Im & 1b\tablenotemark{a} \\
         NGC 4449 & 32.42 $\pm$ 0.01 & 35.68 & 35.46 &  1.77 & 20.57 $\pm$ 0.04 & 0.225 $\pm$ 0.025 & $-$18.14 $\pm$ 0.13 & 0.07 &   3.4 & 33.55 & 0.46 & Im & 1b\tablenotemark{a},2a \\
         NGC 4736 & 32.34 $\pm$ 0.01 & 36.27 & 36.01 &  2.17 & 20.87 $\pm$ 0.04 & 0.214 $\pm$ 0.059 & $-$20.22 $\pm$ 0.13 & 0.06 &   4.8 & 33.76 & 0.46 & Sab & 1b\tablenotemark{a},2a \\
       NGC 4038/9 & 33.27 $\pm$ 0.01 & 37.32 & 37.09 &  1.69 & 22.45 $\pm$ 0.04 & 0.178 $\pm$ 0.025 & $-$21.44 $\pm$ 0.20 & 0.15 &  19.8 & 34.80 & 0.53 & Pec & 1b\tablenotemark{a,d} \\
          NGC 891 & \nodata          & 37.03 & 36.72 &  3.25 & 21.92 $\pm$ 0.04 &  \nd              & $-$20.26 $\pm$ 0.18 & 0.22 &   9.9 & 33.55 & \nd  & Sb & 1b\tablenotemark{a},2a \\
         NGC 4156 & 32.90 $\pm$ 0.09 &  \nd  &  \nd  &  \nd  & 21.98 $\pm$ 0.09 &  \nd              & $-$21.70 $\pm$ 0.05 & 0.09 &  90.0 &  \nd  & \nd  & Sb & 1b\tablenotemark{a},2a \\
         NGC 5253 & 32.03 $\pm$ 0.04 & 35.73 & 35.33 &  0.96 &  \nd             &  \nd              & $-$17.53 $\pm$ 0.12 & 0.18 &   3.6 & 33.47 & 0.61 & Im & 1b\tablenotemark{a} \\
         NGC 2903 & 32.29 $\pm$ 0.03 & 36.53 & 36.23 &  2.81 & 21.33 $\pm$ 0.04 & 0.195 $\pm$ 0.052 & $-$20.09 $\pm$ 0.10 & 0.10 &   6.3 & 33.87 & 0.91 & Sd & 1b\tablenotemark{a},2a \\
         NGC 6090 & 33.44 $\pm$ 0.01 & 37.98 & 37.72 &  1.49 & 22.90 $\pm$ 0.04 &  \nd              & $-$21.34 $\pm$ 0.10 & 0.07 & 117.0 & 35.07 & \nd  & Pec & 1b\tablenotemark{a},2a \\
         NGC 3310 & 33.16 $\pm$ 0.01 & 36.84 & 36.60 &  1.40 & 21.97 $\pm$ 0.04 & 0.047 $\pm$ 0.035 & $-$19.99 $\pm$ 0.10 & 0.07 &  13.9 & 34.48 & \nd  & Sbc & 1b\tablenotemark{a},2a \\
         NGC 4214 & 32.39 $\pm$ 0.02 & 35.52 & 35.30 &  1.63 & 19.91 $\pm$ 0.05 &  \nd              & $-$18.41 $\pm$ 0.15 & 0.07 &   4.2 & 33.48 & 0.23 & Im & 1b\tablenotemark{a},2a \\
           Mrk 66 & 33.29 $\pm$ 0.03 & 36.66 & 36.40 &  1.48 &  \nd             &  \nd              & $-$19.76 $\pm$ 0.10 & 0.04 &  87.0 & 34.09 & \nd  & BCG & 1b\tablenotemark{a} \\
         NGC 4631 & 33.09 $\pm$ 0.03 & 36.91 & 36.66 &  2.52 & 21.92 $\pm$ 0.04 & 0.037 $\pm$ 0.025 & $-$20.49 $\pm$ 0.16 & 0.06 &   8.4 & 34.26 & 0.84 & Sd & 1b\tablenotemark{a},2a \\
  IRAS 08339$+$6517 & 33.82 $\pm$ 0.02 & 37.55 & 37.33 &  1.01 &  \nd           &  \nd              & $-$21.21 $\pm$ 0.40 & 0.30 &  76.0 & 35.05 & \nd  & Pec & 1b\tablenotemark{a} \\
          NGC 925 & 32.96 $\pm$ 0.08 & 36.00 & 35.75 &  3.49 & 20.27 $\pm$ 0.06 & 0.597 $\pm$ 0.031 & $-$19.88 $\pm$ 0.12 & 0.25 &   8.9 & 33.87 & 0.76 & Sd & 1b\tablenotemark{a},2a \\
         NGC 1512 & 32.02 $\pm$ 0.04 & 35.72 & 35.44 &  3.50 &  \nd             &  \nd              & $-$19.67 $\pm$ 0.10 & 0.03 &   9.8 &  \nd  & \nd  & Sab & 1b\tablenotemark{a} \\
         NGC 1291 & 30.99 $\pm$ 0.11 & 35.58 & 35.22 &  5.76 &  \nd             &  \nd              & $-$21.25 $\pm$ 0.04 & 0.04 &   8.6 & 32.66 & \nd  & S0a & 1b\tablenotemark{a} \\
          NGC 253 & 32.21 $\pm$ 0.15 & 36.91 & 36.66 &  1.86 & 21.70 $\pm$ 0.04 & 0.122 $\pm$ 0.024 & $-$19.96 $\pm$ 0.20 & 0.06 &   2.6 & 33.69 & \nd  & Sc & 1b\tablenotemark{a,d} \\
         NGC 1313 & 32.64 $\pm$ 0.04 & 35.85 & 35.63 &  2.56 &  \nd             &  \nd              & $-$19.61 $\pm$ 0.20 & 0.37 &   3.9 & 33.68 & \nd  & Sd & 1b\tablenotemark{a} \\
         NGC 6946 & \nodata          & 36.90 & 36.61 &  2.52 & 21.81 $\pm$ 0.04 & 0.077 $\pm$ 0.013 & $-$21.31 $\pm$ 0.11 & 1.15 &   6.2 & 34.42 & 0.69 & Scd & 1b\tablenotemark{a},2a \\
         NGC 4321 & 33.06 $\pm$ 0.11 & 37.01 & 36.72 &  2.68 & 21.91 $\pm$ 0.04 &  \nd              & $-$21.76 $\pm$ 0.08 & 0.09 &  16.0 & 34.33 & 0.38 & Sbc & 1b\tablenotemark{a},2a \\
         UGC 6697 & 33.64 $\pm$ 0.05 & 37.19 & 36.92 &  1.89 & 22.72 $\pm$ 0.04 &  \nd              & $-$21.25 $\pm$ 0.10 & 0.07 &  90.0 & 34.73 & \nd  & Im & 1b\tablenotemark{a},2a \\
         NGC 3389 & 32.81 $\pm$ 0.01 & 36.48 & 36.23 &  2.53 & 21.47 $\pm$ 0.05 &  \nd              & $-$20.09 $\pm$ 0.06 & 0.09 &  24.0 & 34.03 & \nd  & Sc & 1b\tablenotemark{a},2a \\
         NGC 4647 & 32.41 $\pm$ 0.04 & 36.62 & 36.26 &  2.99 & 21.42 $\pm$ 0.04 &  \nd              & $-$20.30 $\pm$ 0.08 & 0.09 &  20.0 & 33.87 & \nd  & Sc & 1b\tablenotemark{a},2a \\
         NGC 1317 & 32.02 $\pm$ 0.03 & 36.31 & 36.04 &  2.88 &  \nd             &  \nd              & $-$20.55 $\pm$ 0.06 & 0.07 &  20.0 & 33.41 & \nd  & S0a & 1b\tablenotemark{a} \\
         NGC 2993 & 33.10 $\pm$ 0.02 & 37.09 & 36.86 &  1.59 &  \nd             &  \nd              & $-$20.09 $\pm$ 0.14 & 0.20 &  32.0 & 34.42 & \nd  & Sa & 1b\tablenotemark{a} \\
         NGC 2551 & 31.80 $\pm$ 0.18 & 35.73 & 35.38 &  4.04 &  \nd             &  \nd              & $-$20.44 $\pm$ 0.20 & 0.09 &  31.0 &  \nd  & \nd  & S0a & 1b\tablenotemark{a} \\
          Haro 15 & 33.47 $\pm$ 0.04 & 37.11 & 36.80 &  1.45 &  \nd             &  \nd              & $-$21.76 $\pm$ 0.20 & 0.07 &  86.7 & 34.49 & \nd  & Pec & 1c,3a,4b,5a \\
          IC 1586 & 33.03 $\pm$ 0.04 & 36.96 & 36.62 &  1.76 &  \nd             &  \nd              & $-$20.29 $\pm$ 0.20 & 0.14 &  81.3 & 34.38 & 1.21 & \hii & 1c,3a,4b,5a \\
           IC 214 & 33.50 $\pm$ 0.04 & 37.94 & 37.72 &  1.57 & 23.00 $\pm$ 0.04 &  \nd              & $-$21.83 $\pm$ 0.20 & 0.14 & 125.3 & 34.60 & 1.11 & \nd & 1c,2a,3a,4b,5a \\
          Mrk 499 & 33.17 $\pm$ 0.04 & 37.33 & 36.94 &  2.18 &  \nd             &  \nd              & $-$21.02 $\pm$ 0.20 & 0.05 &  98.6 & 34.25 & 0.94 & Im & 1c,3a,4b,5a \\
         NGC 1510 & 31.62 $\pm$ 0.04 & 35.02 & 34.80 &  1.27 &  \nd             &  \nd              & $-$17.25 $\pm$ 0.20 & 0.03 &  11.1 & 32.68 & 0.17 & Pec &  1c,3a,4b,5a \\
         NGC 1705 & \nd              & 34.62 & 34.40 &  1.89 &  \nd             &  \nd              & $-$16.69 $\pm$ 0.20 & 0.02 &   5.9 & 32.35 & \nd  & Pec & 3a,4b,5a\tablenotemark{b} \\
         NGC 1800 & \nd              & 34.87 & 34.60 &  2.28 &  \nd             &  \nd              & $-$17.09 $\pm$ 0.20 & 0.04 &   8.1 & 31.98 & \nd  & Im & 3a,4b,5a\tablenotemark{b} \\
         NGC 3049 & \nd              & 36.15 & 35.88 &  1.56 & 20.67 $\pm$ 0.09 &  \nd              & $-$19.29 $\pm$ 0.20 & 0.12 &  20.6 & 33.47 & \nd  & Sab & 2a,3a,4b,5a\tablenotemark{b} \\
         NGC 3125 & 32.41 $\pm$ 0.04 & 35.83 & 35.61 &  1.02 &  \nd             &  \nd              & $-$17.96 $\pm$ 0.20 & 0.24 &  12.1 & 33.53 & 0.27 & Sab & 1c,3a,4b,5a \\
         NGC 3256 & \nd              & 38.13 & 37.90 &  1.18 &  \nd             &  \nd              & $-$22.25 $\pm$ 0.20 & 0.38 &  37.7 & 34.89 & \nd  & Pec & 3a,4b,5a\tablenotemark{b} \\
         NGC 4194 & 32.52 $\pm$ 0.04 & 37.51 & 37.26 &  1.07 & 22.22 $\pm$ 0.04 &  \nd              & $-$20.39 $\pm$ 0.20 & 0.05 &  37.0 & 34.54 & 1.67 & Im & 1c,2a,3a,4b,5a \\ 
         NGC 4385 & 32.53 $\pm$ 0.04 & 36.78 & 36.48 &  1.28 & 21.24 $\pm$ 0.07 &  \nd              & $-$20.18 $\pm$ 0.20 & 0.08 &  33.1 & 34.13 & \nd  & S0 & 2a,3a,4b,5a\tablenotemark{b} \\
         UGC 9560 & 32.10 $\pm$ 0.04 & 35.42 & 35.13 &  1.75 & 20.31 $\pm$ 0.13 &  \nd              & $-$17.49 $\pm$ 0.20 & 0.04 &  17.0 & 33.30 & 0.31 & Pec & 1c,2a,3a,4b,5a \\ 
         NGC 5860 & 32.80 $\pm$ 0.04 & 37.03 & 36.77 &  1.84 & 21.70 $\pm$ 0.11 &  \nd              & $-$20.69 $\pm$ 0.20 & 0.06 &  73.5 & 34.39 & 1.44 & \nd & 1c,2a,3a,4b,5a \\ 
         NGC 5996 & 32.50 $\pm$ 0.04 & 36.65 & 36.40 &  1.85 & 21.51 $\pm$ 0.05 &  \nd              & $-$20.21 $\pm$ 0.20 & 0.11 &  30.2 & 33.82 & 0.98 & Sc & 1c,2a,3a,4b,5a \\ 
         NGC 6052 & 33.33 $\pm$ 0.04 & 37.36 & 37.13 &  1.66 & 22.65 $\pm$ 0.04 &  \nd              & $-$21.08 $\pm$ 0.20 & 0.24 &  58.6 & 34.46 & 0.44 & \nd & 1c,2a,3a,4b,5a \\ 
     Tololo 1924$-$416 & 33.63 $\pm$ 0.04 & 36.38 & 36.09 &  0.60 &  \nd        &  \nd              & $-$20.61 $\pm$ 0.20 & 0.27 &  38.7 & 34.50 & 0.04 & Pec & 1c,3a,4b,5a \\  
         NGC 7250 & 33.15 $\pm$ 0.04 & 35.94 & 35.75 &  1.34 & 20.82 $\pm$ 0.06 &  \nd              & $-$19.87 $\pm$ 0.20 & 0.47 &  16.6 & 33.52 & 0.19 & Sdm & 1c,2a,3a,4b,5a \\   
         NGC 7552 & \nd              & 37.69 & 37.45 &  1.42 &  \nd             &  \nd              & $-$21.32 $\pm$ 0.20 & 0.04 &  24.9 & 34.25 & \nd  & Sab & 3a,4b,5a\tablenotemark{b} \\
         NGC 7673 & 33.15 $\pm$ 0.04 & 36.96 & 36.78 &  1.40 & 22.03 $\pm$ 0.05 &  \nd              & $-$21.00 $\pm$ 0.20 & 0.13 &  45.1 & 34.28 & 0.88 & Pec & 1c,2a,3a,4b,5a \\   
         NGC 7714 & \nd              & 36.91 & 36.62 &  1.06 & 21.73 $\pm$ 0.04 &  \nd              & $-$20.04 $\pm$ 0.20 & 0.16 &  26.1 & 34.46 & \nd  & Sab & 2a,3a,4b,5a\tablenotemark{b} \\
         NGC 7793 & \nd              & 35.44 & 35.16 &  2.87 &  \nd             &  \nd              & $-$18.44 $\pm$ 0.20 & 0.06 &   3.0 & 31.42 & \nd  & Sab & 3a,4b,5a\tablenotemark{b} \\
           VV 114 & 33.79 $\pm$ 0.01 & 38.19 & 37.96 &  1.36 & 23.31 $\pm$ 0.04 &  \nd              & $-$21.28 $\pm$ 0.30 & 0.05 &  82.8 &  \nd  & \nd  & ULIRG & 1d,2b,3b,4b \\
  IRAS 08572$+$3915 & 32.96 $\pm$ 0.10 & 38.63 & 38.35 &  0.66 & 22.54 $\pm$ 0.15 &  \nd              & $-$19.92 $\pm$ 0.50 & 0.09 & 243.0 &  \nd  & \nd  & ULIRG & 1d,2c,3c,4b \\
           IC 883 & 32.41 $\pm$ 0.06 & 38.10 & 37.92 &  1.82 & 23.05 $\pm$ 0.04 &  \nd              & $-$21.03 $\pm$ 0.30 & 0.04 &  95.2 & 33.59 & 5.59 & ULIRG & 1d,2a,3d,4b,5b \\
          Mrk 273 & 32.75 $\pm$ 0.05 & 38.62 & 38.47 &  1.01 & 23.68 $\pm$ 0.04 &  \nd              & $-$21.44 $\pm$ 0.30 & 0.03 & 157.0 & 34.40 & 3.09 & ULIRG & 1d,2a,3d,4b,5b \\
  IRAS 15250$+$3609 & 33.09 $\pm$ 0.04 & 38.51 & 38.30 &  0.81 & 22.92 $\pm$ 0.07 &  \nd              &   \nd             & 0.06 & 231.7 &  \nd  & \nd  & ULIRG & 1d,2c,4b \\
          Arp 220 & 31.83 $\pm$ 0.12 & 38.62 & 38.50 &  1.11 & 23.33 $\pm$ 0.04 &  \nd              & $-$21.07 $\pm$ 0.30 & 0.17 &  74.1 & 32.43 & \nd  & ULIRG & 1d,2a,3d,4b,5b \\
  IRAS 19254$-$7245 & 32.77 $\pm$ 0.10 & 38.56 & 38.30 &  1.05 & 24.33 $\pm$ 0.04 &  \nd              &   \nd             & 0.28 & 255.2 &  \nd  & \nd  & ULIRG & 1d,2d,4b \\
         NGC 4592 & 31.87 $\pm$ 0.08 & 35.44 & 35.15 &  2.36 & 19.89 $\pm$ 0.12 &  \nd              & $-$18.43 $\pm$ 0.30 & 0.07 &   9.8 &  \nd  & \nd  & Sdm & 1e,2a,4d \\
       PGC 043701 & 33.60 $\pm$ 0.08 & 37.14 & 36.85 &  2.69 &  \nd             &  \nd              & $-$21.76 $\pm$ 0.05 & 0.31 &  56.4 &  \nd  & \nd  & Sb & 1e,3e,4e \\
         NGC 4930 & 33.05 $\pm$ 0.08 & 36.18 & 35.80 &  5.13 &  \nd             &  \nd              & $-$21.95 $\pm$ 0.30 & 0.36 &  34.5 &  \nd  & \nd  & Sbc & 1e,4e \\
         NGC 4793 & 32.80 $\pm$ 0.08 & 37.28 & 37.00 &  2.24 & 22.17 $\pm$ 0.04 &  \nd              & $-$21.00 $\pm$ 0.22 & 0.04 &  33.1 &  \nd  & \nd  & Sc & 1e,2a,4e \\
          IC 2050 & 34.59 $\pm$ 0.08 & 37.18 & 36.88 &  2.95 &  \nd             &  \nd              & $-$22.44 $\pm$ 0.30 & 0.05 & 164.9 &  \nd  & \nd  & Sbc & 1e,4e \\
         NGC 1536 & 32.16 $\pm$ 0.08 & 35.46 & 35.11 &  3.47 &  \nd             &  \nd              & $-$18.74 $\pm$ 0.14 & 0.07 &  17.3 &  \nd  & \nd  & Sc & 1e,4e \\
          IC 2073 & 33.03 $\pm$ 0.08 & 36.42 & 36.17 &  2.10 &  \nd             &  \nd              & $-$19.87 $\pm$ 0.15 & 0.03 &  53.0 &  \nd  & \nd  & Scd & 1e,4e \\
         NGC 1602 & 32.15 $\pm$ 0.08 & 35.58 & 35.31 &  2.45 &  \nd             &  \nd              & $-$18.20 $\pm$ 0.16 & 0.03 &  17.0 &  \nd  & \nd  & Im & 1e,4e \\
         NGC 5264 & 31.60 $\pm$ 0.08 & 33.96 & 33.68 &  2.54 &  \nd             &  \nd              & $-$16.43 $\pm$ 0.15 & 0.17 &   4.5 &  \nd  & \nd  & Im & 1e,4f \\
       PGC 047958 & 33.05 $\pm$ 0.08 & 36.61 & 36.32 &  2.90 &  \nd             &  \nd              & $-$20.15 $\pm$ 0.20 & 0.17 &  60.9 &  \nd  & \nd  & I & 1e,3f,4e \\
          IC 4275 & 33.35 $\pm$ 0.08 & 36.55 & 36.28 &  2.36 &  \nd             &  \nd              & $-$20.17 $\pm$ 0.20 & 0.19 &  57.5 &  \nd  & \nd  & S & 1e,3f,4e \\
          IC 4248 & 33.02 $\pm$ 0.08 & 36.83 & 36.59 &  1.93 &  \nd             &  \nd              & $-$20.46 $\pm$ 0.20 & 0.21 &  55.1 &  \nd  & \nd  & S & 1e,3f,4e \\
         NGC 3956 & 32.51 $\pm$ 0.08 & 36.07 & 35.77 &  2.89 &  \nd             &  \nd              & $-$19.42 $\pm$ 0.20 & 0.13 &  21.9 &  \nd  & \nd  & Sc & 1e,3f,4e \\
         NGC 4027 & 32.74 $\pm$ 0.08 & 36.89 & 36.63 &  2.21 &  \nd             &  \nd              & $-$20.76 $\pm$ 0.04 & 0.14 &  22.3 &  \nd  & \nd  & Sdm & 1e,4e \\
         NGC 6753 & 33.26 $\pm$ 0.08 & 37.39 & 37.13 &  2.90 &  \nd             &  \nd              & $-$22.18 $\pm$ 0.07 & 0.23 &  41.7 &  \nd  & \nd  & Sb & 1e,4e \\
          IC 4845 & 32.94 $\pm$ 0.08 & 36.86 & 36.50 &  3.79 &  \nd             &  \nd              & $-$22.23 $\pm$ 0.14 & 0.19 &  52.7 &  \nd  & \nd  & Sab & 1e,4e \\
          IC 4836 & 33.02 $\pm$ 0.08 & 37.03 & 36.73 &  2.92 &  \nd             &  \nd              & $-$21.22 $\pm$ 0.13 & 0.18 &  54.8 &  \nd  & \nd  & Sc & 1e,4e \\
          IC 4819 & 32.29 $\pm$ 0.08 & 35.45 & 35.14 &  3.31 &  \nd             &  \nd              & $-$18.47 $\pm$ 0.20 & 0.20 &  24.5 &  \nd  & \nd  & Sd & 1e,4e \\
          IC 4828 & 33.10 $\pm$ 0.08 & 36.16 & 35.83 &  3.69 &  \nd             &  \nd              & $-$19.48 $\pm$ 0.20 & 0.19 &  51.9 &  \nd  & \nd  & S & 1e,4e \\
       PGC 062709 & 33.80 $\pm$ 0.08 & 37.29 & 36.98 &  3.25 &  \nd             &  \nd              & $-$22.41 $\pm$ 0.20 & 0.21 & 138.2 &  \nd  & \nd  & Sbc & 1e,4e \\
          IC 4820 & 32.99 $\pm$ 0.08 & 36.08 & 35.79 &  2.58 &  \nd             &  \nd              & $-$19.12 $\pm$ 0.21 & 0.15 &  52.3 &  \nd  & \nd  & Sd & 1e,4e \\
       PGC 039904 & 32.04 $\pm$ 0.08 & 35.11 & 34.84 &  2.30 &  \nd             &  \nd              & $-$16.72 $\pm$ 0.50 & 0.11 &  20.0 &  \nd  & \nd  & BCD & 1e,4g \\
         NGC 4204 & 31.97 $\pm$ 0.08 & 35.00 & 34.69 &  2.57 &  \nd             &  \nd              & $-$17.70 $\pm$ 0.60 & 0.11 &   9.5 &  \nd  & \nd  & Sdm & 1e,4h \\
       PGC 039194 & 33.28 $\pm$ 0.08 & 36.83 & 36.44 &  2.82 & 21.52 $\pm$ 0.18 &  \nd              & $-$20.59 $\pm$ 0.15 & 0.09 &  83.1 &  \nd  & \nd  & Sc & 1e,2a,3g,4e \\
         NGC 4158 & 32.59 $\pm$ 0.08 & 36.22 & 35.85 &  3.40 & 20.53 $\pm$ 0.25 &  \nd              & $-$20.18 $\pm$ 0.40 & 0.11 &  32.8 &  \nd  & \nd  & Sb & 1e,2a,3h,4e \\
        A 1211$+$16 & 33.44 $\pm$ 0.08 & 36.71 & 36.47 &  1.70 & 21.71 $\pm$ 0.16 &  \nd            & $-$21.11 $\pm$ 0.50 & 0.12 &  95.0 &  \nd  & \nd  & \nd & 1e,2a,4e \\
       PGC 038750 & 33.39 $\pm$ 0.08 &  \nd  &  \nd  &  \nd  & 21.78 $\pm$ 0.13 &  \nd              & $-$20.24 $\pm$ 0.50 & 0.14 &  91.0 &  \nd  & \nd  & E & 1e,2e,4e \\
         NGC 4049 & 32.07 $\pm$ 0.08 & 35.29 & 35.00 &  2.72 &  \nd             &  \nd              & $-$17.89 $\pm$ 0.50 & 0.08 &  20.0 &  \nd  & \nd  & I & 1e,4g \\
         NGC 4032 & 32.35 $\pm$ 0.08 & 35.63 & 35.35 &  2.47 & 20.50 $\pm$ 0.12 &  \nd              & $-$19.37 $\pm$ 0.15 & 0.11 &  20.0 &  \nd  & \nd  & Im & 1e,2a,4g \\
         NGC 4455 & 32.09 $\pm$ 0.08 & 35.01 & 34.69 &  3.32 &  \nd             &  \nd              & $-$18.09 $\pm$ 0.50 & 0.07 &  11.6 &  \nd  & \nd  & Sd & 1e,4i \\
         NGC 4635 & 32.42 $\pm$ 0.08 & 35.75 & 35.44 &  3.25 &  \nd             &  \nd              & $-$19.47 $\pm$ 0.20 & 0.09 &  26.0 &  \nd  & \nd  & Sd & 1e,3i,4d \\
         NGC 4615 & 33.45 $\pm$ 0.08 & 36.90 & 36.60 &  2.89 & 21.71 $\pm$ 0.08 &  \nd              & $-$20.74 $\pm$ 0.15 & 0.05 &  62.9 &  \nd  & \nd  & Scd & 1e,2a,3g,4e \\
          IC 3591 & 32.10 $\pm$ 0.08 & 35.28 & 35.05 &  1.60 &  \nd             &  \nd              & $-$17.96 $\pm$ 0.50 & 0.08 &  21.8 &  \nd  & \nd  & Sm & 1e,4i \\
         NGC 4532 & 33.11 $\pm$ 0.08 & 36.82 & 36.62 &  1.74 & 22.03 $\pm$ 0.04 &  \nd              & $-$20.31 $\pm$ 0.09 & 0.07 &  26.8 &  \nd  & \nd  & Im & 1e,2a,4e \\
          IC 3521 & 31.88 $\pm$ 0.08 & 35.74 & 35.48 &  2.23 & 20.24 $\pm$ 0.19 &  \nd              & $-$18.38 $\pm$ 0.50 & 0.07 &  20.0 &  \nd  & \nd  & Sm & 1e,2a,4g \\
          IC 3414 & 32.05 $\pm$ 0.08 & 35.21 & 34.90 &  3.17 &  \nd             &  \nd              & $-$18.17 $\pm$ 0.50 & 0.06 &  20.0 &  \nd  & \nd  & Sm & 1e,4g \\
         NGC 4423 & 32.06 $\pm$ 0.08 & 35.42 & 35.15 &  2.33 & 20.16 $\pm$ 0.22 &  \nd              & $-$18.17 $\pm$ 0.50 & 0.07 &  20.0 &  \nd  & \nd  & Sm & 1e,2a,4g \\
         NGC 4430 & 32.32 $\pm$ 0.08 & 35.94 & 35.62 &  3.60 & 20.50 $\pm$ 0.12 &  \nd              & $-$19.37 $\pm$ 0.50 & 0.06 &  20.0 &  \nd  & \nd  & Sb & 1e,2a,4g \\
         NGC 4470 & 32.48 $\pm$ 0.08 & 36.02 & 35.74 &  2.45 & 20.87 $\pm$ 0.07 &  \nd              & $-$19.18 $\pm$ 0.50 & 0.08 &  20.0 &  \nd  & \nd  & Sa & 1e,2a,4g \\
         NGC 4376 & 32.22 $\pm$ 0.08 & 35.54 & 35.28 &  2.20 &  \nd             &  \nd              & $-$18.38 $\pm$ 0.50 & 0.08 &  20.0 &  \nd  & \nd  & Im & 1e,4g \\ 
         IC 3322A & 32.23 $\pm$ 0.08 & 36.29 & 35.97 &  3.16 & 20.98 $\pm$ 0.07 &  \nd              & $-$19.07 $\pm$ 0.50 & 0.08 &  25.0 &  \nd  & \nd  & Im & 1e,2a,4d$+$4i \\ 
          IC 3268 & 32.26 $\pm$ 0.08 & 35.59 & 35.32 &  2.40 & 20.33 $\pm$ 0.17 &  \nd              & $-$18.09 $\pm$ 0.50 & 0.08 &  20.0 &  \nd  & \nd  & Sm & 1e,2a,4g \\ 
       PGC 040993 & 32.41 $\pm$ 0.08 & 35.53 & 35.26 &  2.18 & 20.44 $\pm$ 0.20 &  \nd              & $-$18.64 $\pm$ 0.50 & 0.07 &  26.0 &  \nd  & \nd  & Sbc & 1e,2a,4i \\ 
         NGC 4451 & 32.36 $\pm$ 0.08 & 36.43 & 36.14 &  2.71 & 20.89 $\pm$ 0.12 &  \nd              & $-$20.11 $\pm$ 0.13 & 0.06 &  32.0 &  \nd  & \nd  & Sbc & 1e,2a,4i \\ 
         NGC 4276 & 32.52 $\pm$ 0.08 & 36.04 & 35.72 &  3.45 &  \nd             &  \nd              & $-$20.01 $\pm$ 0.50 & 0.09 &  34.9 &  \nd  & \nd  & Sc & 1e,4e \\ 
          IC 3107 & 33.30 $\pm$ 0.08 & 36.97 & 36.68 &  2.80 & 21.86 $\pm$ 0.12 &  \nd              & $-$21.47 $\pm$ 0.50 & 0.13 &  97.2 &  \nd  & \nd  & Sbc & 1e,2a,4e \\ 
         NGC 4383 & 32.62 $\pm$ 0.08 & 36.53 & 36.32 &  1.51 & 21.25 $\pm$ 0.05 &  \nd              & $-$19.46 $\pm$ 0.10 & 0.08 &  20.0 &  \nd  & \nd  & Sa & 1e,2a,4g \\ 
          IC 0800 & 32.15 $\pm$ 0.08 & 35.40 & 35.10 &  2.93 &  \nd             &  \nd              & $-$18.13 $\pm$ 0.50 & 0.12 &  20.0 &  \nd  & \nd  & Sc & 1e,4g \\ 
         NGC 4523 & 32.59 $\pm$ 0.08 & 35.46 & 35.14 &  3.41 & 20.14 $\pm$ 0.23 &  \nd              & $-$17.52 $\pm$ 0.16 & 0.13 &  20.0 &  \nd  & \nd  & Sm & 1e,2a,4g \\ 
         NGC 4396 & 32.35 $\pm$ 0.08 & 35.94 & 35.62 &  3.39 & 20.99 $\pm$ 0.06 &  \nd              & $-$19.01 $\pm$ 0.12 & 0.09 &  20.0 &  \nd  & \nd  & Sd & 1e,2a,4g \\ 
          IC 0797 & 32.18 $\pm$ 0.08 & 35.69 & 35.40 &  2.93 & 20.83 $\pm$ 0.07 &  \nd              & $-$18.61 $\pm$ 0.50 & 0.10 &  20.0 &  \nd  & \nd  & Sc & 1e,2f,4g \\ 
          IC 3476 & 32.43 $\pm$ 0.08 & 35.88 & 35.61 &  2.40 & 20.56 $\pm$ 0.11 &  \nd              & $-$18.90 $\pm$ 0.15 & 0.12 &  20.0 &  \nd  & \nd  & Im & 1e,2a,4g \\ 
         NGC 4670 & 32.20 $\pm$ 0.08 & 35.60 & 35.38 &  1.70 & 20.36 $\pm$ 0.07 &  \nd              & $-$17.72 $\pm$ 0.15 & 0.05 &  11.8 & \nd   & \nd  & S0 & 1e,2a,4j \\ 
         NGC 6744 & 32.80 $\pm$ 0.08 & 36.32 & 35.96 &  3.86 &  \nd             &  \nd              & $-$19.62 $\pm$ 0.20 & 0.14 &   6.5 & \nd   & \nd  & Sbc & 1e,3k,4k \\ 
         NGC 4152 & 32.64 $\pm$ 0.08 & 36.33 & 36.07 &  2.12 & 21.19 $\pm$ 0.05 &  \nd              & $-$19.40 $\pm$ 0.11 & 0.11 &  20.0 & \nd   & \nd  & Sc & 1e,2a,4g \\ 
         NGC 4651 & 32.54 $\pm$ 0.08 & 36.55 & 36.27 &  2.71 & 21.21 $\pm$ 0.05 &  \nd              & $-$20.77 $\pm$ 0.09 & 0.09 &  20.0 & \nd   & \nd  & Sc & 1e,2a,4g \\ 
         NGC 4689 & 32.35 $\pm$ 0.08 & 36.37 & 36.03 &  4.31 & 20.78 $\pm$ 0.08 &  \nd              & $-$20.63 $\pm$ 0.09 & 0.07 &  20.0 & \nd   & \nd  & Sbc & 1e,2a,4g \\ 
         NGC 4535 & 32.95 $\pm$ 0.08 & 36.68 & 36.38 &  2.94 & 21.38 $\pm$ 0.04 &  \nd              & $-$21.12 $\pm$ 0.09 & 0.06 &  16.0 & \nd   & \nd  & Sc & 1e,2a,4l \\
         NGC 4519 & 32.77 $\pm$ 0.08 & 36.31 & 36.05 &  1.71 & 20.97 $\pm$ 0.06 &  \nd              & $-$19.86 $\pm$ 0.07 & 0.07 &  21.0 & \nd   & \nd  & Sd & 1e,2a,4d \\
         NGC 4522 & 32.01 $\pm$ 0.08 & 35.76 & 35.45 &  3.23 & 20.81 $\pm$ 0.05 &  \nd              & $-$18.83 $\pm$ 0.40 & 0.07 &  15.6 & \nd   & \nd  & Scd & 1e,2a,4d \\
       NGC 4567/8 & 32.51 $\pm$ 0.08 & 36.99 & 36.74 &  2.83 & 21.81 $\pm$ 0.04 &  \nd              & $-$21.31 $\pm$ 0.30 & 0.11 &  20.0 & \nd   & \nd  & Int & 1e,2f,4g \\
         NGC 4416 & 32.84 $\pm$ 0.08 & 36.46 & 36.13 &  2.90 & 21.01 $\pm$ 0.15 &  \nd              & $-$20.80 $\pm$ 0.40 & 0.08 &  42.0 & \nd   & \nd  & Scd & 1e,2a,4i \\
        NGC 4411B & 32.79 $\pm$ 0.08 & 35.88 & 35.52 &  4.45 &  \nd             &  \nd              & $-$20.01 $\pm$ 0.11 & 0.10 &  28.0 & \nd   & \nd  & Scd & 1e,4i \\
         NGC 4424 & 31.90 $\pm$ 0.08 & 36.17 & 35.94 &  1.79 & 20.27 $\pm$ 0.18 &  \nd              & $-$19.91 $\pm$ 0.09 & 0.07 &  20.0 & \nd   & \nd  & Sa & 1e,2a,4g \\
         NGC 4380 & 32.09 $\pm$ 0.08 & 35.84 & 35.46 &  4.95 &  \nd             &  \nd              & $-$19.91 $\pm$ 0.10 & 0.08 &  20.0 & \nd   & \nd  & Sb & 1e,4g \\
         NGC 4438 & 32.30 $\pm$ 0.08 & 36.31 & 36.10 &  2.75 & 21.48 $\pm$ 0.04 &  \nd              & $-$21.43 $\pm$ 0.07 & 0.09 &  20.0 & \nd   & \nd  & S0 & 1e,2a,4g \\
         NGC 4413 & 32.37 $\pm$ 0.08 & 35.85 & 35.55 &  3.07 &  \nd             &  \nd              & $-$19.36 $\pm$ 0.16 & 0.10 &  20.0 & \nd   & \nd  & Sab & 1e,4g \\
         NGC 4351 & 32.16 $\pm$ 0.08 & 35.66 & 35.37 &  2.85 &  \nd             &  \nd              & $-$18.97 $\pm$ 0.15 & 0.10 &  20.0 & \nd   & \nd  & Sab & 1e,4g \\
         NGC 4299 & 32.56 $\pm$ 0.08 &  \nd  &  \nd  &  \nd  & 20.92 $\pm$ 0.06 &  \nd              & $-$19.13 $\pm$ 0.13 & 0.11 &  20.0 & \nd   & \nd  & Sdm & 1e,2a,4g \\
         NGC 4178 & 32.62 $\pm$ 0.08 & 36.21 & 35.91 &  3.83 & 21.16 $\pm$ 0.05 &  \nd              & $-$20.19 $\pm$ 0.09 & 0.09 &  20.0 & \nd   & \nd  & Sdm & 1e,2a,4g \\
         NGC 4498 & 32.49 $\pm$ 0.08 & 35.95 & 35.64 &  3.25 & 20.28 $\pm$ 0.18 &  \nd              & $-$19.40 $\pm$ 0.40 & 0.10 &  20.0 & \nd   & \nd  & Sd & 1e,2a,4g \\
         NGC 4595 & 32.32 $\pm$ 0.08 & 35.82 & 35.50 &  3.09 & 20.35 $\pm$ 0.16 &  \nd              & $-$19.33 $\pm$ 0.40 & 0.12 &  20.0 & \nd   & \nd  & Sb & 1e,2a,4g \\
         NGC 4654 & 32.91 $\pm$ 0.08 & 36.93 & 36.65 &  2.67 & 21.77 $\pm$ 0.04 & 0.210 $\pm$ 0.062 & $-$21.09 $\pm$ 0.10 & 0.08 &  20.0 & 34.15 & 0.61 & Scd & 1e,2a,4g,5c \\ 
         NGC 4298 & 32.39 $\pm$ 0.08 & 36.77 & 36.42 &  3.85 &  \nd             &  \nd              & $-$20.28 $\pm$ 0.09 & 0.12 &  20.0 & \nd   & \nd  & Sc & 1e,4g \\
         NGC 4254 & 33.14 $\pm$ 0.08 & 37.34 & 37.04 &  2.67 & 22.31 $\pm$ 0.04 & 0.065 $\pm$ 0.039 & $-$21.76 $\pm$ 0.08 & 0.13 &  20.0 & 34.73 & 0.84 & Sc & 1e,2a,4g,5c$+$5d \\
         NGC 0450 & \nodata          & 36.12 & 35.88 &  2.05 & 20.80 $\pm$ 0.08 &  \nd              & $-$18.97 $\pm$ 0.30 & 0.13 &  21.3 &  \nd  & \nd  & BCD & 2a,3l,4m \\
         NGC 1741 & \nodata          & 37.15 & 36.94 &  1.49 & 22.13 $\pm$ 0.05 &  \nd              & $-$21.45 $\pm$ 0.50 & 0.17 &  59.8 &  \nd  & \nd  & BCD & 2g,4m \\
         NGC 2366 & \nodata          & 34.76 & 34.50 &  1.33 & 19.56 $\pm$ 0.06 &  \nd              & $-$17.23 $\pm$ 0.12 & 0.12 &   3.9 &  \nd  & \nd  & BCD & 2a,4m \\ 
         MRK 0162 & \nodata          & 37.07 & 36.84 &  1.23 & 22.25 $\pm$ 0.06 &  \nd              & $-$20.66 $\pm$ 0.50 & 0.03 &  96.6 &  \nd  & \nd  & BCD & 2g,4m \\ 
           UM 448 & \nodata          & 37.37 & 37.15 &  1.04 & 22.38 $\pm$ 0.05 &  \nd              & $-$20.36 $\pm$ 0.50 & 0.08 &  78.5 &  \nd  & \nd  & BCD & 2a,4m \\ 
        UGC 06850 & \nodata          & 35.37 & 35.17 &  0.95 & 20.31 $\pm$ 0.13 &  \nd              & $-$17.24 $\pm$ 0.50 & 0.06 &  17.2 &  \nd  & \nd  & BCD & 2a,4m \\ 
         NGC 4861 & 33.11 $\pm$ 0.12 & 36.11 & 35.86 &  1.25 & 20.97 $\pm$ 0.08 &  \nd              & $-$19.72 $\pm$ 0.31 & 0.03 &  25.2 &  \nd  & \nd  & BCD & 2a,4m \\ 
     SBS 1533$+$574 & \nodata        & 35.89 & 35.66 &  1.58 & 21.16 $\pm$ 0.17 &  \nd              & $-$17.97 $\pm$ 0.60 & 0.04 &  53.2 &  \nd  & \nd  & BCD & 2g,3m,4m \\ 
         NGC 2146 & 32.32 $\pm$ 0.11 & 37.45 & 37.22 &  1.41 & 22.43 $\pm$ 0.04 & 0.148 $\pm$ 0.040 & $-$20.53 $\pm$ 0.13 & 0.31 &  14.5 & 34.05 & \nd  & Sab & 1f,1b\tablenotemark{c},2a \\
         NGC 2595 & 33.41 $\pm$ 0.07 & 36.74 & 36.43 &  3.29 & 21.65 $\pm$ 0.08 &  \nd              & $-$21.67 $\pm$ 0.14 & 0.13 &  57.7 &  \nd  & \nd  & Sc & 1f,2a,4e \\
         NGC 2976 & 31.55 $\pm$ 0.12 & 35.32 & 35.05 &  2.72 & 19.83 $\pm$ 0.04 &  \nd              & $-$17.67 $\pm$ 0.13 & 0.23 &   3.3 &  \nd  & \nd  & Sc & 1f,2a,4e \\
         NGC 3027 & 32.53 $\pm$ 0.12 & 35.47 & 35.20 &  3.02 &  \nd             &  \nd              & $-$19.10 $\pm$ 0.16 & 0.11 &  14.1 &  \nd  & \nd  & Sd & 1f,4e \\
         NGC 3077 & 31.64 $\pm$ 0.08 & 35.43 & 35.20 &  1.83 & 19.75 $\pm$ 0.05 &  \nd              & $-$18.38 $\pm$ 0.13 & 0.22 &   4.0 &  \nd  & \nd  & I & 1f,2a,4n \\
          NGC 3206 & 32.35 $\pm$ 0.12 & 35.44 & 35.17 &  2.30 &  \nd             &  \nd             & $-$18.98 $\pm$ 0.50 & 0.05 &  15.4 &  \nd  & \nd  & Scd & 1f,4e \\
        A 1029$+$54 & 32.38 $\pm$ 0.12 & 36.29 & 36.03 &  1.14 & 20.91 $\pm$ 0.06 &  \nd            & $-$18.77 $\pm$ 0.50 & 0.04 &  20.2 &  \nd  & \nd  & Im & 1f,2a,4o \\
         NGC 3440 & 32.23 $\pm$ 0.12 & 35.67 & 35.40 &  2.29 & 20.71 $\pm$ 0.12 &  \nd              & $-$18.66 $\pm$ 0.50 & 0.04 &  25.4 &  \nd  & \nd  & Sb & 1f,2a,4e \\
         NGC 3445 & 33.02 $\pm$ 0.12 & 36.26 & 36.05 &  2.03 & 21.33 $\pm$ 0.05 &  \nd              & $-$19.68 $\pm$ 0.20 & 0.02 &  27.6 &  \nd  & \nd  & Sm & 1f,2a,4e \\
         NGC 3448 & 32.54 $\pm$ 0.18 & 36.35 & 36.13 &  1.88 & 21.30 $\pm$ 0.04 &  \nd              & $-$19.26 $\pm$ 0.13 & 0.04 &  18.0 &  \nd  & \nd  & I & 1f,2a,4e \\
         NGC 3488 & 32.70 $\pm$ 0.13 & 36.24 & 35.90 &  3.75 &  \nd             &  \nd              & $-$20.05 $\pm$ 0.50 & 0.04 &  39.9 &  \nd  & \nd  & Sc & 1f,4e \\
         NGC 3556 & 32.27 $\pm$ 0.05 & 36.60 & 36.33 &  2.51 & 21.47 $\pm$ 0.04 & 0.085 $\pm$ 0.049 & $-$19.87 $\pm$ 0.10 & 0.06 &   9.3 & 33.70 & \nd  & Scd & 1f,2a,4e,5d \\
         NGC 3623 & 32.13 $\pm$ 0.20 & 36.18 & 35.80 &  5.11 & 20.30 $\pm$ 0.09 &  \nd              & $-$21.40 $\pm$ 0.05 & 0.08 &  13.5 &  \nd  & \nd  & Sa & 1f,2a,4p \\
         NGC 3646 & 33.66 $\pm$ 0.12 &  \nd  &  \nd  &  \nd  & 22.43 $\pm$ 0.04 &  \nd              & $-$22.71 $\pm$ 0.13 & 0.08 &  56.6 &  \nd  & \nd  & Sbc & 1f,2a,4e \\
         NGC 3726 & 33.05 $\pm$ 0.12 & 36.69 & 36.36 &  3.40 & 21.14 $\pm$ 0.05 & 0.085 $\pm$ 0.049 & $-$21.14 $\pm$ 0.09 & 0.05 &  20.0 & 34.24 & \nd  & Sc & 1f,2a,4h,5e \\
         NGC 3782 & 32.05 $\pm$ 0.12 & 35.20 & 34.90 &  2.85 & 20.05 $\pm$ 0.09 &  \nd              & $-$17.53 $\pm$ 0.50 & 0.06 &   9.9 &  \nd  & \nd  & Scd & 1f,2a,4e \\
        A 1137$+$46 & 32.01 $\pm$ 0.12 & 34.79 & 34.51 &  2.50 &  \nd             &  \nd              & $-$17.76 $\pm$ 0.50 & 0.09 &  11.3 &  \nd  & \nd  & Sm & 1f,4e \\
         NGC 3811 & 32.83 $\pm$ 0.12 & 36.73 & 36.48 &  2.30 & 21.44 $\pm$ 0.07 &  \nd              & $-$20.86 $\pm$ 0.40 & 0.06 &  41.4 &  \nd  & \nd  & Scd & 1f,2a,4e \\
         NGC 3877 & 32.33 $\pm$ 0.12 & 36.73 & 36.41 &  3.00 & 21.29 $\pm$ 0.05 &  \nd              & $-$20.59 $\pm$ 0.10 & 0.08 &  20.0 &  \nd  & \nd  & Sc & 1f,2a,4h \\
         NGC 3888 & 32.75 $\pm$ 0.06 & 36.82 & 36.56 &  2.52 & 21.58 $\pm$ 0.05 &  \nd              & $-$20.45 $\pm$ 0.15 & 0.04 &  32.1 &  \nd  & \nd  & Sc & 1f,2a,4e \\
         NGC 3893 & 33.14 $\pm$ 0.12 & 36.96 & 36.68 &  2.61 & 21.83 $\pm$ 0.04 &  \nd              & $-$21.07 $\pm$ 0.50 & 0.07 &  20.0 &  \nd  & \nd  & Sc & 1f,2a,4h \\
         NGC 3906 & 32.26 $\pm$ 0.12 & 35.52 & 35.22 &  3.01 &  \nd             &  \nd              & $-$18.59 $\pm$ 0.50 & 0.08 &  20.0 &  \nd  & \nd  & Sd & 1f,4h \\
         NGC 3913 & 32.26 $\pm$ 0.12 & 35.49 & 35.20 &  2.90 &  \nd             &  \nd              & $-$18.95 $\pm$ 0.11 & 0.04 &  20.0 &  \nd  & \nd  & Sd & 1f,4h \\
         NGC 3928 & 32.07 $\pm$ 0.06 & 36.13 & 35.87 &  1.81 & 20.77 $\pm$ 0.08 &  \nd              & $-$19.02 $\pm$ 0.13 & 0.06 &  20.0 &  \nd  & \nd  & Sb & 1f,2a,4h \\
         NGC 3938 & 33.18 $\pm$ 0.12 & 36.80 & 36.49 &  2.99 & 21.49 $\pm$ 0.04 &  \nd              & $-$21.19 $\pm$ 0.10 & 0.07 &  20.0 &  \nd  & \nd  & Sc & 1f,2a,4h \\
         NGC 3949 & 32.99 $\pm$ 0.12 & 36.79 & 36.53 &  2.33 & 21.75 $\pm$ 0.04 &  \nd              & $-$20.49 $\pm$ 0.15 & 0.07 &  20.0 &  \nd  & \nd  & Sbc & 1f,2a,4h \\
         NGC 3953 & 32.70 $\pm$ 0.12 & 36.85 & 36.48 &  4.44 & 21.33 $\pm$ 0.05 &  \nd              & $-$21.53 $\pm$ 0.10 & 0.10 &  20.0 &  \nd  & \nd  & Sbc & 1f,2a,4h \\
         NGC 3972 & 32.22 $\pm$ 0.12 & 35.93 & 35.58 &  3.61 & 20.48 $\pm$ 0.12 &  \nd              & $-$19.23 $\pm$ 0.16 & 0.05 &  20.0 &  \nd  & \nd  & Sbc & 1f,2a,4h \\
         NGC 3985 & 32.34 $\pm$ 0.12 & 35.91 & 35.63 &  2.41 & 20.84 $\pm$ 0.07 &  \nd              & $-$18.99 $\pm$ 0.50 & 0.09 &  20.0 &  \nd  & \nd  & Sm & 1f,2a,4h \\
        A 1154$+$49 & 32.55 $\pm$ 0.12 & 35.74 & 35.40 &  3.79 &  \nd             &  \nd              & $-$19.50 $\pm$ 0.50 & 0.10 &  20.0 &  \nd  & \nd  & Sd & 1f,4h \\
         NGC 4010 & 32.03 $\pm$ 0.12 & 36.16 & 35.83 &  4.03 & 20.99 $\pm$ 0.06 &  \nd              & $-$19.08 $\pm$ 0.50 & 0.08 &  20.0 &  \nd  & \nd  & Sd & 1f,2a,4h \\
        A 1156$+$52 & 32.54 $\pm$ 0.12 & 35.47 & 35.10 &  4.59 &  \nd             &  \nd              & $-$18.69 $\pm$ 0.50 & 0.09 &  20.0 &  \nd  & \nd  & Scd & 1f,4h \\
         NGC 4068 & 31.55 $\pm$ 0.12 & 34.09 & 33.77 &  3.66 &  \nd             &  \nd              & $-$16.25 $\pm$ 0.50 & 0.07 &   5.2 &  \nd  & \nd  & Im & 1f,4q \\
         NGC 4189 & 33.24 $\pm$ 0.12 & 36.90 & 36.61 &  2.93 & 21.53 $\pm$ 0.06 &  \nd              & $-$21.38 $\pm$ 0.08 & 0.11 &  40.0 &  \nd  & \nd  & Scd & 1f,2a,4p \\ 
         NGC 4190 & 30.96 $\pm$ 0.12 & 33.92 & 33.60 &  3.30 & 18.89 $\pm$ 0.14 &  \nd              & $-$14.50 $\pm$ 0.31 & 0.10 &   3.5 &  \nd  & \nd  & Im & 1f,2a,4r \\ 
          IC 3061 & 32.93 $\pm$ 0.12 & 36.28 & 35.95 &  3.61 &  \nd             &  \nd              & $-$19.90 $\pm$ 0.09 & 0.12 &  47.0 &  \nd  & \nd  & Sc & 1f,4p \\ 
       A 1212$+$36B & 31.86 $\pm$ 0.12 & 34.77 & 34.55 &  1.54 &  \nd             &  \nd              & $-$17.05 $\pm$ 0.50 & 0.05 &  12.6 &  \nd  & \nd  & Sdm & 1f,4e \\ 
         NGC 4212 & 32.76 $\pm$ 0.12 & 36.78 & 36.47 &  2.61 & 21.27 $\pm$ 0.05 &  \nd              & $-$20.87 $\pm$ 0.08 & 0.11 &  24.2 &  \nd  & \nd  & Sc & 1f,2a,4p \\ 
         NGC 4217 & 32.00 $\pm$ 0.12 & 36.82 & 36.55 &  3.98 & 21.76 $\pm$ 0.04 &  \nd              & $-$20.39 $\pm$ 0.10 & 0.06 &  20.0 &  \nd  & \nd  & Sb & 1f,2a,4h \\ 
         NGC 4234 & 32.39 $\pm$ 0.12 & 35.99 & 35.69 &  2.66 & 20.37 $\pm$ 0.15 &  \nd              & $-$18.87 $\pm$ 0.20 & 0.06 &  20.0 &  \nd  & \nd  & Sm & 1f,2a,4h \\ 
         NGC 4237 & 32.72 $\pm$ 0.12 & 36.71 & 36.40 &  3.32 & 20.87 $\pm$ 0.13 &  \nd              & $-$20.98 $\pm$ 0.11 & 0.10 &  32.0 &  \nd  & \nd  & Sbc & 1f,2a,4p \\ 
         NGC 4274 & 32.12 $\pm$ 0.12 & 36.50 & 36.23 &  3.05 & 20.75 $\pm$ 0.09 &  \nd              & $-$21.32 $\pm$ 0.13 & 0.07 &  21.4 &  \nd  & \nd  & Sab & 1f,2a,4p \\ 
         NGC 4275 & 32.55 $\pm$ 0.12 & 36.34 & 36.03 &  2.32 & 20.92 $\pm$ 0.11 &  \nd              & $-$19.01 $\pm$ 0.50 & 0.07 &  30.8 &  \nd  & \nd  & S & 1f,2a,4e \\ 
         NGC 4273 & 33.15 $\pm$ 0.12 & 37.30 & 37.03 &  2.00 & 22.14 $\pm$ 0.04 &  \nd              & $-$21.11 $\pm$ 0.09 & 0.06 &  38.2 &  \nd  & \nd  & Sc & 1f,2a,4p \\ 
        NGC 4303A & 32.19 $\pm$ 0.12 & 35.20 & 34.95 &  1.89 &  \nd             &  \nd              & $-$18.04 $\pm$ 0.09 & 0.08 &  15.5 &  \nd  & \nd  & Scd & 1f,4d \\  
         NGC 4314 & 32.44 $\pm$ 0.21 & 35.84 & 35.63 &  1.88 & 20.41 $\pm$ 0.07 &  \nd              & $-$20.05 $\pm$ 0.15 & 0.08 &  12.8 &  \nd  & \nd  & Sa & 1f,2a,4e \\  
          IC 3255 & 33.31 $\pm$ 0.06 & 36.71 & 36.46 &  2.10 &  \nd             &  \nd              & $-$19.76 $\pm$ 1.00 & 0.08 &  86.4 &  \nd  & \nd  & Sbc & 1f,4e \\  
          IC 3258 & 31.64 $\pm$ 0.06 & 35.38 & 35.13 &  1.96 &  \nd             &  \nd              & $-$18.52 $\pm$ 0.13 & 0.11 &  20.0 &  \nd  & \nd  & Im & 1f,4h \\  
         NGC 4369 & 32.19 $\pm$ 0.12 & 36.12 & 35.89 &  1.96 & 20.74 $\pm$ 0.05 &  \nd              & $-$19.13 $\pm$ 0.13 & 0.08 &  13.9 &  \nd  & \nd  & Sa & 1f,2a,4e \\  
         NGC 4390 & 32.03 $\pm$ 0.06 & 35.64 & 35.36 &  2.52 &  \nd             &  \nd              & $-$18.91 $\pm$ 0.50 & 0.10 &  20.0 &  \nd  & \nd  & Sbc & 1f,4h \\  
         NGC 4393 & 32.02 $\pm$ 0.12 & 34.89 & 34.54 &  4.25 & 19.98 $\pm$ 0.10 &  \nd              & $-$18.00 $\pm$ 0.50 & 0.09 &  10.1 &  \nd  & \nd  & Sd & 1f,2e,4s \\  
         NGC 4402 & 32.19 $\pm$ 0.06 & 36.81 & 36.47 &  3.22 & 21.30 $\pm$ 0.05 &  \nd              & $-$19.98 $\pm$ 0.50 & 0.09 &  25.0 &  \nd  & \nd  & Sb & 1f,2a,4p$+$4d \\  
         NGC 4412 & 32.39 $\pm$ 0.12 & 36.18 & 35.91 &  1.87 & 20.85 $\pm$ 0.07 &  \nd              & $-$18.96 $\pm$ 0.50 & 0.06 &  20.0 &  \nd  & \nd  & Sb & 1f,2a,4h \\  
         NGC 4420 & 32.34 $\pm$ 0.12 & 35.92 & 35.68 &  2.47 &  \nd             &  \nd              & $-$18.64 $\pm$ 0.50 & 0.06 &  15.0 &  \nd  & \nd  & Sbc & 1f,4t \\  
         NGC 4457 & 31.65 $\pm$ 0.12 & 35.85 & 35.61 &  1.97 & 20.56 $\pm$ 0.05 &  \nd              & $-$19.37 $\pm$ 0.16 & 0.07 &  11.0 &  \nd  & \nd  & S0 & 1f,2a,4t \\  
         NGC 4480 & 32.76 $\pm$ 0.12 & 36.38 & 36.06 &  3.48 & 20.61 $\pm$ 0.23 &  \nd              & $-$20.30 $\pm$ 0.11 & 0.08 &  34.0 &  \nd  & \nd  & Sc & 1f,2a,4t \\  
         NGC 4490 & 32.80 $\pm$ 0.06 & 36.51 & 36.31 &  1.80 & 21.81 $\pm$ 0.04 & 0.049 $\pm$ 0.025 & $-$19.80 $\pm$ 0.06 & 0.07 &   8.0 & 34.28 & \nd  & Sd & 1f,2h,4u,5f \\  
         NGC 4525 & 31.98 $\pm$ 0.06 & 35.55 & 35.17 &  5.28 &  \nd             &  \nd              & $-$19.28 $\pm$ 0.50 & 0.07 &  20.0 &  \nd  & \nd  & Scd & 1f,4h \\  
         NGC 4568 & 32.87 $\pm$ 0.06 & 37.34 & 37.05 &  2.79 & 22.03 $\pm$ 0.04 &  \nd              & $-$21.37 $\pm$ 0.09 & 0.11 &  26.0 &  \nd  & \nd  & Sbc & 1f,2a,4p \\  
         NGC 4605 & 32.05 $\pm$ 0.04 & 35.70 & 35.46 &  2.29 & 20.43 $\pm$ 0.04 &  \nd              & $-$18.30 $\pm$ 0.09 & 0.05 &   5.2 &  \nd  & \nd  & Sc & 1f,2a,4v \\  
         NGC 4618 & 32.20 $\pm$ 0.12 & 35.48 & 35.22 &  2.65 & 20.26 $\pm$ 0.05 &  \nd              & $-$18.35 $\pm$ 0.04 & 0.07 &   6.5 &  \nd  & \nd  & Sm & 1f,2a,4d \\  
         NGC 4625 & 31.39 $\pm$ 0.12 & 35.03 & 34.72 &  2.97 & 19.65 $\pm$ 0.11 &  \nd              & $-$17.00 $\pm$ 0.04 & 0.06 &   7.2 &  \nd  & \nd  & Sm & 1f,2a,4d \\  
         NGC 4632 & 32.30 $\pm$ 0.12 & 36.06 & 35.80 &  2.90 & 20.94 $\pm$ 0.05 &  \nd              & $-$19.01 $\pm$ 0.50 & 0.08 &  14.0 &  \nd  & \nd  & Sc & 1f,2a,4d \\  
         NGC 4634 & 32.08 $\pm$ 0.12 & 36.38 & 36.12 &  2.83 & 21.20 $\pm$ 0.05 &  \nd              & $-$19.21 $\pm$ 0.09 & 0.09 &  20.0 &  \nd  & \nd  & Scd & 1f,2a,4h$+$4i \\  
         NGC 4642 & 32.58 $\pm$ 0.12 & 36.21 & 35.90 &  3.08 & 21.11 $\pm$ 0.10 &  \nd              & $-$19.66 $\pm$ 0.50 & 0.08 &  36.0 &  \nd  & \nd  & Sbc & 1f,2a,4p \\  
         NGC 4653 & 33.01 $\pm$ 0.12 & 36.26 & 35.94 &  3.44 &  \nd             &  \nd              & $-$20.57 $\pm$ 0.09 & 0.08 &  35.0 &  \nd  & \nd  & Scd & 1f,4e \\  
         NGC 4666 & 32.61 $\pm$ 0.12 & 37.30 & 37.04 &  2.22 & 22.32 $\pm$ 0.04 &  \nd              & $-$20.85 $\pm$ 0.10 & 0.08 &  20.0 &  \nd  & \nd  & Sc & 1f,2a,4d \\  
         NGC 4688 & 32.23 $\pm$ 0.12 & 35.33 & 35.07 &  2.13 &  \nd             &  \nd              & $-$18.69 $\pm$ 0.50 & 0.10 &  13.1 &  \nd  & \nd  & Scd & 1f,4e \\  
         NGC 4691 & 32.74 $\pm$ 0.12 & 36.91 & 36.63 &  1.50 &  \nd             &  \nd              & $-$20.72 $\pm$ 0.13 & 0.09 &  22.0 &  \nd  & \nd  & S0 & 1f,4w \\
         NGC 4701 & 33.19 $\pm$ 0.12 & 36.66 & 36.39 &  2.35 & 21.44 $\pm$ 0.06 &  \nd              & $-$20.44 $\pm$ 0.10 & 0.10 &  35.0 &  \nd  & \nd  & Scd & 1f,2a,4p \\
         NGC 4713 & 32.41 $\pm$ 0.12 & 35.76 & 35.55 &  2.39 & 20.39 $\pm$ 0.06 &  \nd              & $-$18.48 $\pm$ 0.11 & 0.09 &  10.5 &  \nd  & \nd  & Sd & 1f,2a,4d \\
         NGC 4747 & 32.07 $\pm$ 0.12 & 35.67 & 35.40 &  2.47 & 20.18 $\pm$ 0.11 &  \nd              & $-$18.25 $\pm$ 0.13 & 0.03 &  13.0 &  \nd  & \nd  & Scd & 1f,2a,4d \\
         NGC 4765 & 31.79 $\pm$ 0.12 & 35.39 & 35.14 &  1.94 & 20.30 $\pm$ 0.06 &  \nd              & $-$17.05 $\pm$ 0.10 & 0.13 &   9.7 &  \nd  & \nd  & S0 & 1f,2a,4e \\
         NGC 4779 & 32.92 $\pm$ 0.12 & 36.55 & 36.28 &  2.12 & 21.15 $\pm$ 0.10 &  \nd              & $-$20.36 $\pm$ 0.50 & 0.07 &  37.7 &  \nd  & \nd  & Sbc & 1f,2a,4e \\
         NGC 4808 & 32.61 $\pm$ 0.12 & 36.58 & 36.30 &  2.23 & 21.46 $\pm$ 0.04 &  \nd              & $-$19.89 $\pm$ 0.10 & 0.12 &  20.0 &  \nd  & \nd  & Scd & 1f,2a,4h \\
         NGC 4868 & 33.90 $\pm$ 0.12 & 37.27 & 37.00 &  2.64 & 22.09 $\pm$ 0.05 &  \nd              & $-$21.62 $\pm$ 0.50 & 0.05 &  62.2 &  \nd  & \nd  & Sab & 1f,2a,4e \\
         NGC 4900 & 32.84 $\pm$ 0.12 & 36.53 & 36.25 &  2.26 & 21.35 $\pm$ 0.04 &  \nd              & $-$20.21 $\pm$ 0.17 & 0.08 &  20.0 &  \nd  & \nd  & Sc & 1f,2a,4j \\
         NGC 4961 & 32.44 $\pm$ 0.12 & 36.27 & 35.99 &  2.74 & 21.09 $\pm$ 0.10 &  \nd              & $-$19.13 $\pm$ 0.23 & 0.04 &  33.8 &  \nd  & \nd  & Scd & 1f,2a,4e \\
        A 1307$+$34 & 31.68 $\pm$ 0.12 & 34.56 & 34.28 &  2.53 &  \nd             &  \nd              & $-$16.21 $\pm$ 0.50 & 0.03 &  10.8 &  \nd  & \nd  & Scd & 1f,4e \\
         NGC 5012 & 33.09 $\pm$ 0.12 & 36.72 & 36.43 &  2.93 & 21.64 $\pm$ 0.05 &  \nd              & $-$20.76 $\pm$ 0.50 & 0.05 &  34.9 &  \nd  & \nd  & Sc & 1f,2a,4e \\
         NGC 5016 & 32.59 $\pm$ 0.12 & 36.43 & 36.11 &  3.01 & 20.87 $\pm$ 0.15 &  \nd              & $-$19.85 $\pm$ 0.50 & 0.04 &  34.8 &  \nd  & \nd  & Sc & 1f,2a,4e \\
        A 1310$+$36 & 32.19 $\pm$ 0.12 & 35.03 & 34.72 &  3.16 &  \nd             &  \nd              & $-$17.43 $\pm$ 0.16 & 0.06 &  12.6 &  \nd  & \nd  & Im & 1f,4e \\
        A 1312$+$35 & 31.71 $\pm$ 0.12 & 34.85 & 34.61 &  1.73 &  \nd             &  \nd              & $-$16.60 $\pm$ 0.31 & 0.05 &  11.5 &  \nd  & \nd  & Im & 1f,4e \\
         NGC 5195 & 33.12 $\pm$ 0.12 & 36.07 & 35.79 &  1.87 & 20.87 $\pm$ 0.04 &  \nd              & $-$19.97 $\pm$ 0.07 & 0.12 &   7.6 &  \nd  & \nd  & S0 & 1f,2a,4x \\
         NGC 5320 & 33.07 $\pm$ 0.12 & 36.38 & 36.06 &  3.45 & 20.97 $\pm$ 0.12 &  \nd              & $-$20.54 $\pm$ 0.50 & 0.02 &  34.9 &  \nd  & \nd  & Sc & 1f,2a,4e \\
         NGC 5350 & 33.25 $\pm$ 0.12 & 36.63 & 36.32 &  3.95 & 21.29 $\pm$ 0.06 &  \nd              & $-$21.15 $\pm$ 0.12 & 0.04 &  30.9 &  \nd  & \nd  & Sb & 1f,2a,4e \\
         NGC 5368 & 32.97 $\pm$ 0.12 & 36.80 & 36.55 &  2.23 & 21.62 $\pm$ 0.09 &  \nd              & $-$20.70 $\pm$ 0.50 & 0.04 &  61.9 &  \nd  & \nd  & Sab & 1f,2a,4e \\
         NGC 5371 & 33.41 $\pm$ 0.12 & 37.11 & 36.75 &  3.36 & 21.93 $\pm$ 0.04 &  \nd              & $-$22.07 $\pm$ 0.14 & 0.03 &  34.0 &  \nd  & \nd  & Sbc & 1f,2a,4e \\
         NGC 5383 & 32.67 $\pm$ 0.09 & 36.82 & 36.55 &  2.52 & 21.50 $\pm$ 0.05 & 0.151 $\pm$ 0.101 & $-$21.01 $\pm$ 0.04 & 0.02 &  30.0 & 34.68 & \nd  & Sb & 1f,2a,4e,5g \\
         NGC 5474 & 32.27 $\pm$ 0.05 & 35.06 & 34.76 &  3.61 & 19.82 $\pm$ 0.08 &  \nd              & $-$18.41 $\pm$ 0.16 & 0.03 &   6.8 &  \nd  & \nd  & Scd & 1f,2a,4e \\
         NGC 5477 & 31.31 $\pm$ 0.12 & 34.29 & 34.05 &  1.76 &  \nd             &  \nd              & $-$15.46 $\pm$ 0.15 & 0.04 &   7.7 &  \nd  & \nd  & Sm & 1f,4v \\
         NGC 5486 & 32.18 $\pm$ 0.12 & 35.37 & 35.10 &  2.29 &  \nd             &  \nd              & $-$18.20 $\pm$ 0.50 & 0.06 &  18.5 &  \nd  & \nd  & Sm & 1f,4e \\
         NGC 5806 & 32.15 $\pm$ 0.12 & 36.15 & 35.87 &  2.93 & 20.78 $\pm$ 0.07 &  \nd              & $-$19.76 $\pm$ 0.22 & 0.17 &  18.1 &  \nd  & \nd  & Sb & 1f,2a,4e \\
         NGC 5874 & 32.50 $\pm$ 0.12 & 36.25 & 35.91 &  3.95 &  \nd             &  \nd              & $-$20.72 $\pm$ 0.17 & 0.07 &  41.7 &  \nd  & \nd  & Sc & 1f,4e \\
         NGC 5879 & 32.02 $\pm$ 0.12 & 35.82 & 35.54 &  3.04 & 20.50 $\pm$ 0.06 &  \nd              & $-$18.78 $\pm$ 0.10 & 0.04 &  11.6 &  \nd  & \nd  & Sbc & 1f,2a,4y \\
         NGC 5907 & 32.71 $\pm$ 0.07 & 36.69 & 36.31 &  5.21 & 21.39 $\pm$ 0.04 & 0.208 $\pm$ 0.052 & $-$20.43 $\pm$ 0.11 & 0.03 &  14.0 & 33.78 & \nd  & Sc & 1f,2a,4z,5d \\
         NGC 6207 & 32.24 $\pm$ 0.12 & 35.89 & 35.65 &  2.74 & 20.74 $\pm$ 0.05 &  \nd              & $-$18.70 $\pm$ 0.10 & 0.05 &  11.4 &  \nd  & \nd  & Sc & 1f,2a,4y \\
         NGC 6503 & 31.63 $\pm$ 0.04 & 35.33 & 35.03 &  2.85 & 19.76 $\pm$ 0.05 &  \nd              & $-$17.66 $\pm$ 0.09 & 0.10 &   3.6 &  \nd  & \nd  & Scd & 1f,2a,4y \\
         NGC 7625 & 32.15 $\pm$ 0.04 & 36.71 & 36.45 &  2.01 & 21.54 $\pm$ 0.04 &  \nd              & $-$19.68 $\pm$ 0.13 & 0.08 &  21.8 &  \nd  & \nd  & Sa & 1f,2a,4e \\
         NGC 7677 & 33.07 $\pm$ 0.12 & 37.00 & 36.74 &  1.50 & 21.66 $\pm$ 0.06 &  \nd              & $-$20.33 $\pm$ 0.13 & 0.14 &  47.4 &  \nd  & \nd  & Sbc & 1f,2a,4e \\

\enddata
\tablerefs{ {\it UV data:} 
	(1a) \citet{bell02}; (1b) \citet{uit}; (1c) Measured directly
	from the IUE spectra; (1d) \citet{gold02}; (1e) \citet{deharveng94};
	(1f) \citet{rifatto3}
	    --- {\it Radio data:}
	(2a) \citet{condon02};  (2b) \citet{c98}; (2c) \citet{becker95};
        (2d) \citet{w94}; (2e) \citet{g99ii}; (2f) \citet{g99i}; 
	(2g) \citet{hopkins02}; (2h) \citet{w92} 
	    --- {\it Optical $V$-band data:}
	(3a) \citet{calzetti95} assuming $B-V = 0.4$; (3b) \citet{maddox90}
	assuming $B-V = 0.5$; (3c) \citet{spin95} assuming $B-K = 4.5$
	(the average for the other ULIRGs); (3d) \citet{rc3} assuming
	$B-V = 0.5$; (3e) \citet{han92}; (3f) \citet{esolv}, by 
	averaging $B$ and $R$; (3g) \citet{aplc}; 
	(3h) \citet{aplc}$+$\citet{rc3}; (3i) \citet{rc3}$+$\citet{taka95};
	(3k) \citet{dev88}; (3l) \citet{mat96} converted from $I$-band;
	(3m) \citet{doublier97}
	    --- {\it Distances:}
	(4a) \citet{lee02}; (4b) \citet{calzetti94}; (4c) \citet{gold02};
	(4d) \citet{teerikorpi92}; (4e) Hubble flow, assuming 
	$H_0 = 75$\,kms$^{-1}$\,Mpc$^{-1}$; (4f) \citet{kara02}; (4g)
	Virgo Cluster and Ursa Major Cluster distances following
	\citet{shanks97} and \citet{sakai00}; (4h) \citet{nfgs}; 
	(4i) \citet{yasuda97}; (4j) \citet{tully84};
	(4k) \citet{bott84}; (4l) \citet{macri99}; (4m) \citet{hopkins02};
	(4n) \citet{tonry01}; (4o) \citet{mas99}; (4p) \citet{ekholm00};
	(4q) \citet{mak97}; (4r) \citet{tik98}; (4s) From membership in
	the Coma {\sc i} cloud at 10 Mpc; (4t) \citet{gav99}; (4u)
	\citet{clemens99}; (4v) \citet{kara96}; (4w) \citet{garcia95};
	(4x) \citet{uit}; (4y) \citet{bott86}; (4z) \citet{zepf00}
	    --- {\it \ha data:}
	(5a) \citet{calzetti95}; (5b) \citet{wu98}; (5c) \citet{kk83};
	(5d) \citet{young96}; (5e) \citet{roman90}; (5f) \citet{k87};
	(5g) \citet{sheth00}
}
\tablenotetext{a}{\citet{uit} presents all of the data except for the radio luminosity.}
\tablenotetext{b}{This galaxy has a major axis optical diameter 
$\ge 1.5${\arcmin}, so its UV data is ignored to minimize aperture bias.}
\tablenotetext{c}{\citet{uit} presents all of the data for this
	galaxy except for the UV flux and radio luminosity: the UV flux is taken from \citet{rifatto3}.}
\tablenotetext{d}{The radio data are also taken from \citet{uit} for NGC 4038/9 and NGC 253.}
\tablecomments{The IR data are taken from (in order of preference)
	\citet{rice88}, \citet{soifer89}, and \citet{moshir90}.
	The optical data are taken from the RC3 \citep{rc3} or 
	the ESO-LV catalog \citep{esolv} unless otherwise
	stated.  Thermal radio fractions are taken from 
	(in order of preference) \citet{uit} and \citet{niklas97}.  
	Note that \citet{uit} obtains thermal fractions from a variety
	of sources, with the majority from \citet{niklas97}.
	Balmer decrements are taken from \citet{uit} for normal
	galaxies (and are averages of individual \hii region Balmer
	decrements) and from \citet{calzetti94} and \citet{wu98} for 
	starbursting galaxies and ULIRGs respectively.
	}
\end{deluxetable}


\begin{thebibliography}{}

\bibitem[Adelberger \& Steidel(2000)]{adel00}
	Adelberger, K.\ L., \& Steidel, C.\ C. 2000, \apj, 544, 218 


\bibitem[Becker, White \& Helfand(1995)]{becker95}
	Becker, R.\ H., White, R.\ L., Helfand, D.\ J. 1995, \apj, 450, 559

\bibitem[Bell(2002)]{bell02}
	Bell, E.\ F. 2002, \apj, 577, 150

\bibitem[Bell \& de Jong(2000)]{papii}
	Bell, E.\ F., de Jong, R.\ S. 2000, \mnras, 312, 497

\bibitem[Bell \& Kennicutt(2001)]{uit} 
	Bell, E.\ F., \& Kennicutt Jr., R.\ C. 
	2001, \apj, 548, 681

\bibitem[Bell et al.(2002)]{lmc}
	Bell, E.\ F., Gordon, K.\ D., Kennicutt Jr., R.\ C., Zaritsky,
	D. 2002, \apj, 565, 994


\bibitem[Blain et al.(1999)]{blain99}
	Blain, A.\ W., Smail, I., Ivison, R.\ J., \& Kneib, J.-P. 1999, 
	\mnras, 302, 632

\bibitem[Boissier et al.(2001)]{boissier01}
	Boissier, S., Boselli, A., Prantzos, N., Gavazzi, G. 2001, \mnras,
	321, 733

\bibitem[Bothun, Lonsdale \& Rice(1989)]{bothun89}
	Bothun, G.\ D., Lonsdale, C.\ J., Rice, W. 1989, \apj, 341, 129

\bibitem[Bottinelli et al.(1984)]{bott84}
	Bottinelli, L., Gougenheim, L., Paturel, G., de Vaucouleurs, G.
	1984, \aaps, 56, 381

\bibitem[Bottinelli et al.(1986)]{bott86}
	Bottinelli, L., Gougenheim, L., Paturel, G., Teerikorpi, P.
	1986, \aap, 156, 157

\bibitem[Bressan et al.(2002)]{bressan02}
	Bressan, A., Silva, L., Granato, G.\ L. 2002, \aap, 392, 377

\bibitem[Buat(1992)]{buat92}
	Buat, V. 1992, \aap, 264, 444

\bibitem[Buat \& Xu(1996)]{buat96}
	Buat, V., Xu, C., 1996, \aap, 306, 61

\bibitem[Buat, Deharveng \& Donas(1989)]{buat89}
	Buat, V., Deharveng, J.\ M., Donas, J. 1989, \aap, 223, 42

\bibitem[Buat et al.(1999)]{buat99}
	Buat, V., Donas, J., Milliard, B., Xu, C. 1999, \aap, 352, 371

\bibitem[Buat et al.(2002)]{buat02}
	Buat, V., Boselli, A., Gavazzi, G., Bonfanti, C. 2002, \aap, 383, 801

\bibitem[Calzetti et al.(1994)]{calzetti94}
	Calzetti, D., Kinney, A.\ L., Storchi-Bergmann, T. 1994, \apj, 429, 582

\bibitem[Calzetti et al.(1995)]{calzetti95}
	Calzetti, D., Bohlin, R.\ C., Kinney, A.\ L., Storchi-Bergmann, T., 
	Heckman, T.\ M. 1995, \apj, 443, 136

\bibitem[Caplan \& Deharveng(1986)]{caplan86}
   Caplan, J., Deharveng, L. 1986, \aap, 155, 297

\bibitem[Charlot \& Fall(2000)]{charlot00}
	Charlot, S., Fall, S. M., 2000, \apj, 539, 718

\bibitem[Chi \& Wolfendale(1990)]{chi90}
	Chi, X., Wolfendale, A.\ W. 1990, \mnras, 245, 101

\bibitem[Clemens, Alexander \& Green(1999)]{clemens99}
	Clemens, M.\ S., Alexander, P., Green, D.\ A. 1999, \mnras, 307, 481



\bibitem[Condon(1992)]{condon92} 
	Condon, J.\ J., 1992, \araa, 30, 575





\bibitem[Condon et al.(1991)]{condon91}
	Condon, J.\ J., Anderson, M.\ L., Helou, G. 1991, \apj, 375, 95

\bibitem[Condon et al.(1998)]{c98}
	Condon, J.\ J., Cotton, W.\ D., Greisen, W., Yin, Q.\ F., 
	Perley, R.\ A., Taylor, G.\ B., Broderick, J.\ J. 1998, \aj, 115, 1693

\bibitem[Condon et al.(2002)]{condon02}
	Condon, J.\ J., Cotton, W.\ D., Broderick, J.\ J. 2002, \aj, 124, 675

\bibitem[Cox et al.(1988)]{cox88}
	Cox, M.\ J., Eales, S.\ A., Alexander, P., Fitt, A.\ J.
	1988, \mnras, 235, 1227

\bibitem[Dale et al.(2001)]{dale01}
	Dale, D.\ A., Helou, G., Neugebauer, G., Soifer, B.\ T., 
	Frayer, D.\ T., \& Condon, J.\ J. 2001, \aj, 122, 1736

\bibitem[Deharveng et al.(1994)]{deharveng94}
	Deharveng, J. M., Sasseen, T. P., Buat, V., Bowyer, S., 
	Lampton, M., Wu, X., 1994, \aap, 289, 715

\bibitem[de Jong \& Lacey(2000)]{dejong00}
	de Jong, R.\ S., Lacey, C. 2000, \apj, 545, 781

\bibitem[de Jong et al.(1985)]{dejong85}
	de Jong, T., Klein, U., Wielibinski, R., Wunderlich, E. 1985, 
	\aap, 147, L6

\bibitem[de Vaucouleurs \& Longo(1988)]{dev88} 
	de Vaucouleurs, A., Longo, G. 1988, Catalogue of visual and 
	infrared photometry of galaxies from 0.5 micrometer to 
	10 micrometer (1961--1985) (Austin, University of Texas)

\bibitem[de Vaucouleurs et al.(1991)]{rc3} 
	de Vaucouleurs, G., de Vaucouleurs, A., Corwin Jr., H. G.
	Buta, R. J., Paturel, G., Fouque, P., 1991, Third Reference
	Catalogue of Bright Galaxies, version 3.9 (New York, Springer Verlag)
	(RC3)

\bibitem[Devereux \& Eales(1989)]{devereux89}
	Devereux, N.\ A., Eales, S.\ A. 1989, \apj, 340, 708

\bibitem[Dohm-Palmer et al.(1998)]{dohm98}
	Dohm-Palmer, R.\ C., et al. 1998, \aj, 116, 1227

\bibitem[Doublier et al.(1997)]{doublier97}
	Doublier, V., Comte, G., Petrosian, A., Surace, C., Turatto, M.
	1997, \aaps, 124, 405


\bibitem[Ekholm et al.(2000)]{ekholm00}
	Ekholm, T., Lanoix, P., Teerikorpi, P., Fouqu\'e, P., Paturel, G.
	2000, \aap, 355, 835

\bibitem[Ferreras \& Silk(2001)]{ferreras01}
	Ferreras, I., Silk, J. 2001, \apj, 557, 165


\bibitem[Fitt et al.(1988)]{fitt88}
	Fitt, A.\ J., Alexander, P., Cox, M.\ J. 1988, \mnras, 233, 907

\bibitem[Flores et al.(1999)]{flores99}
	Flores, H., et al. 1999, \apj, 517, 148

\bibitem[Garcia-Barreto et al.(1995)]{garcia95}
	Garcia-Baretto, J.\ A., Franco, J., Guichard, J., Carrillo, R.
	1995, \apj, 451, 156

\bibitem[Gavazzi \& Boselli(1996)]{aplc}
	Gavazzi, G., Boselli, A. 1996, ApL\&C, 35, 1

\bibitem[Gavazzi \& Boselli(1999a)]{g99i}
	Gavazzi, G., Boselli, A. 1999a, \aap, 343, 86

\bibitem[Gavazzi \& Boselli(1999b)]{g99ii}
	Gavazzi, G., Boselli, A. 1999b, \aap, 343, 93

\bibitem[Gavazzi et al.(1999)]{gav99}
	Gavazzi, G., Boselli, A., Scodeggio, M., Pierini, D., Belsole, E.
	1999, \mnras, 304, 595

\bibitem[Goldader et al.(2002)]{gold02}
	Goldader, J.\ D., Meurer, G., Heckman, T.\ M., Seibert, M., 
	Sanders, D.\ B., Calzetti, D., Steidel, C.\ C. 2002, \apj, 
	568, 651

\bibitem[Gordon et al.(2000)]{fluxrat}
	Gordon, K.\ D., Clayton, G.\ C., Witt, A.\ N., Misselt, K.\ A. 2000, 
		\apj, 533, 236

\bibitem[Haarsma et al.(2000)]{haarsma00}
	Haarsma, D.\ B., Partridge, R.\ B., Windhorst, R.\ A., Richards, 
	E.\ A. 2000, \apj, 544, 641

\bibitem[Han(1992)]{han92}
	Han, M. 1992, \apjs, 81, 35


\bibitem[Helou \& Bicay(1993)]{helou93}
	Helou, G., \& Bicay, M.\ D. 1993, \apj, 415, 93

\bibitem[Helou et al.(1988)]{helou88}
	Helou, G., Khan, I.\ R., Malek, L., Boemher, L. 1988, \apjs, 68, 151

\bibitem[Hopkins et al.(2001)]{hopkins01}
	Hopkins, A.\ M., Connolly, A.\ J., Haarsma, D.\ B, Cram, L.\ E.
	2001, \aj, 122, 288

\bibitem[Hopkins et al.(2002)]{hopkins02}
	Hopkins, A.\ M., Schulte-Ladbeck, R.\ E., Drozdovsky, I.\ O.
	2002, \aj, 124, 862


\bibitem[Inoue(2002)]{inoue02}
	Inoue, A.\ K. 2002, \apj, 570, 97L

\bibitem[Isobe et al.(1990)]{isobe90}
        Isobe, T., Feigelson, E.\ D., Akritas, M.\ G., Babu, G.\ J.
        1990, \apj, 364, 104

\bibitem[Jansen et al.(2000)]{nfgs}
	Jansen, R.\ A., Franx, M., Fabricant, D., Caldwell, N. 2000, \apjs,
	126, 271

\bibitem[Karachentsev(2002)]{kara02}
	Karachentsev, I.\ D., et al. 2002, \aap, 385, 21

\bibitem[Karachentsev \& Makarova(1996)]{kara96}
	Karachentsev, I.\ D., Makarova, D.\ A. 1996, \aj, 111, 794

\bibitem[Kauffmann et al.(2003a)]{kauffmann02}
	Kauffmann, G., et al. 2003a, \mnras, submitted (astro-ph/0204055)

\bibitem[Kauffmann et al.(2003b)]{sd}
	Kauffmann, G., et al. 2003b, \mnras, submitted (astro-ph/0204070)

\bibitem[Kennicutt(1998)]{k98}
	Kennicutt Jr., R.\ C. 1998, \araa, 36, 189

\bibitem[Kennicutt \& Kent(1983)]{kk83}
	Kennicutt Jr., R.\ C., Kent, S.\ M., 1983, \aj, 88, 1094

\bibitem[Kennicutt et al.(1987)]{k87} 
	Kennicutt Jr., R. C., Keel, W. C., van der Hulst, 
	J. M., Hummel, E., Roettiger, K. A., 1987, \aj, 93, 1011

\bibitem[Kinney et al.(1993)]{kinney93}
	Kinney, A.\ L., Bohlin, R.\ C., Calzetti, D., Panagia, N., 
	Wyse, R.\ F.\ G. \apjs, 86, 5


\bibitem[Klein(1991)]{klein91}
	Klein, U. 1991, PASAu, 9, 253

\bibitem[Klein, Wielebinski \& Thuan(1984)]{klein84}
	Klein, U., Wielebinski, R., Thuan, T.\ X. 1984, \aap, 141, 241

\bibitem[Klein et al.(1991)]{klein91b}
	Klein, U., Weiland, H., Brinks, E. 1991, \aap, 246, 323

\bibitem[Lauberts \& Valentijn(1989)]{esolv}
	Lauberts, A., Valentijn, E. A., 1989, The Surface Photometry 
		Catalogue of the ESO-Uppsula Galaxies 
		(Garching bei M\"unchen, ESO)

\bibitem[Lee et al.(2002)]{lee02}
	Lee, M.\ G., Kim, M., Sarajendini, A., Geisler, D., Gieren, W.
	2002, \apj, 565, 959

\bibitem[Li \& Draine(2001)]{li01}
	Li, A., Draine, B.\ T. 2001, \apj, 554, 778

\bibitem[Lisenfeld et al.(1996)]{lisenfeld96}
	Lisenfeld, U., V\"olk, H.\ J., Xu, C. 1996, \aap, 314, 745

\bibitem[Lonsdale Persson \& Helou(1987)]{lonsdale87}
	Lonsdale Persson, C.\ J., Helou, G. 1987, \apj, 314, 513


\bibitem[Macri et al.(1999)]{macri99}
	Macri, L.\ M., et al. 1999, \apj, 521, 155

\bibitem[Maddox et al.(1990)]{maddox90}
	Maddox, S.\ J., Sutherland, W.\ J., Efstathiou, G., Loveday,
	J. 1990, \mnras, 243, 692

\bibitem[Makarova, Karachentsev \& Georgiev(1997)]{mak97}
	Makarova, L.\ N., Karachentsev, I.\ D., Georgiev, Ts.\ B. 1997,
	AstL, 23, 378

\bibitem[Mann et al.(2002)]{mann02}
	Mann, R.\ G., et al. 2002, \mnras, 332, 549


\bibitem[Mas-Hesse \& Kunth(1999)]{mas99}
	Mas-Hesse, J.\ M., Dunth, D. 1999, \aap, 349, 765

\bibitem[Mathewson \& Ford(1996)]{mat96}
	Mathewson, D.\ S., Ford, V.\ L. 1996, \apjs, 107, 97

\bibitem[Matthews \& Wood(2001)]{matthews}
	Matthews, L.\ D., Wood, K. 2001, \apj, 548, 150

\bibitem[McGaugh \& de Blok(1997)]{mcgaugh97}
	McGaugh, S.\ S., de Blok, W.\ J.\ G. 1997, \apj, 481, 689

\bibitem[Meurer, Heckman, \& Calzetti(1999)]{meurer99}
	Meurer, G.\ R., Heckman, T.\ M., \& Calzetti, D. 1999, \apj, 521, 64

\bibitem[Misiriotis et al.(2001)]{misiriotis01}
	Misiriotis, A., Popescu, C.\ C., Tuffs, R.\ J., Kylafis, N.\ D.
	2001, \aap, 372, 775
	
\bibitem[Moshir et al.(1990)]{moshir90} 
        Moshir, M.,  
        et al., 1990, IRAS Faint Source Catalogue, version 2.0 
        (Pasedena, IPAC)

\bibitem[Niklas(1997)]{nikradfir}
	Niklas, S. 1997, \aap, 322, 29


\bibitem[Niklas et al.(1997)]{niklas97}
	Niklas, S., Klein, U., Wielebinski, R. 1997, \aap, 322, 19

\bibitem[Pagel(1998)]{pagel98}
  Pagel, B.\ E.\ J. 1998, ``Nucleosynthesis and Chemical Evolution
     of Galaxies'' (Cambridge University Press, Cambridge)

\bibitem[Peletier \& de Grijs(1998)]{peletier98}
	Peletier, R.\ F., de Grijs, R. 1998, \mnras, 300, 3L

\bibitem[Popescu et al.(2000)]{popescu00}
	Popescu, C.\ C., Misiriotis, A., Kylafis, N.\ D., Tuffs, R.\ J., 
	Fischera, J. 2000, \aap, 362, 138

\bibitem[Popescu et al.(2002)]{popescu02}
	Popescu, C.\ C., Tuffs, R.\ J., V\"olk, H.\ J., Pierini, D.,
	Madore, B.\ F. 2002, \apj, 567, 221

\bibitem[Press et al.(1992)]{press92}
  Press W.\ H., Teukolsky S.\ A., Vetterling W.\ T.,Flannery B.\ P., 
  1992, ``Numerical Recipes in Fortran 77: The Art of Scientific Computing''
  (Cambridge University Press, Cambridge).

\bibitem[Price \& Duric(1992)]{price92}
	Price, R., Duric, N. 1992, \apj, 401, 81


\bibitem[Rice et al.(1988)]{rice88} 
        Rice, W., Lonsdale, C.\ J., Soifer, B.\ T., Neugebauer, G.,
        Kopan, E.\ L., Lloyd, L.\ A., de Jong, T., Habing, H.\ J. 1988, 
        \apjs, 68, 91

\bibitem[Rifatto, Longo \& Capaccioli(1995a)]{rifatto2}
	Rifatto, A., Longo, G., Capaccioli, M. 1995a, \aaps, 109, 341

\bibitem[Rifatto, Longo \& Capaccioli(1995b)]{rifatto3}
	Rifatto, A., Longo, G., Capaccioli, M. 1995b, \aaps, 114, 527

\bibitem[Romanishin(1990)]{roman90}
	Romanishin, W. 1990, \aj, 100, 373


\bibitem[Sakai et al.(2000)]{sakai00}
        Sakai, S., et al. 2000, \apj, 529, 698

\bibitem[Sanders \& Mirabel(1996)]{sanders96}
	Sanders, D.\ B., Mirabel, I.\ F. 1996, \araa, 34, 749

\bibitem[Sauvage \& Thuan(1992)]{sauvage92}
	Sauvage, M., Thuan T.\ X. 1992, \apj, 396, 69L

\bibitem[Schlegel et al.(1998)]{sfd} 
        Schlegel, D.\ J., Finkbeiner, D.\ P., \& Davis, M., 1998, 
		\apj, 500, 525 

\bibitem[Shanks(1997)]{shanks97}
	Shanks, T., 1997, \mnras, 290, 77L

\bibitem[Sheth et al.(2000)]{sheth00}
	Sheth, K., Regan, M.\ W., Vogel, S.\ N., Teuben, P.\ J. 2000, \apj,
	532, 221

\bibitem[Skillman, Kennicutt \& Hodge(1989)]{skh89}
  Skillman, E.\ D., Kennicutt Jr., R.\ C., Hodge, P.\ W. 1989, \apj, 347, 875 

\bibitem[Sodroski et al.(1997)]{sodroski97}
	Sodroski, T.\ J., Odegard, N., Arendt, R.\ G., Dwek, E., Weiland, 
	J.\ L., Hauser, M.\ G., Kelsall, T. 1997, \apj, 480, 173

\bibitem[Soifer et al.(1989)]{soifer89} 
        Soifer, B.\ T., Boehmer, L., Neugebauer, G., Sanders, D.\ B. 1989, 
        \aj, 98, 766 

\bibitem[Spinoglio et al.(1995)]{spin95}
	Spinoglio, L., Malkan, M.\ A., Rush, B., Carrasco, L., Recillas-Cruz,
	E. 1995, \apjs, 453, 616

\bibitem[Sullivan et al.(2000)]{sullivan00}
	Sullivan, M., Treyer, M.\ A., Ellis, R.\ S., Bridges, T.\ J., 
	Milliard, B., Donas, J. 2000, \mnras, 312, 442

\bibitem[Sullivan et al.(2001)]{sullivan01}
	Sullivan, M., Mobasher, B., Chan, B., Cram, L., Ellis, R., Treyer, M.,
	Hopkins, A. 2001, \apj, 558, 72

\bibitem[Takamiya, Kron \& Kron(1995)]{taka95}
	Takamiya, M., Kron, R.\ G., Kron, G.\ E. 1995, \aj, 110, 1083

\bibitem[Teerikorpi et al.(1992)]{teerikorpi92}
	Teerikorpi, P., Bottinelli, L., Gougenheim, L., Paturel, G.
	1992, \aap, 260, 17

\bibitem[Tikhonov \& Karachentsev(1998)]{tik98}
	Tikhonov, N.\ A., Karachentsev, I.\ D. \aaps, 128, 325

\bibitem[Tonry et al.(2001)]{tonry01}
	Tonry, J.\ L., Dressler, A., Blakeslee, J.\ P., Ajham, E.\ A.,
	Fletcher, A.\ B., Luppini, G.\ A., Metzger, M.\ R., Moore, C.\ B.
	2001, \apj, 546, 681

\bibitem[Tuffs et al.(2002)]{tuffs02}
	Tuffs, R.\ J., et al. 2002, \apjs, 139, 37

\bibitem[Tully \& Shaya(1984)]{tully84}
	Tully, R.\ B., Shaya, E.\ J. 1984, \apj, 281, 31

\bibitem[van Zee et al.(1997)]{vanzee97}
  van Zee, L., Haynes, M.\ P., Salzer, J.\ J. 1997, \aj, 114, 2479

\bibitem[Vila-Costas \& Edmunds(1992)]{ve92}
  Vila-Costas, M.\ B., Edmunds, M.\ G. 1992, \mnras, 259, 121

\bibitem[Walterbos \& Greenawalt(1996)]{walterbos96}
	Walterbos, R.\ A.\ M., Greenawalt, B. 1996, \apj, 460, 696

\bibitem[Wang \& Heckman(1996)]{wang96}
	Wang, B., Heckman, T. M., 1996, \apj, 457, 645

\bibitem[White \& Becker(1992)]{w92}
	White, R.\ L., Becker, R.\ H. 1992, \apjs, 79, 331

\bibitem[Whittet(1992)]{whittet}
	Whittet, D.\ C.\ B. 1992, Dust in the Galactic Environment
	(New York, IOP Publishing)

\bibitem[Wright et al.(1994)]{w94}
	Wright, A.\ E., Griffith, M.\ R., Burke, B.\ F., Ekers, R.\ D. 1994, 
	\apjs, 91, 111

\bibitem[Wu et al.(1998)]{wu98}
	Wu, H., Zou, Z.\ L., Xia, X.\ Y., Deng, Z.\ G. 1998, \aaps, 127, 521

\bibitem[Xu et al.(1997)]{xu97}
	Xu, C., Buat, V., Boselli, A., Gavazzi, G. 1997, \aap, 324, 32

\bibitem[Xu et al.(1994)]{xu94}
	Xu, C., Lisenfeld, U., V\"olk, H.\ J., Wunderlich, E. 1994, \aap, 
	282, 19

\bibitem[Yasuda, Fukugita \& Okamura(1997)]{yasuda97}
	Yasuda, N., Fukugita, M., Okamura, S. 1997, \apjs, 108, 417

\bibitem[Young et al.(1996)]{young96}
	Young, J.\ S., Allen, L., Kenney, J.\ D.\ P., Lesser, A., Rownd, B., 
	1996, \aj, 112, 1903

\bibitem[Yun et al.(2001)]{yun01}
	Yun, M.\ S., Reddy, N.\ A., Condon, J.\ J. 2001, \apj, 554, 803

\bibitem[Zaritsky(1999)]{zaritsky99}
	Zaritsky, D. 1999, \aj, 118, 2824

\bibitem[Zaritsky, Kennicutt \& Huchra(1994)]{zkh94}
  Zaritsky, D., Kennicutt Jr., R.\ C., Huchra, J.\ P. 1994, \apj, 420, 87

\bibitem[Zaritsky et al.(2002)]{zaritsky02}
	Zaritsky, D., Harris, J., Thompson, I.\ B., Grebel, E.\ K.,
	Massey, P. 2002, \aj, 123, 855

\bibitem[Zepf et al.(2000)]{zepf00}
	Zepf, S., Lui, M.\ C., Marleau, F.\ R., Sackett, P.\ D., Graham, J.\
	R. 2000, \aj, 119, 1701

\end{thebibliography}
\end{document}